%% file: main.tex
\documentclass[12pt]{article}

\usepackage{PRIMEarxiv}
\usepackage[a4paper,
            bindingoffset=0.2in,
            left=1in,
            right=1in,
            top=1in,
            bottom=1in,
            footskip=.25in]{geometry}
\usepackage{textcomp}
\usepackage{makecell}
\usepackage{longtable}
\usepackage[utf8]{inputenc} 
\usepackage[T1]{fontenc}    
\usepackage[colorlinks=true,citecolor=blue]{hyperref}       
\usepackage{url}            
\usepackage{booktabs}       
\usepackage{amsfonts}       
\usepackage{nicefrac}       
\usepackage{microtype}      
\usepackage{lipsum}
\usepackage{fancyhdr}       
\usepackage{graphicx}       
\usepackage[pagewise]{lineno}
\graphicspath{{media/}}     

\usepackage{amsmath}
\numberwithin{equation}{section}
\usepackage{natbib}
\newtheorem{alg}{Algorithm}
\newenvironment{customthm}[1]
  {\renewcommand\thealg{#1}\alg}
  {\endalg}
\usepackage{enumitem}
\usepackage{setspace} 
\usepackage{multirow}
\usepackage{bibunits}
\usepackage[table,xcdraw]{xcolor}
\usepackage{float}
\usepackage[draft]{pgf}

\usepackage{stmaryrd}
\DeclareMathOperator*{\argmin}{argmin}
\usepackage{multirow}

\usepackage[section]{placeins}
\usepackage{array}
\usepackage{booktabs}
\usepackage{subcaption}
\usepackage{etoolbox}
\AfterEndEnvironment{enumerate}{\vskip-\lastskip}
\usepackage[noend]{algpseudocode}
\usepackage[ruled]{algorithm}
\usepackage{chngcntr}
\usepackage[noabbrev,nameinlink]{cleveref}
\usepackage{rotating}
\usepackage{tikz}

\makeatletter
\newenvironment{breakablealgorithm}
  {
   \begin{center}
     \refstepcounter{algorithm}
     \hrule height.8pt depth0pt \kern2pt
     \renewcommand{\caption}[2][\relax]{
       {\raggedright\textbf{\fname@algorithm~\thealgorithm} ##2\par}%
       \ifx\relax##1\relax 
         \addcontentsline{loa}{algorithm}{\protect\numberline{\thealgorithm}##2}%
       \else 
         \addcontentsline{loa}{algorithm}{\protect\numberline{\thealgorithm}##1}%
       \fi
       \kern2pt\hrule\kern2pt
     }
  }{
     \kern2pt\hrule\relax
   \end{center}
  }
\makeatother

\usepackage{titlesec}
\titlespacing*{\section}{0pt}{1ex}{1ex}
\titlespacing*{\subsection}{0pt}{1ex}{1ex}
\titlespacing*{\subsubsection}{0pt}{1ex}{1ex}
\usepackage[T1]{fontenc}
\usepackage{newtxmath,newtxtext}
\usepackage{array}  
\usepackage{graphicx}  

\newcolumntype{L}[1]{>{\raggedright\arraybackslash}p{#1}}
\newcolumntype{C}[1]{>{\centering\arraybackslash}p{#1}}
\newcolumntype{R}[1]{>{\raggedleft\arraybackslash}p{#1}}

\defaultbibliography{references}
\defaultbibliographystyle{apalike}

\title{\bf{Bayesian Inference of Vector Autoregressions with Tensor Decompositions}}

\author{
Yiyong Luo\thanks{Corresponding author at Department of Statistical Science, University College London, WC1E 6BT, UK.\vskip 0.1mm \hskip 2.5mm E-mail address: yiyong.luo.20@ucl.ac.uk} 
   \And
  Jim E. Griffin\thanks{E-mail address: j.griffin@ucl.ac.uk}
 \AND{Department of Statistical Science, University College London, WC1E 6BT, UK}
}

\begin{document}
\maketitle

\begin{abstract}
Vector autoregressions (VARs) are popular model for analyzing multivariate economic time series. However, VARs can be over-parameterized if the numbers of variables and lags are moderately large. Tensor VAR, a recent solution to over-parameterization, treats the coefficient matrix as a third-order tensor and estimates the corresponding tensor decomposition to achieve parsimony. In this paper, we employ the Tensor VAR structure with a CANDECOMP/PARAFAC (CP) decomposition and conduct Bayesian inference to estimate parameters. Firstly, we determine the rank by imposing the Multiplicative Gamma Prior to the tensor margins, i.e. elements in the decomposition, and accelerate the computation with an adaptive inferential scheme. Secondly, to obtain interpretable margins, we propose an interweaving algorithm to improve the mixing of margins and identify the margins using a post-processing procedure. In an application to the US macroeconomic data, our models outperform standard VARs in point and density forecasting and yield a summary of the dynamic of the US economy.
\end{abstract}


\setlength{\belowdisplayskip}{2pt} \setlength{\belowdisplayshortskip}{2pt}
\setlength{\abovedisplayskip}{2pt} \setlength{\abovedisplayshortskip}{2pt}
\begin{bibunit}
\section{Introduction}
Vector autoregression (VAR) is a multivariate time series model that describes
the linear interrelationship of data. Since the advocacy of \cite{sims1980macroeconomics}, VAR is a widely used tool for modelling macroeconomic variables, which are known to be temporally dependent with each other. As suggested in \cite{korobilis2019adaptive}, \cite{carriero2019large}, \cite{banbura2010large} and \cite{giannone2015prior}, to name a few, applying VARs to a large set of variables is advantageous for forecasting and structural analysis. However, one must solve over-parameterization, i.e. the number of parameters is relatively high to the sample size, in order to achieve success in modelling with large VARs. Over-parameterization is especially an issue for macroeconomic data due to the low frequency of data collection.

Methodologies to solve over-parameterization in VARs can be divided into \textit{sparse}- and \textit{dense}-modelling streams, according to \cite{ng2013variable}. The sparse stream assumes that only small sets of predictors are important to model the time series of each variable. For example, \cite{hsu2008subset} proposed using the Lasso penalty \citep{tibshirani1996regression} for VARs. The dense stream relies on an opposite assumption to its sparse-modelling counterpart: all predictors could be important, but their corresponding parameters may have small magnitudes. Shrinkage priors, including the Minnesota-type priors \citep{litterman1986forecasting, doan1984forecasting} and global-local shrinkage priors \citep{huber2019adaptive, huber2019should, gruber2022forecasting} dominate the dense-modelling stream in a VAR framework. An alternative methodology in this stream, called reduced-rank VAR \citep{carriero2011forecasting}, assumes that the VAR coefficient matrix has a low rank and one can decompose this matrix to achieve parsimony. A more recent and related technique, referred to as Tensor VAR, treats the coefficient matrix as a third-order tensor and infers this tensor by its low-rank decomposition. \cite{wang2021high} was the first to introduce Tensor VAR and this technique has been developed in \cite{zhang2021bayesian} and \cite{fan2022bayesian}.

In this paper, we contribute to the dense-modelling stream by employing the Tensor VAR structure with a CANDECOMP/PARAFAC (CP) decomposition \citep{kiers2000towards} and conducting Bayesian inference to estimate parameters. The motivation of choosing this methodology to alleviate over-parameterization is fourfold. Firstly, recent work has questioned whether sparse-modelling is appropriate for macroeconomic data, e.g. see \cite{giannone2021economic} for the "illusion of sparsity". Secondly, a Tensor VAR with an appropriate choice of rank is parsimonious without imposing any penalty term or shrinkage prior (although incorporating these techniques results in further parsimony). Thirdly, Tensor VAR is a useful model for explaining macroeconomic data since its reconstruction provides insights to the economy, and elements in its tensor decomposition (usually called \textit{margins}) are interpretable as shown in \cite{wang2021high} and \cite{chen2022factor}. Lastly, tensor structures with Bayesian inference have been successfully applied in time series models apart from VARs. Related work includes time-varying networks \citep{billio2024bayesian} and Autoregressive Tensor Processes (ART) \citep{billio2023bayesian}, among others. 

Two challenges arise when making Bayesian inference in a Tensor VAR with a CP decomposition. The first challenge is about the inference of the rank, which is an important parameter in the CP decomposition because it controls the model flexibility. Unlike finding the rank in a matrix, there is no straightforward algorithm to determine the rank of a third-order tensor.
Although existing literature gives rank values of some specified tensors, see \cite{kolda2009tensor} and references therein, tensors for large VARs have relatively high dimensions, so they normally do not nest in these specified ones. To overcome this challenge, past literature proposed the multiway Dirichlet generalized double Pareto (M-DGDP) prior \citep{guhaniyogi2017bayesian} and the multiway stick breaking shrinkage prior \citep{guhaniyogi2021bayesian}, based on overfitted mixture models \citep{rousseau2011asymptotic}, to induce a low-rank structure in the CP decomposition and inferred the rank \textit{a posteriori}. Despite being a prominent method to resolve the challenge, it is computationally expensive due to the large initialization of the rank. The second challenge is to retain the interpretability of a Tensor VAR. From a Bayesian perspective, a fundamental prerequisite for a Tensor VAR to be interpretable is the convergence of margin Markov chains, but this prerequisite cannot be achieved using the traditional MCMC scheme because the indeterminacy of the CP decomposition can lead to poorly mixing MCMC algorithm and posterior distributions which are hard to interpret. One solution is to impose restrictions to margins so that they are identified \citep{zhou2013tensor}, whereas solutions in unrestricted parameter space have not been explored yet. 

We tackle the above challenges with two contributions. Our first contribution is to infer the rank using an increasing shrinkage prior. We impose the Multiplicative Gamma prior (MGP) \citep{bhattacharya2011sparse} to the margins and use an adaptive inferential scheme to infer these margins, and subsequently the rank. This idea is closely related to the recent work in \cite{fan2022bayesian}, but our prior and the criterion in the adaptive inference are different from theirs.
In our second contribution, we improve the mixing of the MCMC algorithm by introducing a Gibbs sampler including a variant of the Ancillarity-Sufficiency Interweaving Strategy (ASIS) \citep{yu2011center} with three interweaving steps, inspired by the ASIS algorithm for factor models \citep{kastner2017efficient}. Unlike previous methods for tensors, dividing the margins into three blocks during inference reduces the dependence between the margins in the MCMC output. Even if the mixing of margins is not essential in some instances, e.g. one does not interpret margins and only regards the mixing of entries in tensor itself as important, this contribution is still beneficial because achieving good mixing of margins provides a solid foundation for entries in the VAR coefficient matrix to mix well. Additionally, we introduce a post-processing procedure aimed at identifying the margins.

We examine the utility of Tensor VARs through two US macroeconomic data sets with medium and large sizes. We consider two specifications of Tensor VARs that treat the coefficient matrix in two ways: (1) the matricization of a third-order tensor; and (2) a sum of the matricization of a third-order tensor
and a matrix with only non-zero entries for own lags. The first one corresponds to the original Tensor VAR idea \citep{wang2021high}, and the second one accommodates the main feature of Minnesota-type priors, i.e. the own lags of a
variable are more informative than lags of other dependent variables. In point and density forecasting, these two Tensor VARs obtain the best results for joint forecasts and are competitive to standard VARs with a range of standard prior choices. We demonstrate how to interpret margins by applying our model to the whole large-scale data and constructing factors as linear combinations of lagged data. The Tensor VAR can effectively reduce the number of parameters, and the factors constructed can summarize the dynamics of the data set. The additional own-lag matrix in the second Tensor VAR structure introduces more parameters but allows the tensor to focus on exploring the cross-variable and cross-lag effects.

The paper is organized as follows. Section \ref{Tensor VAR} explains the Tensor VAR and its interpretation. Section \ref{Bayesian Inference} provides the MCMC schemes. Section \ref{ppp} introduces the post-processing procedure. Section \ref{sec5} shows results from simulation experiments. Section \ref{sec6} presents the forecasting performance and interpretation of Tensor VARs. Section \ref{sec7} concludes the paper. 
\section{Tensor VAR} \label{Tensor VAR}
\subsection{Model Specification} \label{Model Specification}
Let $\boldsymbol{y}_t\in\mathbb{R}^N$ be the $t$-th observation in a multivariate time series. A $P$-order VAR model, VAR($P$), describes the linear relation between $\boldsymbol{y}_t$ and its lags with coefficient matrices $\boldsymbol{A}_1,\dots,\boldsymbol{A}_P\in\mathbb{R}^{N\times N}$ by
\begin{equation}
       \boldsymbol{y}_t=\boldsymbol{A}_1\boldsymbol{y}_{t-1}+...+\boldsymbol{A}_P\boldsymbol{y}_{t-P}+\boldsymbol{\epsilon}_t=\boldsymbol{A}\boldsymbol{x}_t+\boldsymbol{\epsilon}_t, \label{standard_VAR}
\end{equation}
where $t=1\,...\,T$, $\boldsymbol{A}=(\boldsymbol{A}_1,\dots,\boldsymbol{A}_P)$ is an $N$-by-$NP$ coefficient matrix linearly connecting $\boldsymbol{y}_t$ and its lags, $\boldsymbol{x}_t=(\boldsymbol{y}^\prime_{t-1},...,\boldsymbol{y}^\prime_{t-P})^\prime\in \mathbb{R}^{NP}$. The error term $\boldsymbol\epsilon_t$ follows a multivariate normal distribution with zero mean and a time-varying covariance matrix $\boldsymbol\Omega_t$. In this paper, we factorize $\boldsymbol\Omega_t$ according to \cite{cogley2005drifts}, i.e.
$\boldsymbol\Omega_t=\boldsymbol{H}^{-1}\boldsymbol{S}_t(\boldsymbol{H}^{-1})^\prime$, where $\boldsymbol{H}^{-1}$ is a lower triangular matrix with ones as diagonal entries, and $\boldsymbol{S}_t$ is a time-varying diagonal matrix with diagonal terms ($s_{t,1},...,s_{t,N}$). 

To fit the VAR model, we must estimate the $N^2P$ parameters in $\boldsymbol{A}$ and parameters for the covariance matrix $\boldsymbol{\Omega}_t$. The number of coefficients grows quadratically as the number of time series increases, thus VARs can become easily overparameterized. We address this problem by achieving parsimony of $\boldsymbol{A}$ through tensor decomposition, in the spirit of \cite{wang2021high}. Specifically, rather than modelling $\boldsymbol{A}$ directly, we model a third-order tensor $\boldsymbol{\mathcal{A}}\in\mathbb{R}^{N\times N\times P}$, where $\boldsymbol{\mathcal{A}}_{i_1,i_2,p}$ corresponds to the ($i_1,\,i_2$) entry in $\boldsymbol{A}_p$. The model in \labelcref{standard_VAR} can be written in term of the tensor $\boldsymbol{\mathcal{A}}$ to give 
\begin{align}
\boldsymbol{y}_t=\boldsymbol{\mathcal{A}}_{(1)}\boldsymbol{x}_t+\boldsymbol{\epsilon}_t, \label{tensorVAR}
\end{align}
where $\boldsymbol{\mathcal{A}}_{(1)}=\boldsymbol{A}$ is the mode-1 matricization of $\boldsymbol{\mathcal{A}}$, with the $i_1$-th row as the vectorization of $\boldsymbol{\mathcal{A}}_{(i_1,\cdot,\cdot)}$.

So far, the number of entries in $\boldsymbol{\mathcal{A}}$ is the same as that in $\boldsymbol{A}$, but 
we can decompose $\boldsymbol{\mathcal{A}}$ via a rank-$R$ CP decomposition,
\begin{equation}
\boldsymbol{\mathcal{A}}=\sum_{r=1}^R\boldsymbol{\mathcal{A}}^{(r)}=\sum_{r=1}^R\boldsymbol\beta^{(r)}_1\circ\boldsymbol\beta^{(r)}_3\circ\boldsymbol\beta^{(r)}_3,  \label{CP decomp tensor var}
\end{equation}
where $\boldsymbol{\mathcal{A}}^{(r)}$ is a third-order tensor with the same dimension as $\boldsymbol{\mathcal{A}}$, for $r=1,\dots,R$; $\boldsymbol{\beta}^{(r)}_1$, $\boldsymbol{\beta}^{(r)}_2\in\mathbb{R}^{N}$ and $\boldsymbol{\beta}^{(r)}_3\in\mathbb{R}^{P}$ are called margins of $\boldsymbol{\mathcal{A}}$; $\boldsymbol{\mathcal{A}}^{(r)}=\boldsymbol\beta^{(r)}_1\circ\boldsymbol\beta^{(r)}_3\circ\boldsymbol\beta^{(r)}_3$ is an outer product of three vectors such that the $(i_1,i_2,i_3)$ entry in $\boldsymbol{\mathcal{A}}^{(r)}$ equals to $\boldsymbol\beta^{(r)}_{1,i_1}\boldsymbol\beta^{(r)}_{2,i_2}\boldsymbol\beta^{(r)}_{3,i_3}$ for $i_1,i_2=1,\dots,N$ and $i_3=1,\dots,P$ (the definition of outer product can be found in Appendix \ref{Basic Notations and Operations}). We define the notation $\boldsymbol{B}_j=(\boldsymbol\beta^{(1)}_j,\cdots,\boldsymbol\beta^{(R)}_j)\in\mathbb{R}^{I_j\times R}$, for $j=1,2,3$, $I_1=I_2=N$ and $I_3=P$, then the tensor $\boldsymbol{\mathcal{A}}$ decomposed by $\boldsymbol{B}_1,\boldsymbol{B}_2,\boldsymbol{B}_3$ is written as $\boldsymbol{\mathcal{A}}=\llbracket \boldsymbol{B}_1,\boldsymbol{B}_2,\boldsymbol{B}_3\rrbracket_{\text{CP}}$, for the sake of brevity. Another useful representation of the margins is $\boldsymbol{B}=(\boldsymbol{B}^\prime_1,\boldsymbol{B}^\prime_2, \boldsymbol{B}^\prime_3)^\prime\in\mathbb{R}^{(2N+P)\times R}$ to which we refer as a \textit{tensor matrix}, then $\boldsymbol{\mathcal{A}}^{(r)}$ is constructed by margins in the $r$-th column of $\boldsymbol{B}$. With an upper bound $N^2P/(2N\texttt{+}P)$ of $R$, the number of parameters reduces from $N^2P$ in the coefficient matrix to $(2N\texttt{+}P)R$ in $\boldsymbol{B}$, so a low-rank structure in the CP decomposition alleviates over-parameterization. 

The CP decomposition is only identified up to scaling and permutation because $\boldsymbol{\mathcal{A}}=\llbracket\boldsymbol{B}_1,\, \boldsymbol{B}_2,\, \boldsymbol{B}_3\rrbracket_\text{CP}=\llbracket\boldsymbol{\tilde{B}}_1,\, \boldsymbol{\tilde{B}}_2,\, \boldsymbol{\tilde{B}}_3\rrbracket_\text{CP}$, if $\boldsymbol{\tilde{B}}_j$ comes from the following transformations for $j=1,2,3$:
\begin{enumerate}[noitemsep,topsep=0pt]
    \item Scaling: $\tilde{\boldsymbol{B}}_j=\boldsymbol{B}_j\boldsymbol{R}_j$, and $\boldsymbol{R}_j$ is an $R$-by-$R$ diagonal matrix satisfying $\prod_{j=1}^{J}\boldsymbol{R}_{j,(r,r)}=1$ for $r=1,\,\dots,\, R$, where $\boldsymbol{R}_{j,(r,r)}$ is the $r$-th diagonal term in $\boldsymbol{R}_{j}$.
    \item Permutation: $\tilde{\boldsymbol{B}}_j=\boldsymbol{B}_j\boldsymbol\Pi$ for an arbitrary $R$-by-$R$ column-wise permutation matrix $\boldsymbol\Pi$.
\end{enumerate}
This indeterminacy will play an important role in our algorithm in Section \ref{Interweaving Strategy}. To interpret the margins, we will identify them using a post-processing procedure described in Section \ref{ppp}.

The model in \labelcref{tensorVAR} represents the original Tensor VAR \citep{wang2021high}, which does not distinguish between the own-lag and cross-lag effects. In Section \ref{effect of D}, we empirically find that introducing this distinction allows us to achieve better forecasting performance and interpretability, so we build an extension of \labelcref{tensorVAR}, called Own-lag Tensor VAR, following the assumption of the Minnesota-type priors - the own-lag effects are more powerful than the cross-lag effects. In particular, we add a matrix $\boldsymbol{D}$, the concatenation of $P$ $N$-by-$N$ diagonal matrices, to give
\begin{equation}
\boldsymbol{y}_t=\boldsymbol{\mathcal{A}}_{(1)}\boldsymbol{x}_t+\boldsymbol{D}\boldsymbol{x}_t+\boldsymbol{\epsilon}_t,
\label{tensorVARdiag}
\end{equation}
so $\boldsymbol{D}$ can only affect entries corresponding to own lags.
\subsection{Model Interpretation} \label{Model Interpretation}
The Tensor VAR connects $\boldsymbol{y}^*_t=\boldsymbol{y}_t-\boldsymbol{D}\boldsymbol{x}_t$\footnote[1]{We include $\boldsymbol{D}$ for completion. $\boldsymbol{D}$ is a zero matrix if we apply Equation \labelcref{tensorVAR}.} with past information through $\boldsymbol{B}_1,\boldsymbol{B}_2,\boldsymbol{B}_3$ in the following reconstruction:
\allowdisplaybreaks
\begin{equation}
\boldsymbol{y}^*_t=\boldsymbol{B}_1\boldsymbol{\mathcal{I}}_{(1)}\text{vec}(\boldsymbol{B^\prime}_2\boldsymbol{X}_t\boldsymbol{B}_3)+\boldsymbol{\epsilon}_t = \sum_{r=1}^R \boldsymbol{B}_{1,(\cdot,r)}\sum_{i_2=1}^{N}\sum_{i_3=1}^{P}\boldsymbol\beta^{(r)}_{2,i_2}\boldsymbol\beta^{(r)}_{3,i_3}\boldsymbol{y}_{t-i_3,i_2}+\boldsymbol{\epsilon}_t,
\label{tensorVAR_represent}
\end{equation}
where $\boldsymbol{\mathcal{I}}_{(1)}\in\mathbb{R}^{R\times R^2}$ is the mode-1 matricization of a third-order superdiagonal tensor $\boldsymbol{\mathcal{I}}$ (see Appendix \ref{Basic Notations and Operations} for a detailed description), $\boldsymbol{X}_t=(\boldsymbol{y}_{t-1},\dots,\boldsymbol{y}_{t-P})$, $\text{vec}(\cdot)$ is the vectorization operation which transforms $\boldsymbol{B^\prime}_2\boldsymbol{X}_t\boldsymbol{B}_3\in\mathbb{R}^{R\times R}$ to an $R^2$-dimensional vector, $\boldsymbol{B}_{1,(\cdot,r)}$ is the $r$-th column of $\boldsymbol{B}_1$, $\boldsymbol\beta^{(r)}_{2,i_2}$, $\boldsymbol\beta^{(r)}_{3,i_3}$ are the $(i_2,r)$ and $(i_3,r)$ entries of $\boldsymbol{B}_2$ and $\boldsymbol{B}_3$, respectively, $\boldsymbol{y}_{t-i_3,i_2}$ is the $i_2$-th entry in $\boldsymbol{y}_{t-i_3}$. 

Following \cite{wang2021high}, we can relate \labelcref{tensorVAR_represent} to a factor model \citep{stock2005implications}, where $\boldsymbol{B}_1$ is the factor loading and $\boldsymbol{\mathcal{I}}_{(1)}\text{vec}(\boldsymbol{B^\prime}_2\boldsymbol{X}_t\boldsymbol{B}_3)$ contains $R$ observable factors. Since the $i_1$-th row in $\boldsymbol{B}_1$ describes the linear relationship between $\boldsymbol{y}_{t,i_1}$ and factors, for $i_1=1,\dots,N$, we refer to $\boldsymbol{B}_1$ as "response loading". The formation of factors describes how past information is combined. We look at $\sum_{i_2=1}^{N}\sum_{i_3=1}^{P}\boldsymbol\beta^{(r)}_{2,i_2}\boldsymbol\beta^{(r)}_{3,i_3}\boldsymbol{y}_{t-i_3,i_2}$ in \labelcref{tensorVAR_represent} to understand this formation. If $\boldsymbol{\beta}^{(r)}_{2,i_2}=0$, the $r$-th factor will not contain information from any lagged values of $\boldsymbol{y}_{t,i_2}$. Similarly, $\boldsymbol{\beta}^{(r)}_{3,i_3}=0$ results in no information about the $i_3$-th lag of $\boldsymbol{y}_t$ in the $r$-th factor. Therefore, the $i_2$-th row of $\boldsymbol{B}_2$ contains the effect from the $i_2$-th variable to $\boldsymbol{y}_t$, and the $i_3$-th row of $\boldsymbol{B}_3$ is related to the effect from the $i_3$-th lag to $\boldsymbol{y}_t$. This interpretation was also discussed in \cite{wang2021high}, who called 
$\boldsymbol{B}_2$ and $\boldsymbol{B}_3$ "predictor loading" and "temporal loading", respectively.

Another way to explain the CP decomposition in the Tensor VAR is that it separates the lag effect from the variable-wise effect because it decomposes $\boldsymbol{A}_p$ as $\boldsymbol{A}_p=\sum_{r=1}^R(\boldsymbol\beta^{(r)}_1\circ\boldsymbol\beta^{(r)}_2)\boldsymbol\beta^{(r)}_{3,p}$.
where $\boldsymbol\beta^{(r)}_1\circ\boldsymbol\beta^{(r)}_2\in\mathbb{R}^{N\times N}$ is the outer product of the two vectors such that the $(i_1,i_2)$ entry of this resulting matrix equals to $\boldsymbol\beta^{(r)}_{1,i_1}\boldsymbol\beta^{(r)}_{2,i_2}$. The first two vectors $\boldsymbol\beta^{(r)}_1$ and $\boldsymbol\beta^{(r)}_2$ (for $r=1,\,\cdots,\,R$) do not depend on the index of $\boldsymbol{A}_p$, suggesting that all lagged coefficients matrices share these vectors. The only difference among these transition matrices reflects on the different entries in $\boldsymbol\beta^{(r)}_{3}$.


\section{Bayesian Inference} \label{Bayesian Inference}
\subsection{Prior Specification} \label{Prior Specification}
As mentioned in Section \ref{Model Specification}, we aim to impose a prior on the tensor matrix $\boldsymbol{B}$ which favours low-rank structure. A particular prior choice that meets our requirement is the MGP \citep{bhattacharya2011sparse} because it possesses the increasing shrinkage property, enabling margins with higher column index to have higher degrees of shrinkage. As a result, the rank can be lowered if some columns in $\boldsymbol{B}$ have magnitudes negligibly small. To be specific, a margin $\boldsymbol\beta^{(r)}_{j,i_j}$ (the $(i_j,r)$ entry of $\boldsymbol{B}_j$) follows the prior below for $j=1,2,3$, $r=1,\dots,R$, $i_1, i_2=1,\dots,N$ and $i_3=1,\dots,P$:
\begingroup
\allowdisplaybreaks
\begin{align*}
    \boldsymbol\beta^{(r)}_{j,i_j}\sim\mathcal{N}\left(0,\,\left(\sigma^{(r)}_{j,i_j}\right)^2\right) &,\,\left(\sigma^{(r)}_{j,i_j}\right)^2=\phi^{-1}_{(r,j,i_j)}\tau^{-1}_r,\\
    \phi_{(r,j,i_j)}\sim\text{Gamma}&\left(\nu/2,\nu/2\right), \tau_r=\prod_{l=1}^r\delta_l,\\
    \delta_1\sim\text{Gamma}\left(a_1,1\right),\, \delta_l&\sim\text{Gamma}\left(a_2,1\right),\, 1<l<R,
\end{align*}
\endgroup
where $\phi_{(r,j,i_j)}$ is a local parameter for the margin with the same index. We store all these local parameters in a matrix $\boldsymbol{\Phi}$ in which each entry corresponds to the local parameter of an entry in the tensor matrix $\boldsymbol{B}$ with the same indices. The increasing shrinkage property is induced by $\tau_r$ since $\mathop{\mathbb{E}}(\tau_r)=\prod_{l=1}^r \mathop{\mathbb{E}}(\delta_l)=a_1a^{r-1}_2$ increases with $r$, when $a_2>1$. Hyperparameter $\nu$ is set to be known, and $a_1$ and $a_2$ will be inferred with Gamma priors. \cite{durante2017note} showed that both $\mathop{\mathbb{E}}(\tau_r)$ and $\mathop{\mathbb{E}}(\tau^{-1}_r)$ increase with $r$ when $1<a_2<2$. This result means that the MGP has the increasing shrinkage property only when $a_2>2$. Thus, we set priors for $a_1$ and $a_2$ as Gamma(5,1) to have the increasing shrinkage property with a high probability. Apart from the shrinkage prior for $\boldsymbol{B}$, we follow priors in \cite{huber2019adaptive} for $\boldsymbol{H}$ and $\boldsymbol{S}_t$, see Appendix \ref{Additional Prior Setting} for details. 

In the case of \labelcref{tensorVARdiag}, we impose a normal-gamma prior defined in \cite{huber2019adaptive} to each non-zero entry in $\boldsymbol{D}$. Let $d_{i,p}$ denote the own-lag coefficient for the $p$-th lag of the $i$-th response, then its prior is written as
\begin{equation*}
    d_{i,p}\sim\mathcal{N}\left(0,\left(2/{\lambda^{2}_d}\right)\psi^{(i,p)}_d\right), \, \psi^{(i,p)}_d\sim\text{Gamma}(a_d,a_d) , \text{ for }i=1,\dots,N \text{ and } p =1 ,\dots,P.
\end{equation*}
Priors of hyperparameters are the same as those for lower triangular matrix $\boldsymbol{H}$. All the full conditionals and their derivation can be found in Appendix \ref{Bayesian Inference Appendix}.
\subsection{MCMC Scheme}
\subsubsection{An Overview of Inferential Scheme}
To illustrate the strengths of our inferential scheme, we contrast it with the widely-used inferential scheme for tensor-structured models. In the traditional scheme \citep{guhaniyogi2017bayesian,billio2023bayesian,zhang2021bayesian,fan2022bayesian}, $\boldsymbol{\beta}^{(r)}_j$ is sampled from $p\left(\boldsymbol{\beta}^{(r)}_j\mid \boldsymbol{\beta}^{(r)}_{-j}, \boldsymbol{B}_{(\cdot, -r)},\boldsymbol{y}_{1:T},\left(\boldsymbol{\sigma}^{(r)}_j\right)^2\right)$, for $r=1,\dots, R$ and $j=1,\dots,J$ ($J$ is 3 in our case), where $\boldsymbol{\beta}^{(r)}_{-j}$ contains all $\boldsymbol{\beta}^{(r)}_{j^\prime}$ with $j^\prime\neq j$, $\boldsymbol{B}_{(\cdot, -r)}$ is $\boldsymbol{B}$ discarding its $r$-th column, $\left(\boldsymbol{\sigma}^{(r)}_j\right)^2$ has all prior variance corresponding to  $\boldsymbol{\beta}^{(r)}_j$. These full conditionals are then incorporated into a usual Gibbs sampler, so each $\boldsymbol{\beta}^{(r)}_j$ sampled depends on other margins, and in turn, other margins are sampled given $\boldsymbol{\beta}^{(r)}_j$ and other parameters. The rank $R$ is fixed to a large value during the inference and can be determined to a smaller value \textit{a posteriori}. This inferential scheme neglects the convergence of margin Markov chains because authors are more interested in the tensor itself, so they pay more attention to the convergence of the tensor elements, rather than the margins. The convergence issue arises from the indeterminacy of margins, mentioned in Section \ref{Model Specification}, which leads to poor mixing of the Markov chains, consequently hindering convergence. We consider the convergence of margins as an important aspect for two reasons. First, margins in Tensor VARs are potentially interpretable, as shown in \cite{wang2021high} and \cite{chen2022factor}, and discussed in Section \ref{Model Interpretation}. Second, as the literature on Tensor VARs grows, one cannot guarantee that the Markov chains in a more complex model, e.g. including time-varying margins, still converges.  Apart from the convergence issue, it is computationally expensive to infer the rank using the traditional MCMC scheme since it assumes $R$ to be fixed during the inference. To address the issues aforementioned, we propose three modifications to our inferential framework. Two of these modifications aim to alleviate the poor mixing that contributes to the convergence issue. The third modification enhances the computational efficiency.

Firstly, we reduce the dependence between columns within $\boldsymbol{B}_j$, for $j=1,2,3$, by introducing a block sampler, which divides margins into three blocks according to the three loadings mentioned in Section \ref{Model Interpretation}. This block sampler is feasible because a Tensor VAR can be written as:
\begin{align}
    \boldsymbol{y}^*_t&=\left(\boldsymbol{x}^\prime_t\left(\boldsymbol{B}_3\otimes \boldsymbol{B}_2\right)\boldsymbol{\mathcal{I}}^\prime_{(1)}\otimes \boldsymbol{I}_N\right)\text{vec}(\boldsymbol{B}_1)+\boldsymbol{\epsilon}_t \label{tensorVAR1}\\
    &=\boldsymbol{B}_1\boldsymbol{\mathcal{I}}_{(1)}\left((\boldsymbol{B}^\prime_3\boldsymbol{X}^\prime_t)\otimes \boldsymbol{I}_R\right)\text{vec}(\boldsymbol{B}^\prime_2)+\boldsymbol{\epsilon}_t \label{tensorVAR2}\\
    &=\boldsymbol{B}_1\boldsymbol{\mathcal{I}}_{(1)}\left(\boldsymbol{I}_R\otimes(\boldsymbol{B}^\prime_2\boldsymbol{X}_t)\right)\text{vec}(\boldsymbol{B}_3)+\boldsymbol{\epsilon}_t ,\label{tensorVAR3}
\end{align}
where $\boldsymbol{\mathcal{I}}\in\mathbb{R}^{R\times R\times R}$, $\boldsymbol{I}_R\in\mathbb{R}^{R\times R}$ is an identity matrix, $\otimes$ is the Kronecker product. Therefore, margins in one loading can be sampled jointly to reduce their dependencies on each other.

Secondly, we do \textit{not} use a usual Gibbs sampler to sample loadings. Instead, we introduce a variant of ASIS, containing four different parameterizations, to reduce the parameter autocorrelation during the sampling. Given a rank value in each sample iteration, the interweaving Gibbs sampler interweaves between full conditional distributions under a base parameterization and the other three (one for each loading
).

Lastly, the rank $R$ in our case is adaptively inferred similarly to \cite{bhattacharya2011sparse} to speed up computation. In the following three subsections, we introduce the interweaving Gibbs sampler for a fixed rank in \ref{Interweaving Strategy} and the adaptive inferential scheme of the rank in \ref{Adaptive Inference of Rank}. 
\subsubsection{Interweaving Gibbs Sampler} \label{Interweaving Strategy}
In principle, we could run a standard Gibbs sampler to infer margins and other parameters, but in practice, Markov chains of margins suffer from poor mixing since these chains are highly autocorrelated. We circumvent margins with poor mixing by introducing a variant of ASIS, which unfolds its strategy from its name: sampling the same block of parameters by interweaving two sampling schemes corresponding to two data augmentations - ancillary statistic and sufficient statistic. The benefit of ASIS is that the sampling will be at least as good as the sampling from only one data augmentation; and a low correlation between these two augmentations leads to faster convergence and better mixing, compared to using either augmentation alone. Because of these benefits, ASIS has been applied to many models, including stochastic volatility \citep{kastner2014ancillarity} and factor models \citep{kastner2017efficient}.
 
 Our ASIS parameterizations are more related to those in \cite{kastner2017efficient} for sampling factor loadings and factors, due to the tensor structure. The tensor structure in the Tensor VAR leads to four parameterizations instead of two in \cite{kastner2017efficient}. The first parameterization, which we call the base one, is simply $\boldsymbol{B}_1$, $\boldsymbol{B}_2$ and $\boldsymbol{B}_3$ described in Section \ref{Model Specification}. The remaining three parameterizations come from specifications of scaling indeterminacy. In particular, 
$\boldsymbol{\mathcal{A}}=\llbracket\boldsymbol{B}_1,\, \boldsymbol{B}_2,\, \boldsymbol{B}_3\rrbracket_\text{CP}=\llbracket\boldsymbol{B}^{*}_1,\, \boldsymbol{B}^{*}_2,\, \boldsymbol{B}_3\rrbracket_\text{CP}$ when $\boldsymbol{B}^{*}_1$, $\boldsymbol{B}^{*}_2$ are transformed from 
 \begin{equation}
    \boldsymbol{B}^*_1=\boldsymbol{B}_1\boldsymbol{D}_1^{-1},\boldsymbol{B}^*_2=\boldsymbol{B}_2\boldsymbol{D}_1,\label{para1}
\end{equation}
where $\boldsymbol{D}_1$ is a diagonal matrix with non-zero, non-infinite diagonal entries.

There are infinite choices of $\boldsymbol{D}_1$ to get this equivalence, but since our objective is boosting the mixing of margins, we restrict $\boldsymbol{D}_1$ to be related to $\boldsymbol{B}_1$ and $\boldsymbol{B}_2$. We choose $\boldsymbol{D}_1=\text{diag}\left(\boldsymbol\beta^{(1)}_{1,1},\dots,\boldsymbol\beta^{(R)}_{1,1}\right)$ for further demonstration. This choice constrains the first row of $\boldsymbol{B}^*_1$ to be ones. Other choices of $\boldsymbol{D}_1$ will be investigated in future work. After the transformation, we are able to write the model in terms of $\boldsymbol{B}^*_1$, $\boldsymbol{B}^*_2$ and $\boldsymbol{D}_1$ for the second parameterization. For $i_1,\, i_2=1,\dots,N$, we have
\begin{equation}
    \boldsymbol\beta^{*(r)}_{1,1}=1 ,\,\boldsymbol\beta^{*(r)}_{1,i_1}\sim\mathcal{N}
\left(0,{\left(\frac{\sigma^{(r)}_{1,i_1}}{\boldsymbol\beta^{(r)}_{1,1}}\right)}^2\right),\, \boldsymbol\beta^{*(r)}_{2,i_2}\sim\mathcal{N}\left(0,\left(\sigma^{(r)}_{2,i_2}\boldsymbol\beta^{(r)}_{1,1}\right)^2\right) .\label{parameterization2}
\end{equation}
The above parameterization only improves the mixing of margins in $\boldsymbol{B}_1$ and $\boldsymbol{B}_2$, so we also need a parameterization to improve the mixing of margins in $\boldsymbol{B}_3$. An obvious choice is to pair $\boldsymbol{B}_2$ and $\boldsymbol{B}_3$. At this point, each $\boldsymbol{B}_j$ has been paired at least once, but we conjecture that an additional pair of $\boldsymbol{B}_1$ and $\boldsymbol{B}_3$ would provide better mixing than just considering three parameterizations because the mixing would be improved across margins in each pair of $\boldsymbol{B}_j$'s. Transformations of these two pairs are similar to the one for  $\boldsymbol{B}_1$ and $\boldsymbol{B}_2$,
\begin{equation}
\boldsymbol{B}^{**}_2=\boldsymbol{B}_2\boldsymbol{D}_2^{-1},\boldsymbol{B}^{**}_3=\boldsymbol{B}_3\boldsymbol{D}_2; \,
\boldsymbol{B}^{***}_3=\boldsymbol{B}_3\boldsymbol{D}_3^{-1},\boldsymbol{B}^{***}_1=\boldsymbol{B}_1\boldsymbol{D}_3,
\end{equation}
where $\boldsymbol{D}_2$ and $\boldsymbol{D}_3$ are diagonal matrices with non-zero, non-infinite diagonal entries.

Similarly, we choose the diagonal entries in $\boldsymbol{D}_2$ to be the first row of $\boldsymbol{B}_2$, and likewise for those in $\boldsymbol{D}_3$ (as the first row of $\boldsymbol{B}_3$). These lead to the last two parameterizations which are presented in terms of $\boldsymbol{B}^{**}_2$, $\boldsymbol{B}^{**}_3$, $\boldsymbol{D}_2$ and $\boldsymbol{B}^{***}_3$, $\boldsymbol{B}^{***}_1$, $\boldsymbol{D}_3$, respectively. For $i_1,\, i_2=1,\dots,N$, $i_3=1,\dots,P$, we have
\begin{align}
    \boldsymbol\beta^{**(r)}_{2,1}&=1,\,\boldsymbol\beta^{**(r)}_{2,i_2}\sim\mathcal{N}\left(0,\left(\frac{\sigma^{(r)}_{2,i_2}}{\boldsymbol\beta^{(r)}_{2,1}}\right)^2\right),\, \boldsymbol\beta^{**(r)}_{3,i_3}\sim\mathcal{N}\left(0,\left(\sigma^{(r)}_{3,i_3}\boldsymbol\beta^{(r)}_{2,1}\right)^2\right)\label{parameterization3},\\
    \boldsymbol\beta^{***(r)}_{3,1}&=1,\,\boldsymbol\beta^{***(r)}_{3,i_3}\sim\mathcal{N}\left(0,\left(\frac{\sigma^{(r)}_{3,i_3}}{\boldsymbol\beta^{(r)}_{3,1}}\right)^2\right),\, \boldsymbol\beta^{***(r)}_{1,i_1}\sim\mathcal{N}\left(0,\left(\sigma^{(r)}_{1,i_1}\boldsymbol\beta^{(r)}_{3,1}\right)^2\right) \label{parameterization4}.
\end{align}
We need to sample margins under the four parameterizations described in each iteration. The sampling using the base parameterization is stated in Appendix \ref{full conditionals of margins}, so we focus on sampling margins under the other three parameterizations introduced in this subsection. For $\boldsymbol\beta^{(r)}_{1,1}$, its normal prior implies that $\left(\boldsymbol\beta^{(r)}_{1,1}\right)^2$ has a gamma prior, Gamma$\left(\frac{1}{2},\frac{1}{2\left(\sigma^{(r)}_{1,1}\right)^2}\right)$. The full conditional of $\left(\boldsymbol\beta^{(r)}_{1,1}\right)^2$ under \labelcref{parameterization2} is a Generalized Inverse Gaussian (GIG),
\begin{align}\label{gig1}
    \left(\boldsymbol\beta^{(r)}_{1,1}\right)^2\mid \boldsymbol{B}^*_{1,(\cdot,r)},\, \boldsymbol{B}^*_{2,(\cdot,r)}\sim \text{GIG}\left(0,\sum_{i_2=1}^M\left(\frac{\beta^{*(r)}_{2,i_2}}{\sigma^{(r)}_{2,i_2}}\right)^2,\sum_{i_1=2}^M\left(\frac{\beta^{*(r)}_{1,i_1}}{\sigma^{(r)}_{1,i_1}}\right)^2+\left(\frac{1}{\sigma^{(r)}_{1,1}}\right)^2\right),
\end{align}
where a variable $x\sim$ GIG$\left(\lambda,\chi,\psi\right)$ has probability density function $p(x)\propto x^{\lambda-1}\exp \left(-\left(\chi/x+\psi x\right)/2\right)$. Similarly, we can get full conditionals of $\left(\boldsymbol\beta^{(r)}_{2,1}\right)^2$ under \labelcref{parameterization3} and $\left(\boldsymbol\beta^{(r)}_{3,1}\right)^2$ under \labelcref{parameterization4}:
\begin{align}
\left(\boldsymbol\beta^{(r)}_{2,1}\right)^2&\mid \boldsymbol{B}^{**}_{2,(\cdot,r)},\, \boldsymbol{B}^{**}_{3,(\cdot,r)}\sim \text{GIG}\left(\frac{M-P}{2},\sum_{i_3=1}^P\left(\frac{\beta^{**(r)}_{3,i_3}}{\sigma^{(r)}_{3,i_3}}\right)^2,\sum_{i_2=2}^M\left(\frac{\beta^{**(r)}_{2,i_2}}{\sigma^{(r)}_{2,i_2}}\right)^2+\left(\frac{1}{\sigma^{(r)}_{2,1}}\right)^2\right),\label{gig2}\\
   \left(\boldsymbol\beta^{(r)}_{3,1}\right)^2&\mid \boldsymbol{B}^{***}_{3,(\cdot,r)},\, \boldsymbol{B}^{***}_{1,(\cdot,r)}\sim \text{GIG}\left(\frac{P-M}{2},\sum_{i_1=1}^M\left(\frac{\beta^{***(r)}_{1,i_1}}{\sigma^{(r)}_{1,i_1}}\right)^2,\sum_{i_3=2}^P\left(\frac{\beta^{***(r)}_{3,i_3}}{\sigma^{(r)}_{3,i_3}}\right)^2+\left(\frac{1}{\sigma^{(r)}_{3,1}}\right)^2\right)\label{gig3}.
\end{align} 

Algorithm \ref{alg:2} outlines how to interweave sampling under the base parameterization to the second one described in \labelcref{para1}. Similar algorithms can be applied to the third and fourth parameterizations, incorporating with full conditionals in \labelcref{gig2} and \labelcref{gig3}. Combining these three algorithms leads to a Gibbs sampler of which the full algorithm can be found in Appendix \ref{Full Gibbs Sampler}. If we only sample margins using Step (a), the algorithm is just a standard Gibbs sampler with the base parameterization. Every interweaving step starts at the base parameterization, then switches to an alternative parameterization and swaps back to the base one. Note that $\boldsymbol{B}_2$'s in this algorithm has superscript $\tilde{\text{new}}$. This is because $\boldsymbol{B}_2$ is included in two interweaving steps, but we only store one sample for $\boldsymbol{B}_2$ in each iteration. It will be easier to distinguish between the one stored (with superscript "new") and the one left (with superscript $\tilde{\text{new}}$). One can find the same superscripts in the full algorithm.  

It is worth stressing that the interweaving strategy improves the mixing of entries in $\boldsymbol{B}$ up to column permutations and sign-switching issues. Thus, we also propose a post-processing procedure to identify the margins \textit{a posteriori} in Section \ref{ppp}.

\begin{customthm}{1}\label{alg:2}
\normalfont Interweave between the base parameterization and the one in \labelcref{para1}.
\vspace{-0.3cm}
\begin{enumerate}[label=(\alph*),align=left,leftmargin=*,noitemsep]
\item [Step (a):] Update $\boldsymbol{B}^{\text{old}}_1$ under the base parameterization.
\item[Step (b):] Store the first row of $\boldsymbol{B}^{\text{old}}_1$ into $\boldsymbol{D}_1$ and determine $\boldsymbol{B}^{*}_1$ and $\boldsymbol{B}^{*}_2$.
\item[Step (c):] Sample $\left(\boldsymbol{\beta}^{\text{new}(r)}_{1,1}\right)^2$ for $r=1,\dots,R$ using the corresponding full conditional in \labelcref{gig1} and store sampled values into $\boldsymbol{D}_1$.
\item[Step (d):] Update $\boldsymbol{B}^{\text{new}}_1$ and $\boldsymbol{B}^{\tilde{\text{new}}}_2$ with transformation
$\boldsymbol{B}^{\text{new}}_1= \boldsymbol{B}^{*}_1\boldsymbol{D}_1,\,  \boldsymbol{B}^{\tilde{\text{new}}}_2= \boldsymbol{B}^{*}_2\boldsymbol{D}^{-1}_1.$
\end{enumerate}
\end{customthm}


\subsubsection{Adaptive Inference of Rank} \label{Adaptive Inference of Rank}
We aim to infer the rank by finding inactive columns in $\boldsymbol{B}$, i.e. those columns which do not contribute much to the tensor $\boldsymbol{\mathcal{A}}$. An adaptive algorithm, inspired by \cite{bhattacharya2011sparse} and \cite{legramanti2020bayesian}, is displayed in Algorithm \ref{alg:1}.

In this algorithm, we initialize the rank as $R^* = \lceil5\log{N}\rceil$, which is the same as for the number of factors in \cite{bhattacharya2011sparse}. Empirically, this initialization is large enough to estimate the coefficient matrix. In order to meet diminishing adaptation condition \citep{roberts2007coupling} for the weak law of large number in adaptive MCMC, we discard inactive columns in the $m$-th iteration with probability $p(m)=\text{exp}(\alpha_0+\alpha_1 m)$, where $\alpha_0\leq 0$, $\alpha_1<0$. Since $p(m)$ is getting smaller as $m$ increases, $R$ is less likely to change during the inference. Lastly, we need to set a criterion to decide whether a column in $\boldsymbol{B}$ is active or not. In this paper, this criterion is related to the proportion of small magnitudes in $\boldsymbol{\mathcal{A}}^{(r)}$, for $r=1,\dots,R$. For ease of explanation, we omit $m$ here. We regard an entry in $\boldsymbol{\mathcal{A}}^{(r)}$ to have a small magnitude if its absolute value is smaller than a threshold $\gamma_1$, e.g. $\gamma_1=10^{-3}$. If the proportion of small magnitudes in $\boldsymbol{\mathcal{A}}^{(r)}$ is larger than another threshold $\gamma_2$ set \textit{a-priori}, e.g. $\gamma_2=0.9$, then we regard the $r$-th column in $\boldsymbol{B}$ as inactive. We use the simulation study to determine
$\gamma_1$ and $\gamma_2$ so as to minimize the rank inferred, while simultaneously ensuring accurate inference of the coefficient matrix. More discussion and details about choosing $\gamma_1$ and $\gamma_2$ are available in Appendix \ref{Additional Results Simulation}.

Adaptive inference begins after the $\tilde{m}$-th iteration to stabilize Markov chains and stops at the last iteration during the burn-in period to allow easy interpretation of margins. If the number of inactive columns is greater than 0, we remove these columns in $\boldsymbol{B}$ and remove corresponding parameters in $\boldsymbol\Phi$, $\boldsymbol\delta=(\delta_1,\dots,\delta_{R})$, $\boldsymbol\tau=(\tau_1,\dots,\tau_{R})$. The rank will then be shrunk to a smaller number of active columns. If the algorithm does not detect any inactive column, we first sample a new column in $\boldsymbol\Phi$, a new entry in $\boldsymbol\delta$ and subsequently compute the new entry in $\boldsymbol\tau$. A new column in $\boldsymbol{B}$ will also be sampled using these newly-sampled hyperparameters.

\section{Post-Processing Procedure}\label{ppp}
The interweaving algorithm allows Markov chains to improve mixing, but it does not completely solve the indeterminacy of tensor decomposition, which is the origin of non-convergence of Markov chains. Therefore, we propose a post-processing procedure to identify margins \textit{a posteriori}. Note that there exists methods to identify margins \textit{a priori}. For example, \cite{zhou2013tensor} restricted $\boldsymbol{B}_{1,(1,\dot)}$ and $\boldsymbol{B}_{2,(1,\dot)}$ as ones and sorted elements in $\boldsymbol{B}_{3,(1,\dot)}$ in descending order. We opt to maintain an unrestricted tensor decomposition because it can incorporate the increasing shrinkage property of the MGP and therefore enables us to infer the rank. 

The procedure proposed is inspired by the Match-Sign-Factor (MSF) algorithm in the R package \textbf{infinitefactor} \citep{poworoznek2021efficiently}. The MSF performs a greedy search to rotate factor loadings and factors in factor models, and we apply a variant of this algorithm to Tensor VARs. Our algorithm is presented in \ref{alg:3}, along with a detailed explanation divided into two parts: (1) solve column permutations by the label-matching method (up to line 11); (2) solve sign-switching issues by the sign-matching method. 

Column permutations in $\boldsymbol{B}$ are equivalent to those in $\boldsymbol{B}_3$, so if we solve the equivalent issue in $\boldsymbol{B}_3$, we will automatically solve column permutations in $\boldsymbol{B}$. There are analogous equivalences related to $\boldsymbol{B}_1$ and $\boldsymbol{B}_2$, but the empirical finding in Figure \ref{fig: mixing insight different pivots} shows that the label matching related to $\boldsymbol{B}_3$ gives the best mixing results in the simulation study. The label matching needs a \textit{pivot} matrix $\boldsymbol{B}^{(\text{pivot})}_3$ as a template to align $\boldsymbol{B}_3$ sampled in each iteration, i.e. columns in $\boldsymbol{B}_3$ after label being matched will have the same order as that of columns in $\boldsymbol{B}^{(\text{pivot})}_3$. Following \cite{poworoznek2021efficiently}, $\boldsymbol{B}^{(\text{pivot})}_3$ is the one with the median of the condition number $\kappa=\sigma_{\text{max}}(\boldsymbol{B}_3)$, where $\sigma_{\text{max}}(\boldsymbol{B}_3)$ is the maximal singular value of $\boldsymbol{B}_3$. 

After choosing the pivot, we compute the Euclidean distance between columns in $\boldsymbol{B}_3$ in each iteration and $\left(\boldsymbol{B}^{(\text{pivot})}_3, -\boldsymbol{B}^{(\text{pivot})}_3\right)$, and store the distances into an $R$-by-$2R$ distance matrix $\boldsymbol{\Theta}$ with row and column indices corresponding to columns in $\boldsymbol{B}_3$ and $\left(\boldsymbol{B}^{(\text{pivot})}_3, -\boldsymbol{B}^{(\text{pivot})}_3\right)$, respectively. As shown in Algorithm \ref{alg:3}, a greedy algorithm then starts from the lowest Euclidean distance to align the corresponding column in $\boldsymbol{B}_3$ to that in  $\boldsymbol{B}^{(\text{pivot})}_3$ or -$\boldsymbol{B}^{(\text{pivot})}_3$, and these columns will not be matched again. The label matching is finished after repeating the procedure for $R$ times. 

Next, we explain the sign-matching method. For $j=1,2$, $r=1,\dots, R$, we determine whether to flip the sign of $\boldsymbol{B}_{j,(\cdot,r)}$ by comparing its distances to both $\boldsymbol{B}^{(\text{pivot})}_{j,(\cdot,r)}$ and -$\boldsymbol{B}^{(\text{pivot})}_{j,(\cdot,r)}$. The general guideline for flipping signs in $\boldsymbol{B}_{3,(\cdot,r)}$ is to do so only if this procedure identifies the tensor, i.e. the tensors before and after sign-matching are the same. If not, we leave the sign unflipped.

\addtocounter{algorithm}{2}
\addtocounter{algorithm}{-1}
\alglanguage{pseudocode}
\begin{breakablealgorithm}
\setstretch{1.2}
\caption{Match Labels and Signs}\label{alg:3}
\begin{algorithmic}[1]
\State Find a pivot matrix $\boldsymbol{B}^{(\text{pivot})}_3$ and its corresponding tensor matrix $\boldsymbol{B}^{(\text{pivot})}$
\For{each iteration}
\State Compute the $R$-by-$2R$ distance matrix $\boldsymbol\Theta$
\For{$r=1,\dots,R$}
\State Find $\left(r^*_1,\, r^*_2\right)=\mathop{\argmin}\limits_{r_1,\,r_2} \boldsymbol\Theta_{r_1,\,r_2}$
\If{$r^*_2\leq R$}
\State Match the $r^*_1$-th column in $\boldsymbol{B}_3$ to the $r^*_2$-th column in $\boldsymbol{B}^{(\text{pivot})}_3$.
\State Change the $r_1$-th row, $r_2$-th and $\left(R+r_2\right)$-th columns in $\boldsymbol{\Theta}$ to infinity.
\Else
\State Match the $r^*_1$-th column in $\boldsymbol{B}$ to the $\left(r^*_2-R\right)$-th column in $\boldsymbol{B}^{(\text{pivot})}_3$.
\State Change the $r_1$-th row, $(r_2-R)$-th and $r_2$-th columns in $\boldsymbol{\Theta}$ to infinity.\EndIf
\For{$j=1,2$}
\State Compute distance $d_1=d\left(\boldsymbol{B}_{j,(\cdot,r)},\boldsymbol{B}^{(\text{pivot})}_{j,(\cdot,r)}\right)$ and $d_2=d\left(\boldsymbol{B}_{j,(\cdot,r)},-\boldsymbol{B}^{(\text{pivot})}_{j,(\cdot,r)}\right)$
\If {$d_1\leq d_2$}
\State Keep signs in $\boldsymbol{B}_{j,(\cdot,r)}$. Record $\text{ind}_{j,r}=1$
\Else
\State Flip signs in $\boldsymbol{B}_{j,(\cdot,r)}$. Record $\text{ind}_{j,r}=\texttt{-}1$
\EndIf
\EndFor
\If{$\text{ind}_{1,r}\text{ind}_{2,r}=1$}
\State Keep the signs in $\boldsymbol{B}_{3,(\cdot,r)}$ 
\Else
\State Flip the signs in $\boldsymbol{B}_{3,(\cdot,r)}$
\EndIf
\EndFor
\EndFor
\end{algorithmic}
\end{breakablealgorithm}
\section{Simulation Results} \label{sec5}
\subsection{Data and Implementation}
We assess the merits of inferring ranks using the MGP and the adaptive inferential scheme in \ref{Rank Selection}, compared to the M-DGDP \citep{guhaniyogi2017bayesian} prior commonly used in tensor-structured models. Section \ref{mixing} shows that interweaving strategy can improve the mixing of margins, and the post-processing procedure identifies the margins. We will leave comparison of predictive performance to the real data example. The following two subsections use the same simulated data, which includes three scenarios with different combinations of the number of time series and rank ($N$, $R$):  $(10,3)$, $(20,5)$ and $(50,10)$. The lag order is $P=3$. We assume that the true rank increases with the number of time series. \cite{kolda2009tensor} and the reference therein summarise ranks of some specific third-order tensors, but the rank of a tensor applied in a VAR with lag order exceeding 2 is not specified. Only an upper bound of the rank is available, which is min$(N^2, NP)$.

In each scenario, we generate 25 data sets following VAR(3) models with independently generated parameters. The coefficient matrix of each model is the 1-mode matricization of a tensor from a CP decomposition, and the covariance matrix is an identity matrix. Margins of the CP decomposition follow uniform distributions with different parameters, see Table \ref{DGP} for more details. All time series are checked for stationarity via the Dickey-Fuller test and the Kwiatkowski–Phillips–Schmidt–Shin (KPSS) tests with significance level set as 5\%. All data sets are consistent with stationarity.

\begin{table}[!htb]
\vspace{-0.4cm}
\centering
\small
\begin{tabular}{llll}
\hline
          & (10,3)   & (20,5)   & (50,10)  \\
          \hline
$B_1$        & U(-1,1)     & U(-1,1)     & U(-1,1)     \\
$B_2$        & U(-1,1)     & U(-1,1)     & U(-0.6,0.6) \\
$B_{3,(1,\cdot)}$ & U(-1,1)     & U(-1,1)     & U(-0.6,0.6) \\
$B_{3,(2,\cdot)}$ & U(-0.5,0.5) & U(-0.2,0.2) & U(-0.2,0.2) \\
$B_{3,(3,\cdot)}$ & U(-0.1,0.1) & U(-0.1,0.1) & U(-0.1,0.1) \\
\hline
\end{tabular}
\caption{Uniform distributions of margins in different locations indicated by rows and different combination of $N$ and $R$ indicated by columns.}
\label{DGP}
\end{table}
\vskip -0.6cm
We apply the MGP to both simulation experiments by setting $\nu=3$ as shown in \cite{bhattacharya2011sparse}, $\gamma_1=10^{-3}$, and $\gamma_2=0.9$. A table illustrating the sensitivity to the choice of $\gamma_1$ and $\gamma_2$ is available in Appendix \ref{Additional Results Simulation}. Our chosen combination of $\gamma_1$ and $\gamma_2$ gives the most parsimonious model and the narrowest 90\% credible interval of inferred rank. Apart from the MGP, we briefly introduce the M-DGDP prior, which is a global-local shrinkage prior proposed for tensor margins with the following expression:
\begin{align*}
\boldsymbol{\beta}^{(r)}_j\sim\mathcal{N}\left(\boldsymbol{0},(\phi_r\tau)\boldsymbol{W}_{jr}\right),\;& w_{jr,k}\sim\text{Exp}\left(\lambda_{jr}/2\right), \; \lambda_{jr}\sim\text{Gamma}\left(a_\lambda,b_\lambda\right),\\
    \boldsymbol{\Phi}=\left(\phi_1,\dots,\phi_R\right)^\prime\sim &\text{ Dirichlet}\left(\alpha,\dots,\alpha\right),\; \tau\sim\text{Gamma}\left(a_\tau,b_\tau\right),
\end{align*}
where $\boldsymbol{W}_{jr}=\text{diag}\left(w_{jr,1},\dots,w_{jr,I_j}\right)$, $I_j=N$ when $j=1,2$ and $I_j=P$ when $j=3$ in our case. $\alpha$ is uniformly distributed on a grid with values equally placed on $\left[R^{-3},R^{-0.01}\right]$, and $R$ is the rank set in advance. We follow the same setting of hyperparameters as in \cite{guhaniyogi2017bayesian}, i.e. $a_\lambda=3$ and $b_\lambda=\sqrt[6]{a_\lambda}$, $a_\tau=R\alpha$, $b_\tau=\alpha \sqrt[3]R$. 

For both priors, the initialization of rank is $\lceil 5\log (N)\rceil$, but the adaptive inferential scheme is only applied when using the MGP after iteration reaches 200 in the burn-in period. For the M-DGDP, the rank is determined \textit{a posteriori} by removing negligible margins as in Algorithm \ref{alg:1}. We implement all simulations with Intel(R) Xeon(R) Gold 6140 CPU 2.30GHzr and R 4.2.0.
\subsection{Rank Selection} \label{Rank Selection}
The first simulation assesses our approach to infer the rank $R$. Both samplers with MGP and M-DGDP were run for 10,000 iterations after 10,000 burn-in and incorporated the interweaving strategy. We record the performance of MGP and M-DGDP in Table \ref{rank inference table} including four metrics: (1) mean squared error (MSE) of the coefficient matrix for coefficient accuracy; (2) averaged effective sample size (ESS) of coefficients for sampling efficiency; (3) averaged rank inferred ($R$) for rank accuracy; and (4) approximate running time for computational efficiency. 

According to Table \ref{rank inference table}, both models estimate coefficient matrices with similar accuracy under the MSE. The MGP is able to infer ranks equal to or lower than the true ones, while MDGDP can infer the true ranks after deleting redundant columns which the corresponding averaged proportions of small magnitude ($\gamma_1=10^{-3}$) is greater than $\gamma_2=0.9$. The MGP also explores coefficient posteriors more efficiently as suggested by ESS results from the first two scenarios. The adaptive shrinkage algorithm accelerates computation since the running time of the MGP grows more slowly with $N$ and $R$ in comparison to the growth rate of the M-DGDP. This lead to a large difference if $N=50$ and $R=10$ where the inference with the MGP runs more than 5 times faster than the M-DGDP.
\begin{table}[!htbp]
\vspace{-2ex}
\centering
\small
\begin{tabular}{llllll}
\hline
($N$, $R$)   & method & MSE& $R$& ESS & running time (hr)   \\
\hline
\addlinespace[0.2cm]
(10,3)   & MGP    &   0.006   &   4   &   3977.539  &  0.45  \\
\addlinespace[0.1cm]
         & M-DGDP   &   0.006 &  3   & 3938.573 &  1.16  \\
\addlinespace[0.1cm]         
(20, 5)  & MGP    &   0.008   &   4  &   2657.043  &  0.585   \\
\addlinespace[0.1cm]  
          & M-DGDP   &  0.008  &   5  &  2644.262  & 2.60 \\
\addlinespace[0.1cm]
(50, 10) & MGP    &   0.006   &   7  &   2125.425  &   2.52  \\
\addlinespace[0.1cm]
         & M-DGDP  &  0.006  &  10   & 2315.662 & 13.34   \\
         \hline
\end{tabular}
\vskip 0.5cm
\caption{Performance of MGP and M-DGDP in 25 simulations for different dimensionality combinations.}
\label{rank inference table}
\end{table}
\subsection{Quality of Markov Chains} \label{mixing}
The second simulation investigate the quality of Markov chains, i.e. whether the interweaving strategy and the post-processing procedure contribute to mixing and convergence of Markov chains. We choose three prior settings (standard normal, MGP, M-DGDP) to infer margins with/without interweaving. The burn-in period still has 10,000 iterations, but we change the number of iterations after burn-in to 100,000 to demonstrate results with longer chains. 

\begin{figure}[!htbp]
     \centering
     \begin{subfigure}[b]{0.48\textwidth}
         \centering
         \includegraphics[width=\textwidth, height=0.6\textwidth]{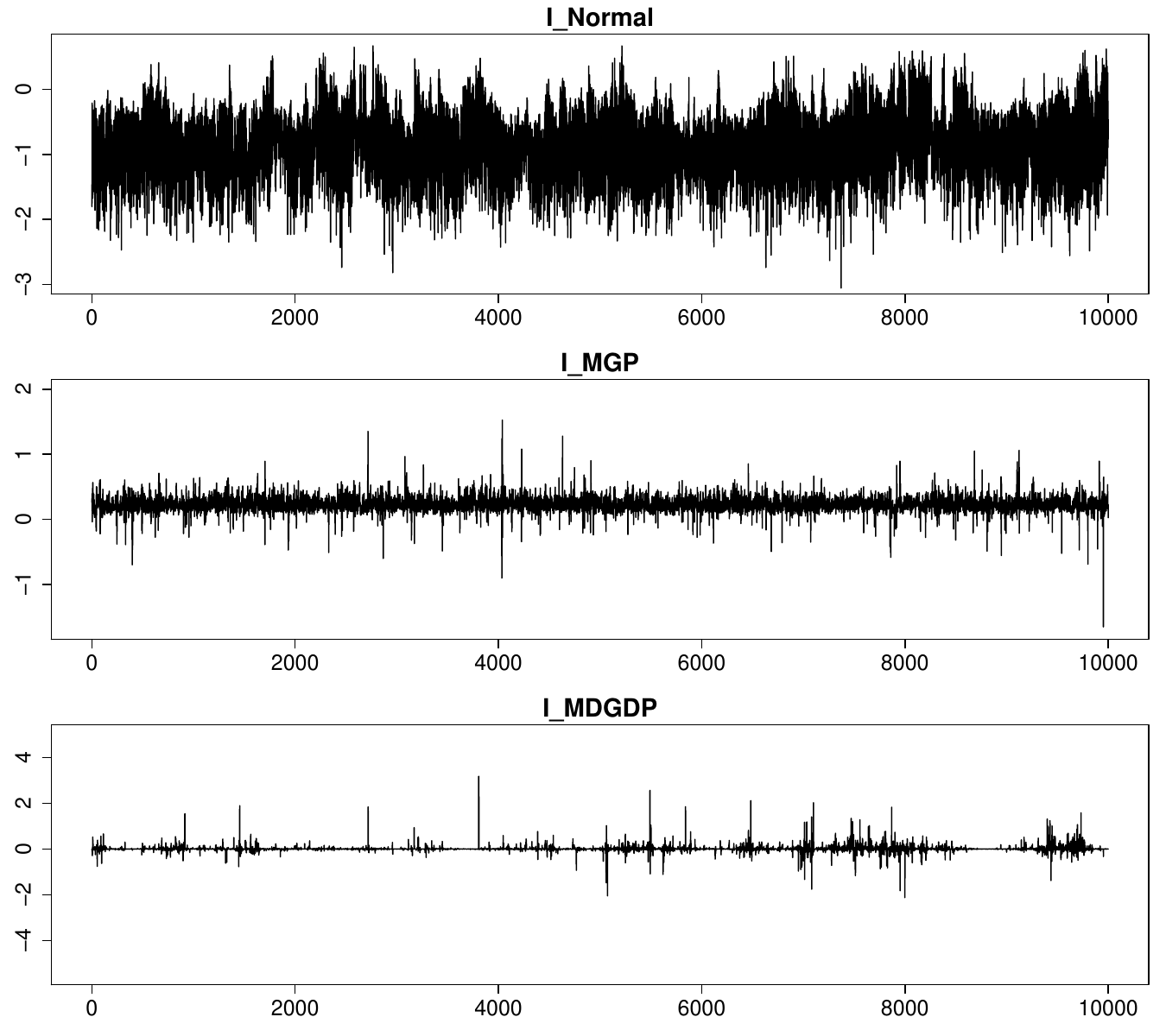}
         \caption{With interweaving}
         \label{mixing insight inweaving}
     \end{subfigure}
     \hfill
     \begin{subfigure}[b]{0.48\textwidth}
         \centering
         \includegraphics[width=\textwidth, height=0.6\textwidth]{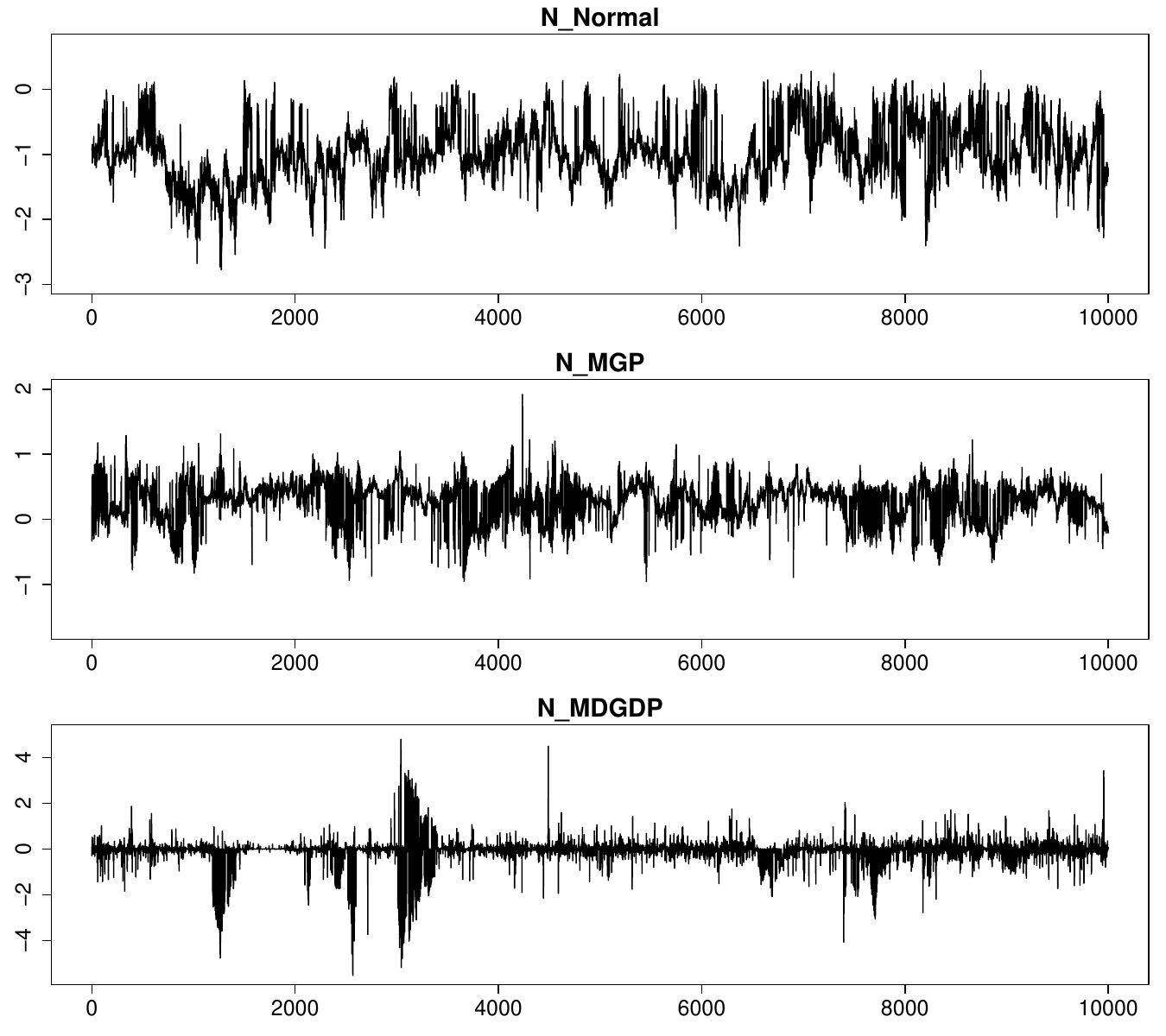}
         \caption{Without interweaving}
         \label{mixing insight no inweaving}
     \end{subfigure}
        \caption{Trace plots of the first 10,000 draws of $\boldsymbol{\beta}^{(1)}_{1,1}$ in $N=10,$ $R=3$ scenario after burn-in period. The inferential scheme adopts standard normal (top), MGP (middle) and M-DGDP (bottom) as priors and applies with (left panel) and without (right panel) interweaving strategy. }
        \label{mixing insight}
\end{figure}

We first focus on the interweaving strategy by conducting the post-processing procedure to both samples with/without interweaving. To give an insight into the effect of interweaving, Figure \ref{mixing insight} shows trace plots of the margin $\boldsymbol{\beta}^{(1)}_{1,1}$ when $N=10$ and $R=3$ based on different prior settings with/without interweaving. Even though we used the label- and sign-matching methods, trace plots without interweaving still suffer from the mixing problem, while the interweaving strategy substantially improves mixing. The autocorrelations (acfs) of all draws of $\boldsymbol{\beta}^{(1)}_{1,1}$ after the burn-in period, see Figure \ref{acf insight}, also support the merit of the interweaving strategy.

\begin{figure}[!htbp]
     \centering
     \begin{subfigure}[b]{0.32\textwidth}
         \centering
         \includegraphics[width=\textwidth,height=1.1\textwidth]{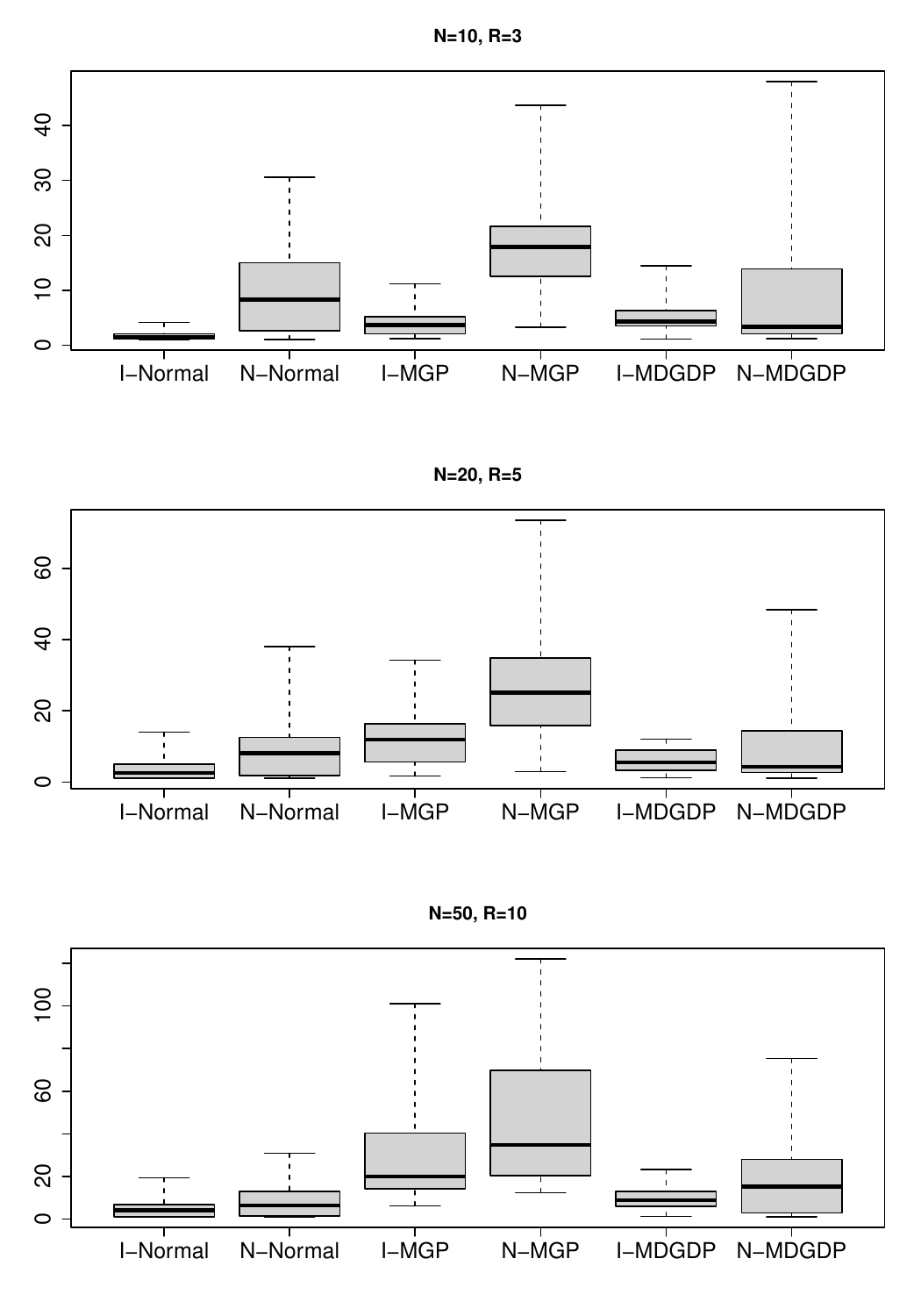}
         \caption{$\boldsymbol{\beta}^{(1)}_{1,1}$}
         \label{fig:M=10}
     \end{subfigure}
     \hfill
     \begin{subfigure}[b]{0.32\textwidth}
         \centering
         \includegraphics[width=\textwidth,height=1.1\textwidth]{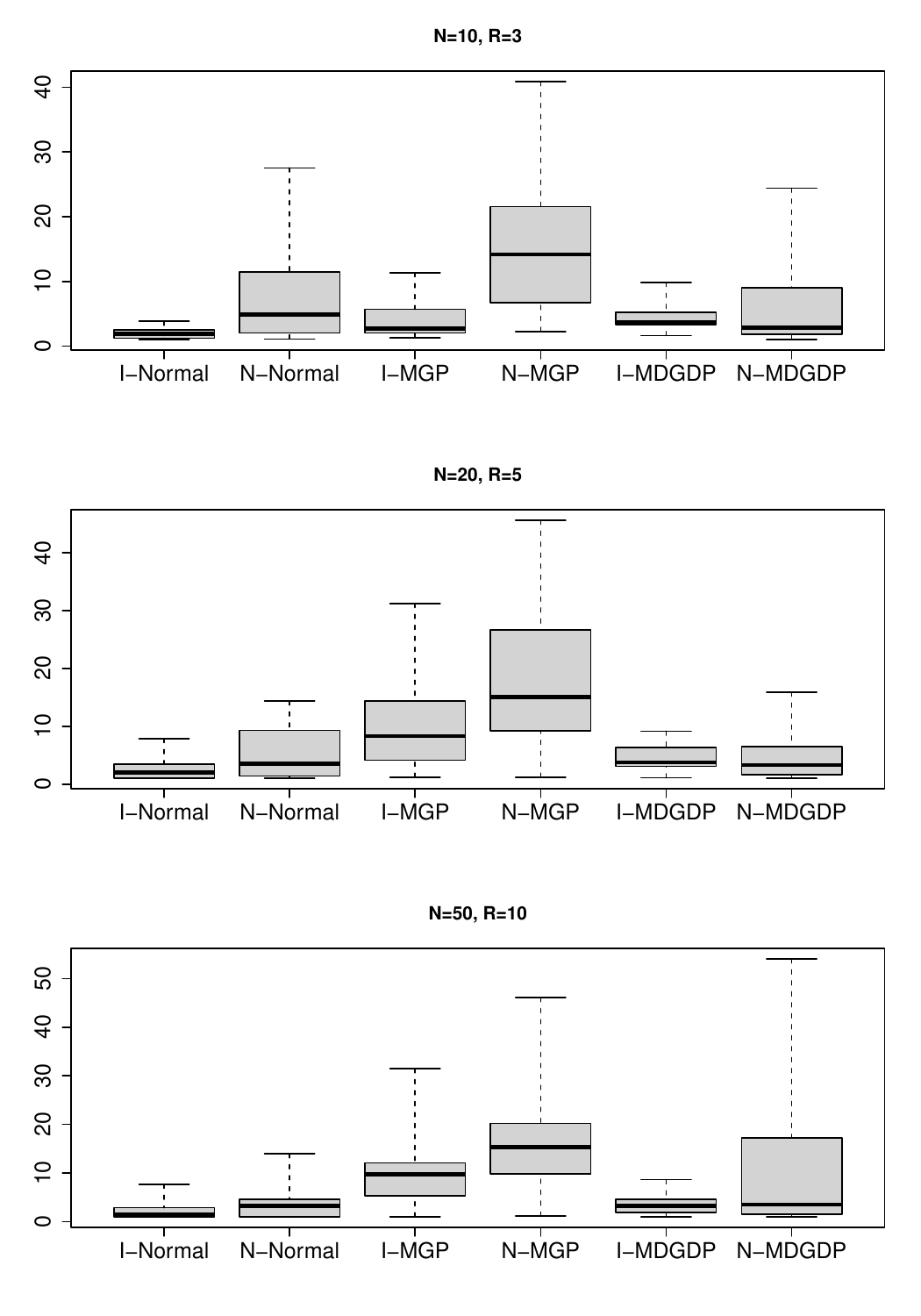}
         \caption{$\boldsymbol{\beta}^{(1)}_{2,1}$}
         \label{fig:M-20}
     \end{subfigure}
     \hfill
     \begin{subfigure}[b]{0.32\textwidth}
         \centering
         \includegraphics[width=\textwidth,height=1.1\textwidth]{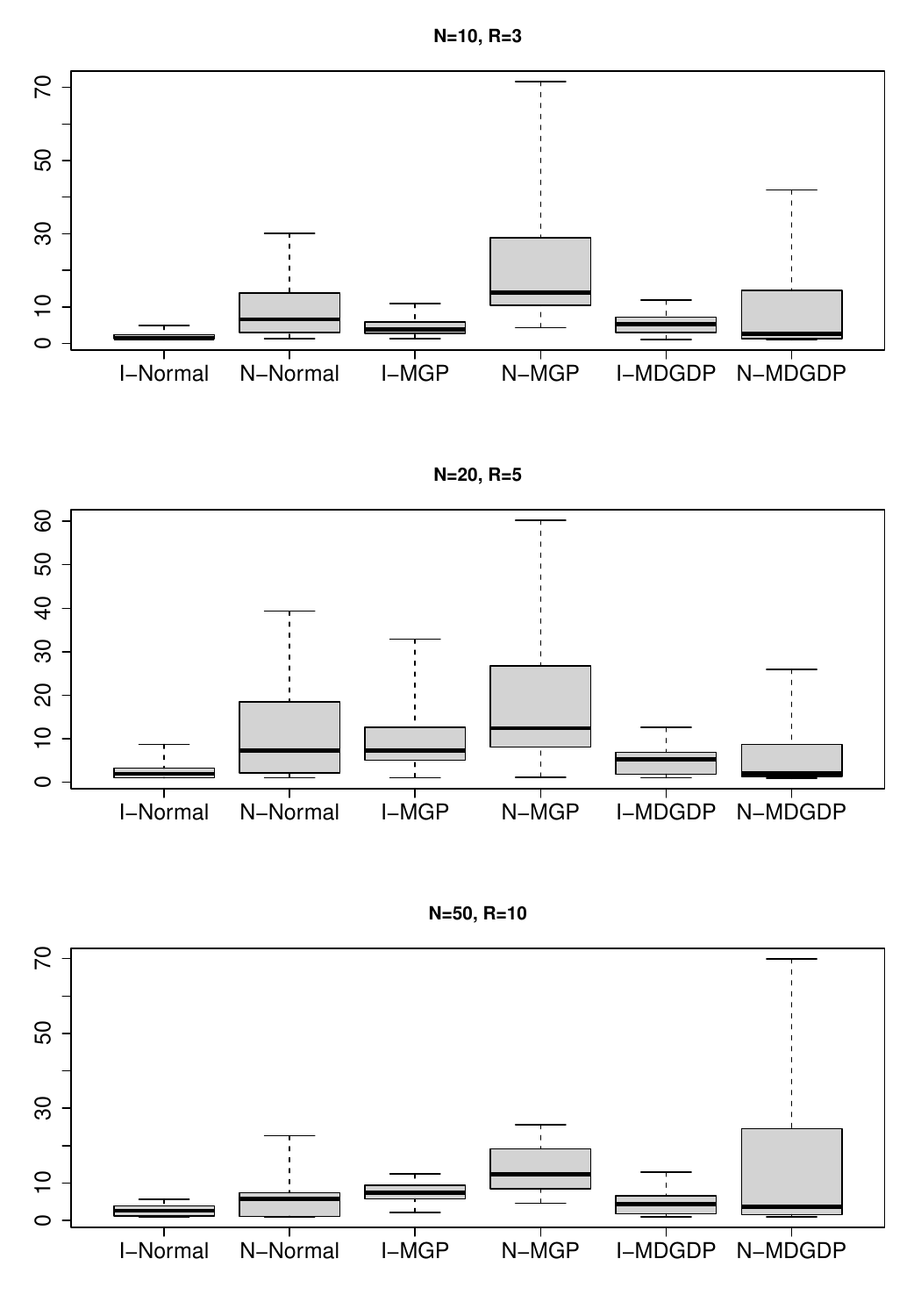}
         \caption{$\boldsymbol{\beta}^{(1)}_{3,1}$}
         \label{fig:M-50}
     \end{subfigure}
        \caption{Boxplots of inefficiency factor of the 1-1 entry of $\boldsymbol{B}_1$ (left), $\boldsymbol{B}_2$ (middle) and $\boldsymbol{B}_3$ (right) from different scenarios: $(N,R)=(10,3)$ (top), $(N,R)=(20,5)$ (middle) and $(N,R)=(50,10)$ (bottom). Inferential schemes with and without interweaving are represented as "I-" and "N-", respectively, followed by a prior setting.}
        \label{fig:three graphs}
\end{figure}

We follow the procedure in \cite{kastner2017efficient} to compute the inefficiency factor (IF) of each margin in different scenarios and prior settings. A smaller IF means that the sampling of a parameter is more efficient. Figure \ref{fig:three graphs} displays boxplots of IFs where each panel corresponds to a scenario with a combination of $(N,R)$ and $\boldsymbol{B}_j$. Each boxplot contains 25 data points from the 25 simulation data sets. Each data point in a boxplot is the IF of the 1-1 entry of $\boldsymbol{B}_j$, for $j=1,2,3$, inferred from one data set. We exclude outliers because there are only a handful of them and this exclusion allows us to focus on the medians and quantiles of IFs. Overall, most IFs with interweaving have lower median values and less variation, compared to their counterparts without interweaving. 

We then use the Stable Gelman-Rubin method \citep{vats2021revisiting} to diagnose the convergence of the margin Markov chains. The reason why we apply the Stable Gelman-Rubin instead of the Gelman-Rubin \citep{gelman1992inference} is twofold: (1) the Gelman-Rubin is suitable when the simulation has multiple Markov chains for each parameter, while our simulation only has one Markov chain for each parameter. The Stable Gelman-Rubin can be applied to both multiple and single Markov chains; (2) The conventional Gelman-Rubin threshold of 1.1 implies an approximation of ESS of 5 according to \cite{vats2021revisiting}, and the authors propose a threshold depending on the parameter dimension and a significance level. The results are presented in Table \ref{convergence test} where each cell is the averaged proportion of margins of which the Markov chains are determined as convergent. Overall, the algorithm with interweaving achieves over 90\%  convergent Markov chains in all scenarios and with all prior choices. All proportions are higher based on the results from interweaving algorithm, compared to the non-interwoven one. We also include the Geweke diagnostic \citep{geweke1991evaluating} in Appendix \ref{Additional Results Simulation} with most interweaving results having a better convergence performance. 
\begin{table}[]
\scriptsize
\centering
\begin{tabular}{lllllllll}
\hline
\textbf{N=10\_R=3} &  Interweaving & Non-interwoven & \textbf{N=20\_R=5} &    Interweaving & Non-interwoven   & \textbf{N=50\_R=10} &    Interweaving & Non-interwoven\\
\hline
Normal                            & 1.000                 & 0.847                & Normal             & 0.996 & 0.916 & Normal              & 0.996 & 0.978 \\
MGP                                        & 0.998                 & 0.866                & MGP                & 0.986 & 0.740 & MGP                 & 0.940 & 0.770 \\
MDGDP                                      & 0.996                 & 0.871                & MDGDP              & 0.998 & 0.819 & MDGDP               & 0.989 & 0.858\\
\hline
\end{tabular}
\vskip 0.5cm
\caption{Averaged proportions of margins which are convergent according to stable Gelman Rubin Statistics.}
\label{convergence test}
\end{table}

\begin{figure}[!htbp]
    \centering
     \begin{subfigure}[b]{0.48\textwidth}
         \centering
         \includegraphics[width=\textwidth,height=0.6\textwidth]{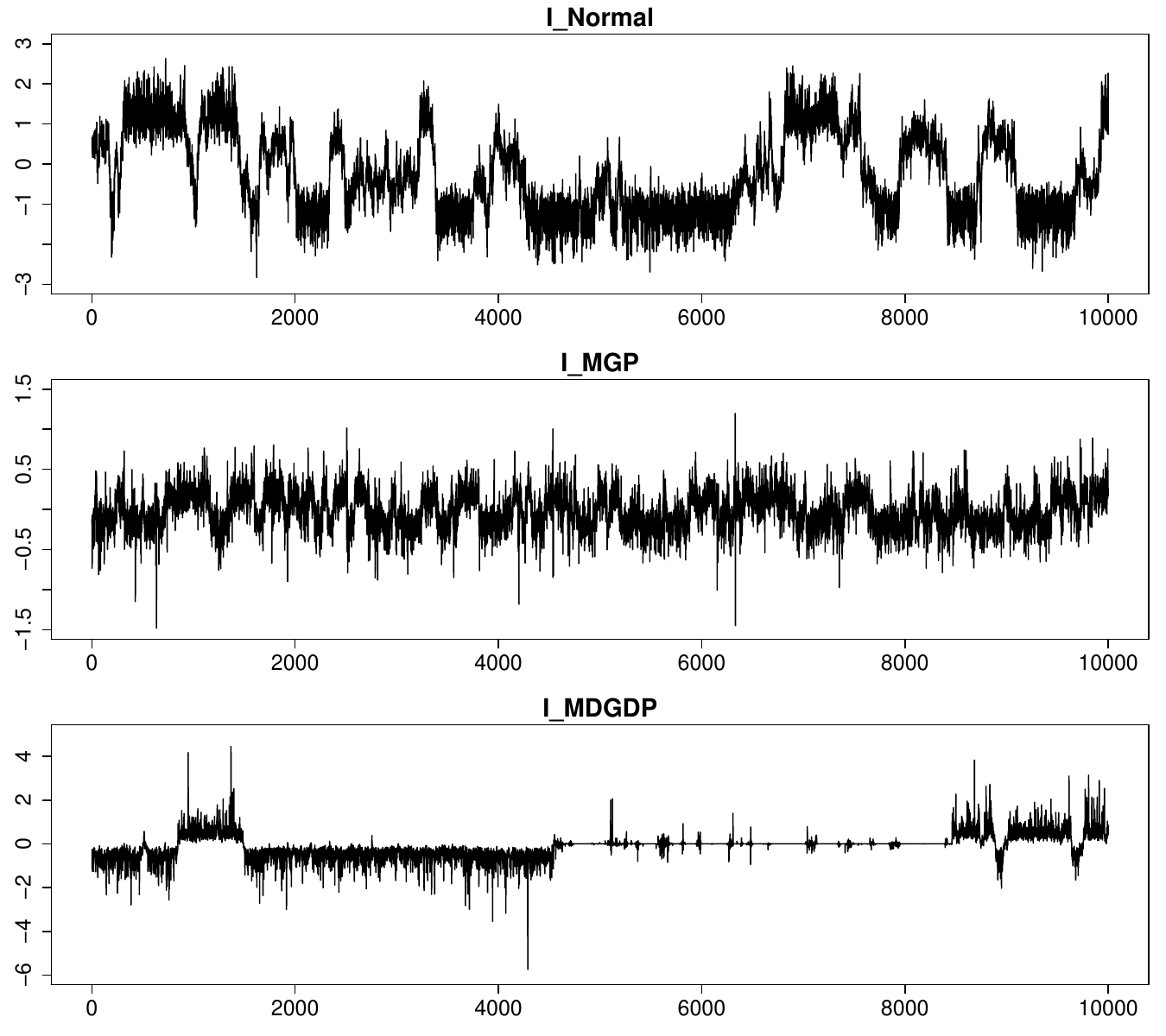}
         \caption{Scenario $(N,R)=(10,3)$}
         \label{mixing sign}
     \end{subfigure}
     \hfill
     \begin{subfigure}[b]{0.48\textwidth}
         \centering
         \includegraphics[width=\textwidth,height=0.6\textwidth]{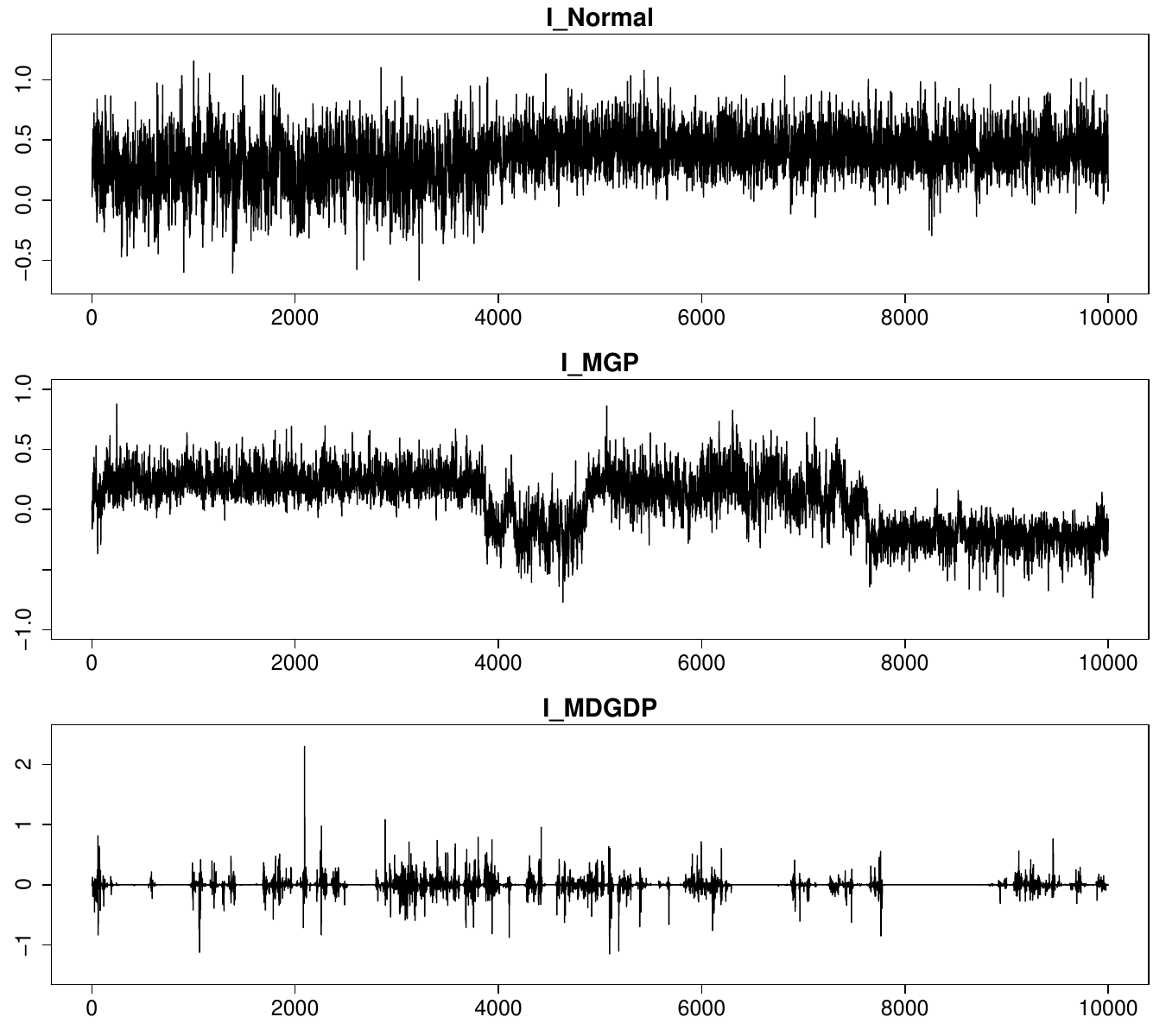}
         \caption{Scenario $(N,R)=(20,5)$}
          \label{mixing label}
     \end{subfigure}
    \caption{Trace plots of $\boldsymbol{\beta}^{(1)}_{1,1}$ in $N=10,$ $R=3$ scenario (left) and $\boldsymbol{\beta}^{(1)}_{1,2}$  in $N=20,$ $R=5$ scenario (right) after burn-in period. The inferential scheme adopts standard normal (top), MGP (middle) and M-DGDP (bottom) as priors and applies with the interweaving strategy.}
    \label{mixing no matching}
\end{figure}

Lastly, we demonstrate the necessity of the post-processing procedure. Figure \ref{mixing no matching} displays trace plots of the whole draws (with thinning of 10) of two selected margins inferred with interweaving strategy, and we exclude the post-processing procedure at this time. All three panels in Figure \ref{mixing sign} and the middle panel in Figure \ref{mixing label} have sign-switching issues. If we do not match signs, the interpretation of margins will be infeasible because the posterior mode or mean of some margins would be zero, but they should be non-zero. The top panel in Figure \ref{mixing label} provides evidence of column permutations, with the sample mean moving from 0 to 0.5. The bottom panel in Figure \ref{mixing label} has neither sign switching nor column permutations, but the M-DGDP does not guarantee convergence only with the interweaving strategy due to the evidence provided in Figure \ref{mixing sign}.

\section{Real Data Application} \label{sec6}
\subsection{Data and Implementation}
We use the US macroeconomic data to assess the utility of Tensor VARs. The data contains 124 quarterly variables from Federal Reserve Economic Data (FRED) \citep{mccracken2020fred} and spans from 1959Q1 to 2019Q4\footnote[2]{The data is available at https://research.stlouisfed.org/econ/mccracken/fred-databases/.}. All time series are transformed to stationarity and standardized to have mean zero and variance one, to avoid any scaling issues. We construct medium-scale and large-scale data sets by selecting 20 and 40 variables, respectively, as referred to in \cite{korobilis2019adaptive}. The selected 40 variables can be divided into 8 categories: (i) output and income, (ii) consumption, orders and inventories, (iii) labour market, (iv) prices, (v) interest rate, (vi) money and credit, (vii) stock market and (viii) exchange rate. Since no variables in the categories of money and credit, and stock market were selected into the medium-scale data set, we also construct an alternative 20-variable data set that contains variables from all 8 categories. We use this alternative data set to examine the robustness of forecasting performance with results available in Appendix \ref{Additional Results Real}. A full description of the variables selected and their transformations can be found in Appendix \ref{data}. Since we have a lower triangular matrix $\boldsymbol{H}$ in the model, the order of time series matters. We follow \cite{bernanke2005measuring} by splitting time series to slow, fast groups and Federal Funds Rate (FEDFUNDS). The slow group contains variables that respond to a shock of FEDFUNDS with a lag, and variables in the fast group respond to it contemporaneously. The order is slow variables, FEDFUNDS, and fast variables. 

For each data set, we estimate various VAR models with 5 lags. Tensor VARs with and without the additional own-lag matrix $\boldsymbol{D}$ are denoted as Tensor MGP Own-lag and Tensor MGP, respectively. For these two Tensor VARs, we use the same choice of $\gamma_1$ and $\gamma_2$ as in the simulation study. Implementation of the MGP is the same as in Section \ref{sec5} and the prior for $\boldsymbol{D}$ is described in Section \ref{Prior Specification}. For competitors, we include standard VARs with hierarchical Minnesota \citep{giannone2015prior}, Horseshoe \citep{carvalho2009handling} and a specification of normal-gamma (NG) prior introduced to VARs by \cite{huber2019adaptive}. All of these three priors can be written as $\boldsymbol{A}_{p,(i,j)}\sim\mathcal{N}\left(0,\underline{V}_{p,(i,j)}\right)$ for $(i,j)$ entry in $\boldsymbol{A}_p$, where $i,\,j=1,\dots, N$ and $p=1,\dots,5$. For hierarchical Minnesota, $\underline{V}_{p,(i,j)}=\begin{cases}
    \frac{\lambda^2_1}{p^2}, & \text{if }i=j\\
    \frac{\lambda^2_1\lambda_2}{p^2}\frac{\hat{\sigma}_i}{\hat{\sigma}_j}, &\text{if }i\neq j
    \end{cases},$
where $\hat{\sigma}^2_i$ is the variance estimate of $\boldsymbol{y}_{t,i}$ sequence modelled by an AR(5) process. $\lambda_1$ and $\lambda_2$ have prior Gamma(0.01,0.01) and are inferred using a random walk Metropolis-Hastings step. For Horseshoe prior, $\underline{V}_{p,(i,j)}=\lambda^2_{p,(i,j)}\tau^2$, where $\lambda^2_{p,(i,j)}$ and $\tau$ are local and global parameters, respectively, following a half Cauchy prior. We apply the NG described in Section \ref{Prior Specification} to the coefficient matrix. Priors for $\boldsymbol{H}$ and stochastic volatility $\boldsymbol{S}_t$, for $t=1,\dots,T$, are the same for all models. The MCMC sampler runs 10,000 iterations after the 10,000 burn-in period. 

Note that the decomposition of $\boldsymbol{\Omega}_t$, the variance-covariance matrix of the VARs, employs a triangular system (i.e. the lower triangular matrix $\boldsymbol{H}_t$), which might lead to the ordering issue when estimating the parameters. This issue has been discussed in \cite{carriero2019large, chan2024large, arias2023macroeconomic}, among others. Thus, we also provide the forecasting performance of which we apply a non-restrictive matrix $\boldsymbol{H}_t$ as defined in \cite{chan2024large}. The results and further discussion about this ordering-invariant model is available in Appendix \ref{Additional Results Real}.

\subsection{Forecasting Results} \label{Forecasting Results}
Before delving into the evaluation of forecasting performance, we compare Tensor VARs and standard VARs with the NG prior in computational time and number of parameters (margins or coefficients) inferred. As shown in Table \ref{computation}, fewer parameters were inferred within the Tensor VAR framework, leading to the reduced computing time of this framework compared to standard VARs. For the medium-scale data set, Tensor VARs require at least six times fewer parameters than standard VARs. Similarly, for the large-scale data set, Tensor VARs only need to infer fewer than 10\% of the parameters compared to those inferred with standard VARs. In term of running time, Tensor and standard VARs take similar amount of time to infer the medium-scale data set, but the former require approximately one-third of the time taken by the latter when we switch to the large-scale data set. The inference using Tensor MGP is faster than Tensor MGP Own-lag because the latter necessitates additional time to infer the own-lag matrix. Note that the code for both VAR frameworks has been accelerated by Rcpp package. 
\begin{table}[!htbp]
\footnotesize
\centering
\begin{tabular}{{cllll}}
\hline
\input{computational_cost.tex}
\end{tabular}
\caption{Averaged number of parameters and running time of Tensor MGP, Tensor MGP Own-lag and standard VARs with the NG prior.}
\label{computation}
\end{table}

We follow the expanding window procedure to assess the forecasting performance of our models. Specifically, we first fit each VAR model with the historical data from 1959Q1 to 1984Q4, then get 1-, 2- and 4-step-ahead forecasts for 1985Q1, 1985Q2 and 1985Q4, respectively. Next, we expand the historical data with the endpoint at 1985Q1 and conduct the multi-step-ahead forecasting again. This procedure is repeated iteratively and stops after conducting the 1-step-ahead forecast of 2019Q4.

We evaluate the forecasting performance of Tensor VARs and standard VARs with both joint and marginal results. For the marginal ones, we select 7 variables which are salient to the US economy, as shown in Table \ref{ALPL medium} and \ref{ALPL large}. The metrics for the forecasting evaluation are mean squared forecast error (MSFE), mean absolute error (MAE) and averaged log predictive likelihood (ALPL), see Appendix \ref{Additional Results Real} for mathematical expressions. All marginal metrics are relative to a standard VAR with a flat prior, taking the 7 time series selected as responses. 

Results about point forecasts evaluated by MSFE and MAE can be found in Appendix \ref{Additional Results Real}. Overall, Tensor VARs achieve better joint and marginal performance than standard VARs. Table \ref{ALPL medium} and Table \ref{ALPL large} present density forecasting performance from the medium and large data sets. Tensor VARs have competitive performance when making joint density forecasts. They also outperform standard VARs in marginal forecasts since they are the best models in 11 and 13 out of 21 cases for medium and large data sets, respectively. Forecasts of FEDFUNDS, GDP, and UNRATE are more favourable when using Tensor VARs, while standard VARs have better performance in forecasting PAYEMS, GDPDEFL. In comparing the performance of the two models within Tensor VARs,  Tensor MGP Own-Lag demonstrates superior results to Tensor MGP. If we focus on individual models in standard VARs, the hierarchical Minnesota prior is the best among these three priors. The superior performance of Tensor MGP Own-lag and the hierarchical Minnesota highlights the importance of own-lag effect in economic data. When comparing each marginal evaluation in these two tables, most results inferred from the large data set are smaller than those inferred from the medium data set, indicating that the large amount of information is advantageous for the marginal forecasting.  

\begin{table}[!htbp]
\scriptsize
\centering
\begin{tabular}{llcccccccl}
\hline
\input{forecast_ALPL_medium.tex}
\end{tabular}
\caption{ALPL of joint and marginal variables using the medium-scale data set. The best forecasts are in bold. }
\label{ALPL medium}
\end{table}
\begin{table}[!htbp]
\scriptsize
\centering
\begin{tabular}{llcccccccl}
\hline
\input{forecast_ALPL_large.tex}
\end{tabular}
\caption{ALPL of joint and marginal variables using the large-scale data set. The best forecasts are in bold. }
\label{ALPL large}
\end{table}

\subsection{Interpretation} \label{interpretation}
Since Tensor MGP Own-lag achieves better performance than Tensor MGP, we demonstrate how to interpret a Tensor VAR by fitting it with the whole large-scale data set ($N$=40). The Tensor VAR infers a rank of 3 which reduces the number of parameters in the coefficient matrix from 8,000 (standard VAR(5)) to 455.

\begin{figure}[!htbp]
     \centering
         \centering
         \includegraphics[width=0.6\textwidth, height=0.45\textwidth]{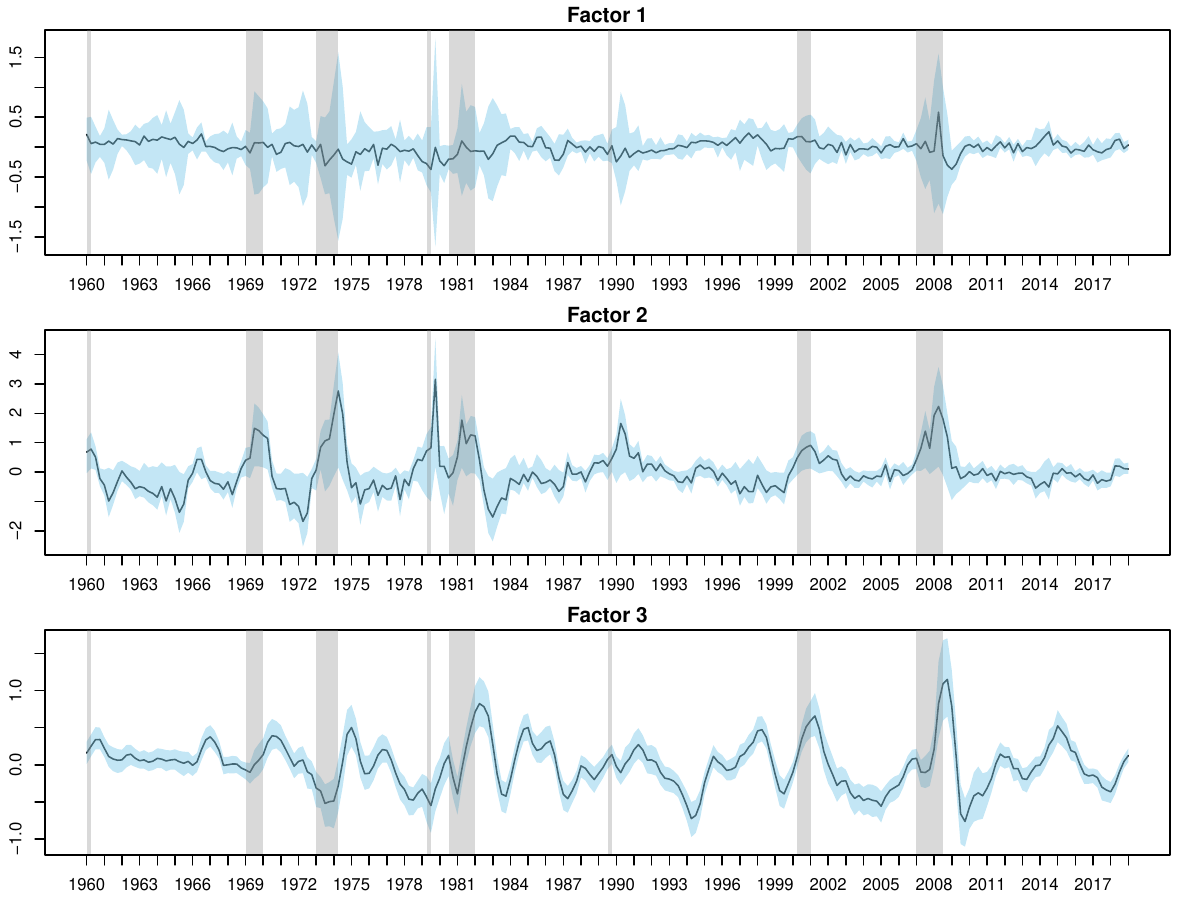}
        \caption{Time series plots of factors with median (solid line) and 80\% credible interval (dashed line). The factors are derived from the inferential results of Tensor MGP Own-lag.}
        \label{fig:factor traceplot}
\end{figure}
\begin{figure}[!htbp]
    \centering
\includegraphics[width=0.5\textwidth,height=0.7\textwidth]{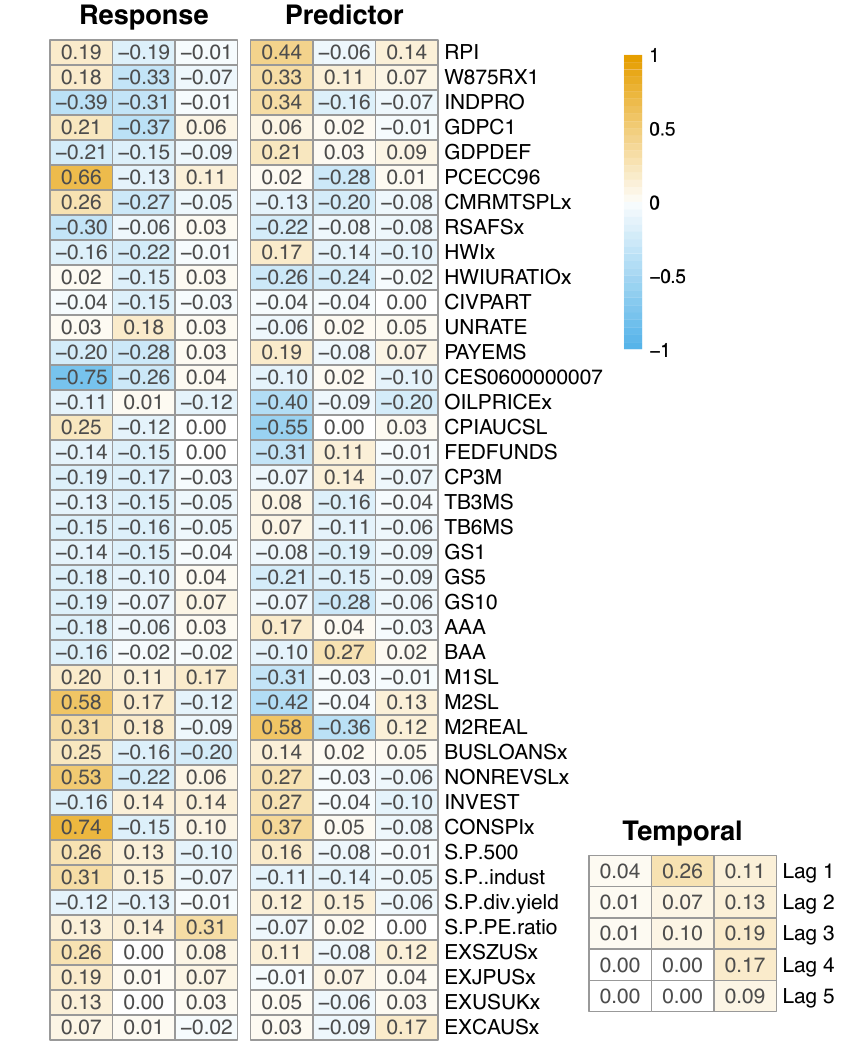}
    \caption{Posterior mean of response, predictor and temporal loadings inferred from Tensor MGP Own-lag.}
    \label{fig:large tensor MGP diag}
\end{figure}

\begin{table}[!htbp]
\scriptsize
\centering
\begin{tabular}{lll}
\hline
Factor 1 & Factor 2           & Factor 3             \\
\hline
M2REAL (0.44)                 & PAYEMS (-0.84)     & S\&P PE ratio (0.62) \\
NONREVSLx (0.41)                 & UNRATE (0.76)      & M2REAL (0.41)        \\
CONSPIx (0.40)                 & INDPRO (-0.72)     & BUSLOANSx (-0.31)    \\
BUSLOANSx (0.32)                    & HWIURATIOx (-0.69) & INVEST (0.30)        \\
PCECC96 (0.26)                   & HWIx (-0.62)       & M2SL (0.28)       \\
\hline
\end{tabular}
\caption{Variables with the top 5 correlations with the factors.}
\label{tab: corr}
\end{table}
According to \labelcref{tensorVAR_represent}, a Tensor VAR can be interpreted as a factor model with observable factors and loadings with different effects. Figure \ref{fig:factor traceplot} shows these factors are consistent with recession periods reported by the National Bureau of Economic Research (NBER) (available on \url{https://fred.stlouisfed.org/series/USRECQ}). The first factor has wider credible intervals during or after the NBER recession periods. The second factor peaks at these recession periods and has relatively high values during the recession of 1960-1961 and the dot-com bubble in the early 2000s. The third factor peaks after recession periods, and the reason will be explained later according to Figure \ref{fig:large tensor MGP diag}. Furthermore, we present the variables that exhibit the five highest magnitudes of correlation with these three factors in Table \ref{tab: corr}. The first factor shows a high correlation with variables from the money and credit category, while the second factor is highly correlated to the variables from the labour market and industrial production. The correlations associated with PAYEMS and UNRATE are reversed indicating that the second factor is positively linked to the unemployment. Proceeding to the third factor, M2REAL and BUSLOANx are both found in the first and third columns in Table \ref{tab: corr}, but we consider the third factor to bear a connection with the financial market, due to its high correlation with the S\&P price earning ratio. It may seem surprising that none of the factors shows a strong connection with interest rates, but all three factors have non-negligible correlation to interest rates according to the full correlation displayed in Appendix \ref{Additional Results Ordering}.

Next, we use Figure \ref{fig:large tensor MGP diag} to answer two questions: 1) which lagged time series contribute to the factors; 2) what is the effect from factors to responses. Figure \ref{fig:large tensor MGP diag} depicts the posterior mean of response, predictor and temporal loadings. Larger margin magnitudes are associated with more deeply saturated hues. 

The first question is answered by the predictor and temporal loading. The columns with the same index in these two loadings reveal how the corresponding factor is constructed. For the first factor representing money and credit, we inspect the top 5 margin magnitudes (M2REAL, CPIAUCSL, RPI, M2SL and OILORICEx) in the first column of the predictor loading, and show that price is the main category contributing to this factor. The negative margins of CPIAUCSL and OILORICEx indicate that prices have a negative effect on the first factor. This conclusion is further strengthened by the opposite signs of M2REAL and M2SL margins, since M2SL drops while M2REAL rises with decreasing prices. Additionally, the positive margin of RPI margin, which is adjusted by inflation, supports this conclusion. In the first column of temporal loading, the first lag suggests to be the most important one because its magnitude is the largest within the corresponding column. Combine with the findings from predictor and temporal loading, the first factor is formed by the prices one quarter ago. We follow a similar method to investigate the formation of the second factor and get the following finding: Firstly, A decline in real M2 money supply (M2REAL) and personal consumption expenditures (PCECC96) contributes to an increase in this factor about unemployment. Secondly, the factor grows with the increase of credit risk because the opposite signs of BAA and GS10, representing the spread between the corresponding two yields. Akin to the formation of the first factor, the first lag exhibits the most significant contribution to the formation of the second factor. Lastly, we focus on the columns corresponding to the third factor and find two differences compared to other columns: 1) margins with relatively high magnitudes are related to financial market, for example, oil price (OILORICEx) in the commodity market, exchange rates (EXSZUsx and EXCAUSx) in the FX market; 2) the column in the temporal loading spans in all five lags, which explains why the third factor peaks after the recession periods.

The second question is answered by the response loading, which has the same definition as the factor loading in a factor model if one considers the factors as factor scores. Each column of the response loading shows how each factor impacts the responses. In the first column, margins corresponding to variables in the money and credit category have high magnitudes, which follows expectation because the first factor represents this category. Assume that the first factor to be positively associated with money supply given the evidence in Table \ref{tab: corr}, we can explain the negative margins of interest rates: during economic downturns, both rate cut and quantitative easing are applied as part of the monetary policy toolkit to boost economic activity. Similarly, the positive margins in exchange rate category suggests the depreciation of US dollars when money supply increases in the US. Moving to the second column, 
the negative margins in the income and output category have high magnitudes, suggesting an increase of this unemployment factor (the second factor) results to the slowdown of economic activities. Negative margins of interest rates show the expectation of interest rate reduction given that the second factor rises. If we look at the loading corresponding to the third factor, it is unsurprising that the largest margin corresponds to S\&P PE ratio because the third factor is highly correlated to this variable. 

\subsection{Effect of $\boldsymbol{D}$} \label{effect of D}
This subsection compares the Tensor VARs with and without the own-lag matrix $\boldsymbol{D}$. Firstly, we do not find a strong own-lag effect in the last subsection because the variables with high margin magnitudes in the response loading do not coincide those counterparts in the predictor loading. Secondly, we use Tensor MGP (without $\boldsymbol{D}$) to conduct the same experiment as in Section \ref{interpretation}. After the inference, we apply Welch's t test to check whether margins inferred from these two Tensor VAR models are significantly different. Only 4 out of 255 margins cannot reject the null hypothesis that no significant difference between the two posterior samples with 0.1\% significance level. Figure \ref{fig:large tensor MGP} depicts the posterior mean of the loadings without $\boldsymbol{D}$. As shown in this figure, the same variable (PAYEMS) is associated with the largest margin magnitudes in the first columns of response and predictor loadings. This pattern holds true for the second and third columns as well, with corresponding variables being M2REAL and BUSLOANS. This finding indicates the additional own-lag matrix $\boldsymbol{D}$ allows the tensor to explore more cross-lag effects. In addition, these large margins in PAYEMS, M2REAL and BUSLOANS have the potential to distort the coefficients in such a manner that the rows and columns corresponding to these three variables in the coefficient matrix exhibit a higher proportion of large magnitudes compared to their counterparts associated with other variables. Table \ref{distort} gives a detailed analysis in Appendix \ref{Additional Results Ordering}.

\vspace{-0.2cm}
\section{Conclusion and Discussion} \label{sec7}
In this paper, we apply the Multiplicative Gamma Prior (MGP) to margins and use an adaptive inferential scheme to infer the rank. To overcome the convergence issue, we introduce an interweaving Gibbs sampler to allow better mixing of Markov chains, and match labels and signs after the inference. 

The Tensor VAR is closely related to the reduced-rank VAR \citep{carriero2011forecasting, geweke1996bayesian}. A detailed discussion of these two structures is available in the introduction section of \cite{wang2021high}. In short, reduced-rank VAR only apply the low-rank assumption to the mode-1 matricization of the tensor, but Tensor VAR make the same assumption to all three matricizations (model-1, -2 and -3). Following this connection, we find that reduced-rank VAR is a special case of Tensor VAR with the following expression:
\begin{equation*}
\boldsymbol{\mathcal{A}}=\sum_{r=1}^R\boldsymbol{\mathcal{A}}^{(r)}=\sum_{r=1}^R\boldsymbol\beta^{(r)}_1\circ \boldsymbol{C}^{(r)},  \label{reduced-rank var}
\end{equation*}
where $\boldsymbol{C}^{(r)}$ is an $N$-by-$P$ matrix, and $\circ$ is the outer product of a vector and a matrix such that $\boldsymbol\beta^{(r)}_{1,i_1}\boldsymbol{C}^{(r)}$ equals to $\boldsymbol{\mathcal{A}}^{(r)}_{i_1,\cdot,\cdot}$, the $i_1$-th matrix on the first dimension of $\mathcal{A}^{(r)}$, for $i_1 = 1,\dots,N$. If we decompose $\boldsymbol{C}^{(r)}$ to $\boldsymbol{\beta}^{(r)}_2\circ \boldsymbol{\beta}^{(r)}_3$, then we retain Equation \ref{CP decomp tensor var}. We leave the comparison between Tensor VAR and reduced-rank VAR to future work.

 Several extensions can also be investigated. Firstly, it will be interesting to adopt time-varying margins and rank to the Tensor VAR. A related work is studied by  \cite{zhang2021bayesian}, who kept margins time-invariant and switched each column of the tensor matrix $\boldsymbol{B}$ on or off with a prior. Secondly, we can modify the MGP to include a local parameter corresponding to each row of the loadings so as to provide more interpretability. Lastly, a similar MCMC scheme can be applied to Tucker decomposition \citep{tucker1966some}, another popular tensor decomposition with a more flexible structure compared to the CP decomposition.

\putbib
\end{bibunit}
\begin{bibunit}
    
\appendix
\counterwithin{figure}{section}
\counterwithin{table}{section} 
\counterwithin{algorithm}{section} 
\section{Basic Notations and Operations}\label{Basic Notations and Operations}
\setlength{\belowdisplayskip}{2pt} \setlength{\belowdisplayshortskip}{2pt}
\setlength{\abovedisplayskip}{2pt} \setlength{\abovedisplayshortskip}{2pt}
We follow the convention in tensor literature to introduce some basic notations and operations. See \cite{kolda2009tensor} for a review. A tensor $\boldsymbol{\mathcal{A}}$ is a $J$th-order tensor if $\boldsymbol{\mathcal{A}}\in\mathbb{R}^{I_1\times\cdots\times I_J}$ with entries $\boldsymbol{\mathcal{A}}_{i_1,\dots,i_J}$ for $i_j=1,\dots,I_j$. first- and second-order tensors are simply vectors and matrices. Similar to the definition of a diagonal matrix, a tensor is called a $J$th-order \textit{superdiagonal} tensor if $I_1=\cdots =I_J=I$ and only its ($i,\, \dots,\,i$) entries are non-zero for $i=1,\,\dots,\,I$. Extracting entries by a selected index from a matrix $\boldsymbol{A}$ and a tensor $\boldsymbol{\mathcal{A}}$ is akin. We denote $\boldsymbol{A}_{(i_1,\cdot)}$, $\boldsymbol{A}_{(\cdot,i_2)}$ and $\boldsymbol{\mathcal{A}}_{(\dots,i_j,\dots)}$ as the $i_1$-th row, the $i_2$-th column in $\boldsymbol{A}$ and a $(J-1)$th-order tensor with entries having the index $i_j$ on the $j$-th dimension in $\boldsymbol{\mathcal{A}}$.

Matricization of a $J$th-order tensor ($J$\textgreater 2) is an operation that transforms the tensor into a matrix. 
There are $J$ possible matricizations for a $J$th-order tensor $\boldsymbol{\mathcal{A}}$, and the matricization to the $j$-th dimension is called the \textit{mode-j matricization} with notation $\boldsymbol{\mathcal{A}}_{(j)}\in\mathbb{R}^{I_j\times \prod_{j^\prime\neq j}^J I_{j^\prime}}$, where the $i_j$-th row corresponds to the vectorization of $\boldsymbol{\mathcal{A}}_{(\dots,i_j,\dots)}$. The \textit{$j$-mode product} of $\boldsymbol{\mathcal{A}}$ and a matrix $\boldsymbol{B}\in\mathbb{R}^{K\times I_j}$ is denoted as $\boldsymbol{\mathcal{A}}\times _j \boldsymbol{B}$, which gives a $J$th-order tensor $\boldsymbol{\mathcal{C}}\in\mathbb{R}^{I_1\times\cdots\times I_{j-1}\times K \times I_{j+1}\times \cdots \times I_J}$ with the $(i_1,\dots, i_{j-1},k,i_{j+1},\dots,i_J)$ entry as 
\begin{equation*}
    \boldsymbol{\mathcal{C}}_{i_1,\dots, i_{j-1},k,i_{j+1},\dots,i_J}=\sum_{i_j=1}^{I_j}\boldsymbol{\mathcal{A}}_{i_1,\dots,i_j,\dots,i_J}\boldsymbol{B}_{k,i_j}.
\end{equation*}

The following are some preliminaries for tensor decompositions. The outer (or tensor) product of two vectors $\boldsymbol\beta_1\in\mathbb{R}^{I_1}$ and $\boldsymbol\beta_2\in\mathbb{R}^{I_2}$ is $\boldsymbol\beta_1\circ\boldsymbol\beta_2$, yielding an $I_1$-by-$I_2$ matrix with its ($i_1$, $i_2$) entry as $\boldsymbol\beta_{1,i_1}\boldsymbol\beta_{2,i_2}$. Assume we have $J$ vectors with $\boldsymbol\beta_j\in\mathbb{R}^{I_j}$ denoting the $j$-th vector, then the \textit{$J$-way outer product} of these vectors, $\boldsymbol\beta_1\circ\cdots\circ\boldsymbol\beta_J$, is a $I_1\times\cdots\times I_J$ tensor, with ($i_1,\, \dots ,\, i_j$) entry as $\boldsymbol\beta_{1,i_1}\cdots\boldsymbol\beta_{J,i_J}$. Note that the outer product can also apply to a pair of a vector and a matrix. Assume that the vector is $\boldsymbol{a}\in\mathbb{R}^{K}$ and the matrix is $\boldsymbol{B}\in\mathbb{R}^{M\times N}$. The their outer product, $\boldsymbol{a}\circ\boldsymbol{B}$, is a third-order tensor with dimension $K$ \texttimes $M$ \texttimes $N$, and the $(k, m, n)$ entry equals to $\boldsymbol{a}_k\boldsymbol{B}_{m.n}$.

The most prominent tensor decompositions are CANDECOMP/PARAFAC (CP) decomposition \citep{kiers2000towards} and Tucker decomposition \citep{tucker1966some}. A rank-$R$ CP decomposition of a $J$th-order tensor is 
 \begin{equation}
     \boldsymbol{\mathcal{A}}=\sum_{r=1}^R\boldsymbol\beta^{(r)}_1\circ\cdots\circ\boldsymbol\beta^{(r)}_J=\sum_{r=1}^R\boldsymbol{\mathcal{A}}^{(r)}\label{CP_decomp},\\
 \end{equation}
where entries in $\boldsymbol\beta^{(r)}_j\in\mathbb{R}^{I_j}$ for $j=1,\dots,J$ and $r=1,\dots,R$, are \textit{margins} of the tensor, $\boldsymbol{\mathcal{A}}^{(r)}=\boldsymbol\beta^{(r)}_1\circ\cdots\circ\boldsymbol\beta^{(r)}_J\in\mathbb{R}^{I_1\times\dots\times I_J}$.

Instead of having only one rank, Tucker decomposition has $J$ ranks, $R_1,\dots,R_J$, to decompose a $J$th-order tensor as
\begin{align}
\boldsymbol{\mathcal{A}}=\sum_{r_1=1}^{R_1}\sum_{r_2=1}^{R_2}\cdots\sum_{r_J=1}^{R_J}\boldsymbol{\mathcal{G}}_{r_1,\,\cdots,\,r_J}\boldsymbol\beta^{(r_1)}_1\circ\cdots\circ\boldsymbol\beta^{(r_J)}_J,\label{tucker}
\end{align}
where $\boldsymbol{\mathcal{G}}$ is a $J$th-order tensor with dimension $R_1\times \cdots \times R_J$. Tucker decomposition is the generalization of the CP decomposition because they are equivalent when $\boldsymbol{\mathcal{G}}$ is a superdiagonal tensor.

\section{Bayesian Inference} 
\label{Bayesian Inference Appendix}
\subsection{Additional Prior Setting}
\label{Additional Prior Setting}
Each non-zero off-diagonal entry in $\boldsymbol{H}$ follows a variant of the normal-gamma prior \citep{brown2010inference},
\begin{equation*}
    \boldsymbol{H}_{i,j}\sim\mathcal{N} \left( 0,\left(2/\lambda^{2}_h\right)\psi^{(i,j)}_h \right), \, \psi^{(i,j)}_h\sim\text{Gamma}(a_h,a_h), \text{ for }i=1,\dots,N \text{ and }j<i,
\end{equation*}
where $\lambda^{2}_h$ is the global parameter which controls the overall shrinkage and follows Gamma(0.01,0.01) prior, $\psi^{(i,j)}_h$ allows flexibility locally, and hyperparameter $a_h$ follows an exponential prior with parameter 1. 

The sequence $s_{1,n}, \dots, s_{T,n}$ evolves with a stochastic volatility model \citep{jacquier2002bayesian, kim1998stochastic}. The logarithm of $s_{t,n}$ follows
\begin{align*}
    \ln(s_{t,n})\mid \ln(s_{t-1,n}), \mu_n, \psi_n, \sigma_{n} &\sim \mathcal{N}\left(\mu_n+\psi_n\left( \ln(s_{t-1,n})-\mu_n\right),\sigma^2_{n}\right),\\
    \ln(s_{0,n})\mid \mu_n, \psi_n, \sigma_{n}&\sim \mathcal{N}\left(\mu_n, \sigma^2_{n}/\left(1-\psi^2_n\right)\right).
\end{align*}
Priors of hyperparameters, $\mu_n,\, \psi_n, \, \sigma^2_n$, are the same as those in \cite{kastner2014ancillarity}, for $n=1,\dots,N$. We impose $\mathcal{N}(0,100)$ to $\mu_n$, $\text{Beta}(5,1.5)$ to $\frac{1+\psi_n}{2}$ and $\text{Gamma}(1/2,1/2)$ to $\sigma^2_n$. The prior for $\sigma^2_n$ implies that $\sigma_n$ follows a standard normal distribution.

\subsection{Full Conditionals of $\boldsymbol{B}_1$, $\boldsymbol{B}_2$, $\boldsymbol{B}_3$ and $\boldsymbol{D}$}
\label{full conditionals of margins}
In this subsection, we consider a Tensor VAR with the additional own-lag matrix $\boldsymbol{D}$. The inference of a Tensor VAR without $\boldsymbol{D}$ is simply to treat $\boldsymbol{D}$ as a zero matrix. Recall a Tensor VAR in terms of $\boldsymbol{B}_1$, $\boldsymbol{B}_2$ and $\boldsymbol{B}_3$:
\begin{align}
    \boldsymbol{y}^*_t&=\left(\boldsymbol{x}^\prime_t\left(\boldsymbol{B}_3\otimes \boldsymbol{B}_2\right)\boldsymbol{\mathcal{I}}^\prime_{(1)}\otimes \boldsymbol{I}_N\right)\text{vec}(\boldsymbol{B}_1)+\boldsymbol{\epsilon}_t \label{tensorVAR_1}\\
    &=\boldsymbol{B}_1\boldsymbol{\mathcal{I}}_{(1)}\left((\boldsymbol{B}^\prime_3\boldsymbol{X}^\prime_t)\otimes \boldsymbol{I}_R\right)\text{vec}(\boldsymbol{B}^\prime_2)+\boldsymbol{\epsilon}_t \label{tensorVAR_2}\\
    &=\boldsymbol{B}_1\boldsymbol{\mathcal{I}}_{(1)}\left(\boldsymbol{I}_R\otimes(\boldsymbol{B}^\prime_2\boldsymbol{X}_t)\right)\text{vec}(\boldsymbol{B}_3)+\boldsymbol{\epsilon}_t. \label{tensorVAR_3}
\end{align}
We assume the terms before vectorizations of $\boldsymbol{B}_1$, $\boldsymbol{B}^\prime_2$ and $\boldsymbol{B}_3$ as $\boldsymbol{Z}_{t,1}$, $\boldsymbol{Z}_{t,2}$ and $\boldsymbol{Z}_{t,3}$, respectively. Given other parameters, the full conditional of $\text{vec}(\boldsymbol{B}_j)$ for $j=1, \, 3$ or that of $\text{vec}(\boldsymbol{B}^\prime_j)$ for $j=2$ is $\mathcal{N}\left(\overline{\boldsymbol{\mu}}_j,\overline{\boldsymbol\Sigma}_{j} \right)$ with
\allowdisplaybreaks
\begin{align*}
    \overline{\boldsymbol\Sigma}^{-1}_{j}&= \underline{\boldsymbol\Sigma}^{-1}_{j}+\sum_{t=1}^T\boldsymbol{Z}^\prime_{t,j}\boldsymbol{H}^\prime \boldsymbol{S}^{-1}_t\boldsymbol{H}\boldsymbol{Z}_{t,j},\\ \overline{\boldsymbol{\mu}}_j&= \overline{\boldsymbol\Sigma}^{-1}_{j}\sum_{t=1}^T\boldsymbol{Z}^\prime_{t,j}\boldsymbol{H}^\prime \boldsymbol{S}^{-1}_t\tilde{\boldsymbol{y}}^{*}_t,
\end{align*}
where $\tilde{\boldsymbol{y}}^{*}_t=\boldsymbol{H}\boldsymbol{y}^{*}_t=\boldsymbol{H}(\boldsymbol{y}_t-\boldsymbol{D}\boldsymbol{x}_t)$, $\underline{\boldsymbol\Sigma}_{j}$ is the prior covariance matrix of the corresponding vector.

Given $\boldsymbol{B}_1$, $\boldsymbol{B}_2$, $\boldsymbol{B}_3$ and other parameters, we can infer $\boldsymbol{D}$ in a similar way as \cite{carriero2022corrigendum}. Assume that $\boldsymbol{D}=\left(\text{diag}\left(d_{1,1},\dots,d_{N,1}\right),\dots,\text{diag}\left(d_{1,P},\dots,d_{N,P}\right)\right)$, and $\boldsymbol{y}^{**}_t=\boldsymbol{y}_t-\boldsymbol{\mathcal{A}}_{(1)}\boldsymbol{x}_t=\boldsymbol{D}\boldsymbol{x}_t+\boldsymbol{\epsilon_t}$, if we multiply both sides of the equation aforementioned with $\boldsymbol{H}$, 
we get $\tilde{\boldsymbol{y}}^{**}_t=\boldsymbol{H}\boldsymbol{y}^{**}_t=\boldsymbol{H}\boldsymbol{D}\boldsymbol{x}_t+\boldsymbol{u}_t$, where $\boldsymbol{u}_t\sim\mathcal{N}\left(\boldsymbol{0},\boldsymbol{S}_t \right)$. This equation can be expanded to 
\allowdisplaybreaks
\begin{align}
    \tilde{\boldsymbol{y}}^{**}_{t,1}&=(\boldsymbol{y}^{(1)}_t)^\prime\boldsymbol{d}_1+\boldsymbol{u}_{t,1}, \nonumber\\
    \tilde{\boldsymbol{y}}^{**}_{t,1}&=h_{2,1}(\boldsymbol{y}^{(1)}_t)^\prime\boldsymbol{d}_1+(\boldsymbol{y}^{(2)}_t)^\prime\boldsymbol{d}_2+\boldsymbol{u}_{t,2}, \nonumber\\
    &\vdots \nonumber\\
    \tilde{\boldsymbol{y}}^{**}_{t,N}&=h_{N,1}(\boldsymbol{y}^{(1)}_t)^\prime\boldsymbol{d}_1+\dots+h_{N,N-1}(\boldsymbol{y}^{(N-1)}_t)^\prime\boldsymbol{d}_{N-1}+(\boldsymbol{y}^{(N)}_t)^\prime\boldsymbol{d}_{N}+\boldsymbol{u}_{t,N}, \label{diag_inference}
\end{align}
where $\boldsymbol{y}^{(j)}_t$ is a vector that contains the $P$ lagged values of $\boldsymbol{y}_{t,j}$, $\boldsymbol{d}_j=(d_{j,1},\dots,d_{j,P})^\prime$, for $j=1,\dots,N$, $h_{i,j}$ is the $(i,j)$ entry of $\boldsymbol{H}$.

It is noteworthy that \labelcref{diag_inference} is similar to Equation (12) in \cite{carriero2022corrigendum}. 
An important difference is that they multiplied the same $\boldsymbol{x}_t$ to each row of the coefficient matrix, whereas we multiply $(\boldsymbol{y}^{(j)}_t)^\prime$ to each $\boldsymbol{d}_j$. After slightly modifying Equations (13) - (15) in \cite{carriero2022corrigendum}, we get the full conditional posterior $\boldsymbol{d}_j \mid \boldsymbol{B}_1,\boldsymbol{B}_2,\boldsymbol{B}_3, \boldsymbol{d}_{-j}, \boldsymbol{H},\boldsymbol{S}_{1:T}\sim \mathcal{N}\left(\overline{\boldsymbol{\mu}}_{\boldsymbol{d}_j}, \overline{\boldsymbol{\Sigma}}_{\boldsymbol{d}_j}\right)$, with
\allowdisplaybreaks
\begin{align}
\overline{\boldsymbol{\Sigma}}^{-1}_{\boldsymbol{d}_j}&=\underline{\boldsymbol\Sigma}^{-1}_{\boldsymbol{d}_j}+\sum_{i=j}^{N}h^2_{i,j}\sum_{t=1}^T s^{-1}_{t,i}\boldsymbol{y}^{(j)}_t (\boldsymbol{y}^{(j)}_t)^\prime, \label{infer_diag_mu}\\
\overline{\boldsymbol{\mu}}_{\boldsymbol{d}_j}&=\overline{\boldsymbol{\Sigma}}_{\boldsymbol{d}_j}\left(\sum_{i=j}^{N}h^2_{i,j}\sum_{t=1}^Ts^{-1}_{t,i}\boldsymbol{z}^{(j)}_{t,i}\boldsymbol{y}^{(j)}_t\right),\label{infer_diag_sigma}
\end{align}
where $\boldsymbol{z}^{(j)}_{t,j+l}=\tilde{\boldsymbol{y}}^{**}_{t,j+l}-\sum_{i\neq j, i=1}^{j+l}h_{j+l,i}(\boldsymbol{y}^{(i)}_t)^\prime \boldsymbol{d}_i$, $\boldsymbol{d}_{-j}$ represents $\boldsymbol{D}$ without $\boldsymbol{d}_j$, $\underline{\boldsymbol\Sigma}_{\boldsymbol{d}_j}$ is the prior covariance matrix of $\boldsymbol{d}_j$.

A more efficient way is to rewrite the system in \labelcref{diag_inference} to 
\begin{align}
    \left(\boldsymbol{Y}^{**}-\boldsymbol{X}\left(\boldsymbol{D}^{[j=0]}\right)^\prime\right)\boldsymbol{H}^\prime_{j:N,1:N}=\boldsymbol{Y}^{(j)}\boldsymbol{d}_j\boldsymbol{H}^\prime_{j:N,k}+\boldsymbol{U}_{j:N}, \label{system_repre}
\end{align}
where $\boldsymbol{Y}^{**}=(\boldsymbol{y}^{**}_1,\dots,\boldsymbol{y}^{**}_T)^\prime$, $\boldsymbol{X}=(\boldsymbol{x}_1,\dots, \boldsymbol{x}_T)^\prime$, $\boldsymbol{H}_{j:N,1:N}$ is the block of $\boldsymbol{H}$ composed of $j$- to $N$-th rows and all $N$ columns, $\boldsymbol{Y}^{(j)}=(\boldsymbol{y}^{(j)}_1,\dots, \boldsymbol{y}^{(j)}_T)^\prime$, $\boldsymbol{D}^{[j=0]}$ is the same as $\boldsymbol{D}$ except the $j$-th row as zeros, $\boldsymbol{U}_{j:N}=(\boldsymbol{u}_{1,j:N},\dots, \boldsymbol{u}_{T,j:N})^\prime$.

If we vectorize both sides of \labelcref{system_repre}, the new equation is 
\begin{equation*}
    \text{vec}\left(\left(\boldsymbol{Y}^{**}-\boldsymbol{X}\left(\boldsymbol{D}^{[j=0]}\right)^\prime\right)\boldsymbol{H}^\prime_{j:N,1:N}\right)=\left(\boldsymbol{H}_{j:N,k}\otimes \boldsymbol{Y}^{(j)}\right)\boldsymbol{d}_j+\text{vec}\left(\boldsymbol{U}_{j:N}\right).
\end{equation*}
Let 
\begin{align*}
    \tilde{\boldsymbol{Y}}^{(j)}&= \text{vec}\left(\left(\boldsymbol{Y}^{**}-\boldsymbol{X}\left(\boldsymbol{D}^{[j=0]}\right)^\prime\right)\boldsymbol{H}^\prime_{j:N,1:N}\right)./\text{vec}\left(\boldsymbol{S}^{1/2}_{1:T,k:N}\right)\\
    \tilde{\boldsymbol{X}}^{(j)}&=\left(\boldsymbol{H}_{j:N,k}\otimes \boldsymbol{Y}^{(j)}\right)./\text{vec}\left(\boldsymbol{S}^{1/2}_{1:T,j:N}\right),
\end{align*}
where $./$ is Matlab element-by-element division operation, $\boldsymbol{S}_{1:T,j:N}$ is a $T$-by-$(N\texttt{-} J\texttt{+}1)$ matrix with the $t$-th row has entries $(s_{t,j},\dots, s_{t,N})$.

Then \labelcref{infer_diag_mu} and \labelcref{infer_diag_sigma} are simplified to
\begin{align}
\overline{\boldsymbol{\Sigma}}^{-1}_{\boldsymbol{d}_j}&=\underline{\boldsymbol\Sigma}^{-1}_{\boldsymbol{d}_j}+ \left(\tilde{\boldsymbol{X}}^{(j)}\right)^\prime  \tilde{\boldsymbol{X}}^{(j)}\\
\overline{\boldsymbol{\mu}}_{\boldsymbol{d}_j}&=\overline{\boldsymbol{\Sigma}}_{\boldsymbol{d}_j}\left(\tilde{\boldsymbol{X}}^{(j)}\right)^\prime \tilde{\boldsymbol{Y}}^{(j)}.
\end{align}
\subsection{Full Conditionals Related to Multiplicative Gamma Prior}
Posteriors of hyperparameters in the MGP are similar to those in \cite{bhattacharya2011sparse}.
Since $\phi_{(r,j,i_j)}$ is a local hyperparameter of $\boldsymbol{\beta}^{(r)}_{j,i_j}$, the derivation of its conditional posterior given $\boldsymbol{\beta}^{(r)}_{j,i_j}$ and $\tau_r$ is 
\allowdisplaybreaks
\begin{align*}
    p\left(\phi_{(r,j,i_j)}\mid \boldsymbol{\beta}^{(r)}_{j,i_j}, \tau_r\right)&\propto \left(\phi^{-1}_{(r,j,i_j)}\right)^{-1/2}\exp\left(-\frac{\left(\boldsymbol{\beta}^{(r)}_{j,i_j}\right)^2}{2\phi^{-1}_{(r,j,i_j)}\tau^{-1}_2}\right)\left(\phi_{(r,j,i_j)}\right)^{\frac{\nu}{2}-1}\exp\left(-\frac{\nu}{2}\phi_{(r,j,i_j)}\right)\\
    &=\left(\phi_{(r,j,i_j)}\right)^{\frac{\nu+1}{2}-1}\exp\left(-\frac{\tau_r\left(\boldsymbol{\beta}^{(r)}_{j,i_j}\right)^2+\nu}{2}\phi_{(r,j,i_j)}\right).
\end{align*}
Thus, the conditional posterior of $\phi_{(r,j,i_j)}$ is a Gamma distribution
\begin{align*}
    \phi_{(r,j,i_j)}\mid \boldsymbol{\beta}^{(r)}_{j,i_j}, \tau_r \sim \text{Gamma}\left(\frac{\nu+1}{2},\frac{\nu+\tau_r\left(\boldsymbol{\beta}^{(r)}_{j,i_j}\right)^2}{2}\right).
\end{align*}

$\delta_1$ involves in all $\tau_r$'s, for $r=1,\dots, R$, so the sampling $\delta_1$ is conditional to all margins and corresponding hyperparameters, denoted as $\cdot$. Combining likelihood and prior, we get
\allowdisplaybreaks
\begin{align}
    p(\delta_1\mid \cdot)&\propto \delta_1^{a_1-1}\exp (-\delta_1) \prod_{r=1}^R\prod_{j=1}^3\prod_{i_j=1}^{I_j}\delta_1^{\frac{1}{2}}\exp\left( -\frac{\phi_{(r,j,k)}\delta_1\tau^{(1)}_l\left(\boldsymbol{\beta}^{(r)}_{j,i_j}\right)^2}{2}\right)\label{derive_delta1}\\
    &=\delta_1^{a_1+\frac{(2N+P)R}{2}-1}\exp \left(-\left(1+\sum_{r=1}^R\tau^{(1)}_l\sum_{j=1}^3\sum_{i_j=1}^{I_j} \frac{\phi_{(r,j,i_j)}\left(\boldsymbol{\beta}^{(r)}_{j,i_j}\right)^2}{2}\right)\delta_1\right),\nonumber
\end{align}
where $\tau^{(r)}_l=\prod_{i=1,i\neq r}^{l}\delta_i$, $p_j$ is the number of rows in $\boldsymbol{B}_j$.

The derivation leads to a Gamma conditional posterior of $\delta_1$
\allowdisplaybreaks
\begin{equation*}
    \delta_1\mid\cdot \sim \text{Gamma}\left(a_1+\frac{(2N+P)R}{2},1+\frac{1}{2}\sum_{l=1}^R\tau^{(1)}_l\sum_{d=1}^3\sum_{k=1}^{p_j} \phi_{(r,j,i_j)}\left(\boldsymbol{\beta}^{(l)}_{j,i_j}\right)^2\right).
\end{equation*}
The derivation of the conditional posterior of $\delta_r$, for $r>1$, is similar to the above derivation, but the prior and likelihood are slightly different. We first need to change $a_1$ in \ref{derive_delta1} to $a_2$, and since $\delta_r$ is only related to $\beta^{(l)}_{j,i_j}$'s and their corresponding hyperparameters, where $l\geq r$, the starting value of $l$ is $r$ rather than 1, and we amend $\tau^{(1)}_l$ to $\tau^{(r)}_l$. This results to a Gamma conditional posterior of $\delta_r$
\begin{equation*}
    \delta_r \mid \cdot \sim \text{Gamma}\left(a_2+\frac{(2N+P)(R-r+1)}{2},1+\frac{1}{2}\sum_{l=r}^R\tau^{(r)}_l\sum_{d=1}^3\sum_{i_j=1}^{p_j} \phi_{(r,j,i_j)}\left(\boldsymbol\beta^{(l)}_{j,i_j}\right)^2\right),
\end{equation*}
where we keep $\cdot$ as conditions for brevity. $\tau_r$ is updated as the product of $\delta_1,\dots,\delta_r$ in each iteration. 
\subsection{Details for Other Full Conditionals}
\label{Details for Other Full Conditionals}
Conditional posteriors related to the normal-gamma prior (hyperparameters of $\boldsymbol{D}$ and $\boldsymbol{H}$) are almost identical to those in \cite{huber2019adaptive}. The only difference is that these posteriors are conditional on entries of $\boldsymbol{D}$ and $\boldsymbol{H}$, instead of the coefficient matrix. 

The conditional posterior of $\boldsymbol{H}$ can also be found in \cite{huber2019adaptive}. For stochastic volatility, we use an ASIS algorithm proposed in \cite{kastner2014ancillarity} and implement it with an R package called \textbf{stochvol} \citep{JSSv069i05}.

\section{Algorithms}
\label{Full Gibbs Sampler}
\begin{breakablealgorithm}
\caption{Adaptive Inference of Rank}\label{alg:1}
\begin{algorithmic}
\State Initialize $R^*,\, \alpha_0,\, \alpha_1$ and set a criterion
\While{iteration $\tilde{m}<m\leq m_{\text{burn-in}}$}
\State Sample $u$ from Uniform(0,1) and let $R^{(m)}$ be the rank at iteration $m$
\If{$p(m)\geq u$}
\State{$k$=\#\{inactive columns\}}
    \If{$k>0$}
        \State{Remove inactive columns in $\boldsymbol{B}$ and related parameters}
        \State{$R^{(m)}=R^{(m-1)}-k$}
    \Else
        \State{Add one column to $\boldsymbol{B}$ and expand related parameters}
        \State{$R^{(m)}=R^{(m-1)}+1$}
    \EndIf
\EndIf 
\EndWhile
\end{algorithmic}
\end{breakablealgorithm}

\vskip 1cm
\begin{customthm}{C.2}\label{alg:2_1}
\normalfont Initialize unknown parameters and repeat the following steps in each iteration:
\begin{enumerate}[label=(\alph*),align=left,leftmargin=*]
\setstretch{1.1}
\item [Step (a):] Update $\boldsymbol{B}^{\text{old}}_1$ under the base parameterization.
\item[Step (b*):] Store the first row of $\boldsymbol{B}^{\text{old}}_1$ into $\boldsymbol{D}_1$ and determine $\boldsymbol{B}^*_1,\boldsymbol{B}^*_2$.
\item[Step (b**):] Sample $\left(\boldsymbol{\beta}^{\text{new}(r)}_{1,1}\right)^2$ for $r=1,\dots,R$ using the second parameterization and store corresponding values in $\boldsymbol{D}_1$.
\item[Step (b***):] Update $\boldsymbol{B}^{\text{new}}_1$ and $\boldsymbol{B}^{\tilde{\text{new}}}_2$ with transformation
\begin{align*}
    \boldsymbol{B}^{\text{new}}_1= \boldsymbol{B}^{*}_1\boldsymbol{D}_1,\,  \boldsymbol{B}^{\tilde{\text{new}}}_2= \boldsymbol{B}^{*}_2\boldsymbol{D}^{-1}_1.
\end{align*}
\item [Step (c):] Update $\boldsymbol{B}^{\text{old}}_2$ under the base parameterization.
\item[Step (d*):] Store the first row of $\boldsymbol{B}^{\text{old}}_2$ into $\boldsymbol{D}_2$ and determine $\boldsymbol{B}^{**}_2,\boldsymbol{B}^{**}_3$.
\item[Step (d**):] Sample $\left(\boldsymbol{\beta}^{\text{new}(r)}_{2,1}\right)^2$ for $r=1,\dots,R$ using the third parameterization and store corresponding values in $\boldsymbol{D}_2$.
\item[Step (d***):] Update $\boldsymbol{B}^{\text{new}}_2$ and $\boldsymbol{B}^{\tilde{\text{new}}}_3$ with transformation
\begin{align*}
    \boldsymbol{B}^{\text{new}}_2= \boldsymbol{B}^{**}_2\boldsymbol{D}_2,\,  \boldsymbol{B}^{\tilde{\text{new}}}_3= \boldsymbol{B}^{**}_3\boldsymbol{D}^{-1}_2.
\end{align*}
\item [Step (e):] Update $\boldsymbol{B}^{\text{old}}_3$ under the base parameterization.
\item[Step (f*):] Store the first row of $\boldsymbol{B}^{\text{old}}_3$ into $\boldsymbol{D}_3$ and determine $\boldsymbol{B}^{***}_3,\boldsymbol{B}^{***}_1$.
\item[Step (f**):] Sample $\left(\boldsymbol{\beta}^{\text{new}(r)}_{3,1}\right)^2$ for $r=1,\dots,R$ using the fourth parameterization and store corresponding values in $\boldsymbol{D}_3$.
\item[Step (f***):] Update $\boldsymbol{B}^{\text{new}}_3$ and $\boldsymbol{B}^{\tilde{\text{new}}}_1$ with transformation
\begin{align*}
    \boldsymbol{B}^{\text{new}}_3= \boldsymbol{B}^{***}_3\boldsymbol{D}_3,\,  \boldsymbol{B}^{\tilde{\text{new}}}_1= \boldsymbol{B}^{***}_1\boldsymbol{D}^{-1}_3.
\end{align*}
\item [Step (g):] Sample other unknown parameters from their full conditionals.
\end{enumerate}
\end{customthm}


\section{Additional Results} \label{Additional Results}
\subsection{Additional Results in Simulation Study} \label{Additional Results Simulation}
\begin{figure}[!htbp]
     \centering
     \begin{subfigure}[b]{0.48\textwidth}
         \centering
         \includegraphics[width=\textwidth, height=0.6\textwidth]{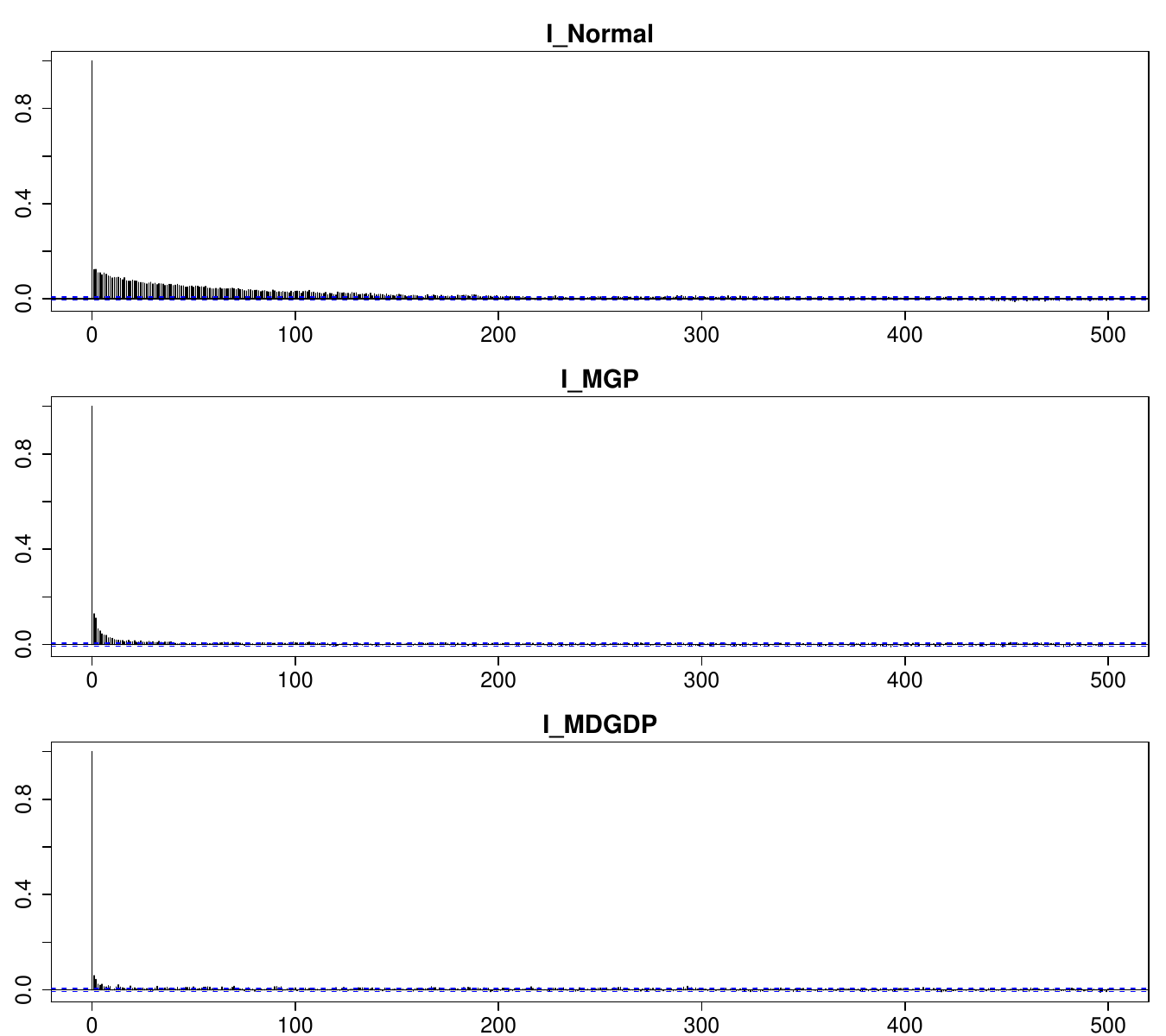}
         \caption{With interweaving}
         \label{acf insight inweaving}
     \end{subfigure}
     \hfill
     \begin{subfigure}[b]{0.48\textwidth}
         \centering
         \includegraphics[width=\textwidth, height=0.6\textwidth]{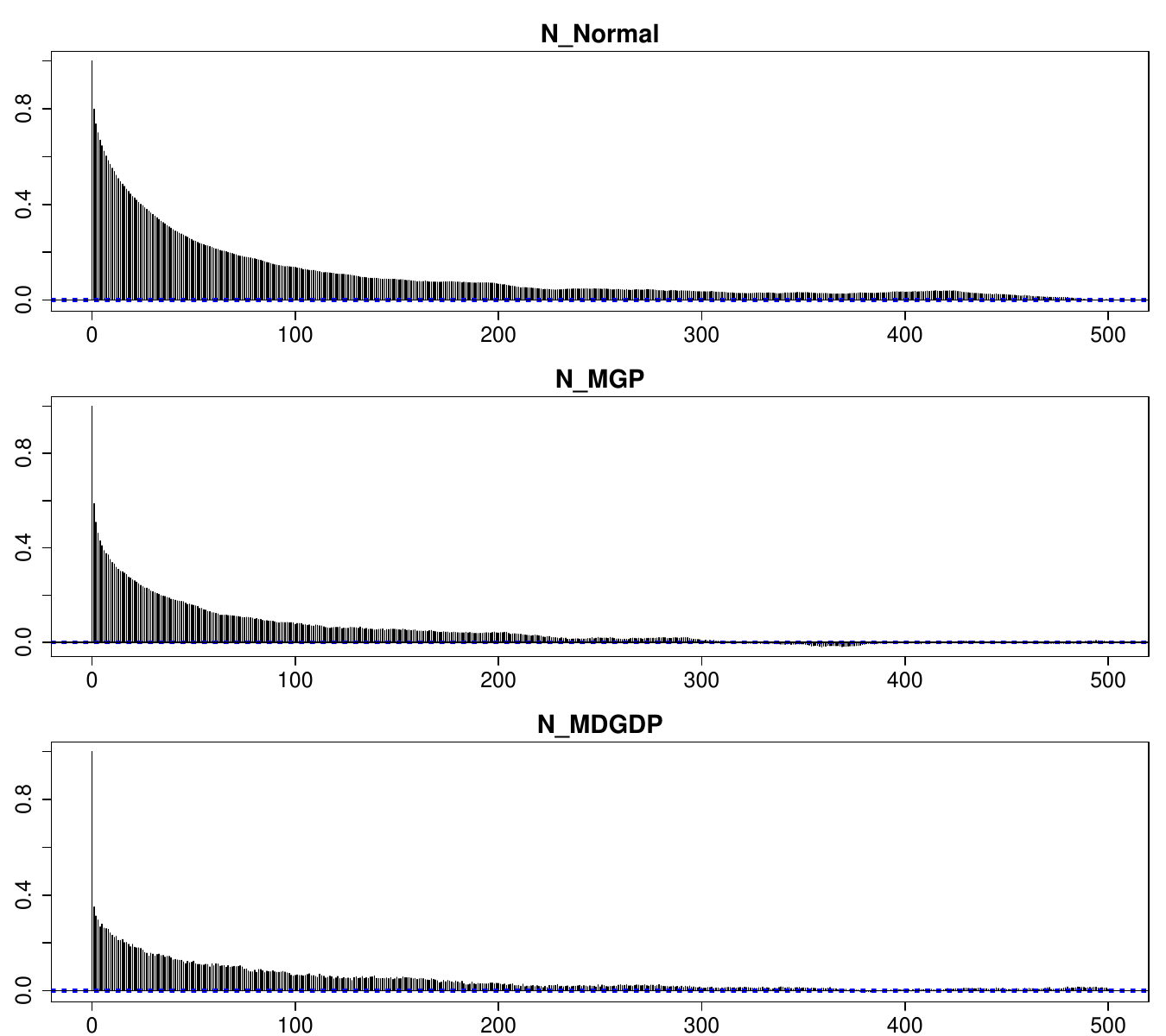}
         \caption{Without interweaving}
          \label{acf insight no inweaving}
     \end{subfigure}
        \caption{Autocorrelations of $\boldsymbol{\beta}^{(1)}_{1,1}$ in $N=10,$ $R=3$ scenario after the burn-in period. The inferential scheme adopts standard normal (top), MGP (middle) and M-DGDP (bottom) as priors and applies with (left panel) and without (right panel) interweaving strategy.}
        \label{acf insight}
\end{figure}

\begin{figure}[!htbp]
     \centering
         \centering
         \includegraphics[width=0.6\textwidth, height = 0.4\textwidth]{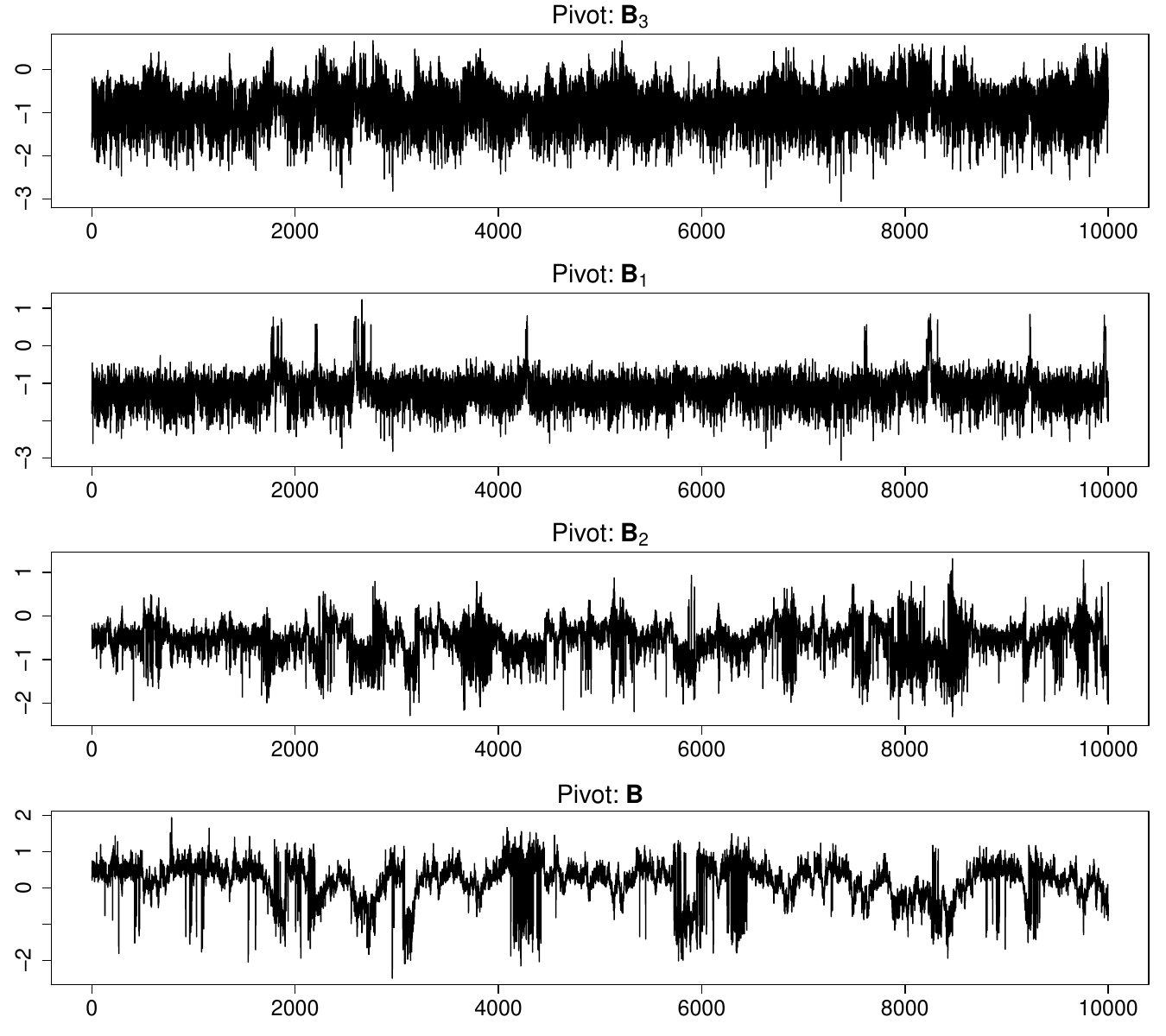}
        \caption{Trace plots of the first 10,000 draws of $\boldsymbol{\beta}^{(1)}_{1,1}$ in $N=10,$ $R=3$ scenario after burn-in period. The inferential scheme adopts standard normal prior with interweaving strategy, and the post-processing procedure in each panel chooses different pivot indicated in the title. }
        \label{fig: mixing insight different pivots}
\end{figure}

\begin{figure}[!htbp]
     \centering
\includegraphics[width=0.6\textwidth,height=0.4\textwidth]{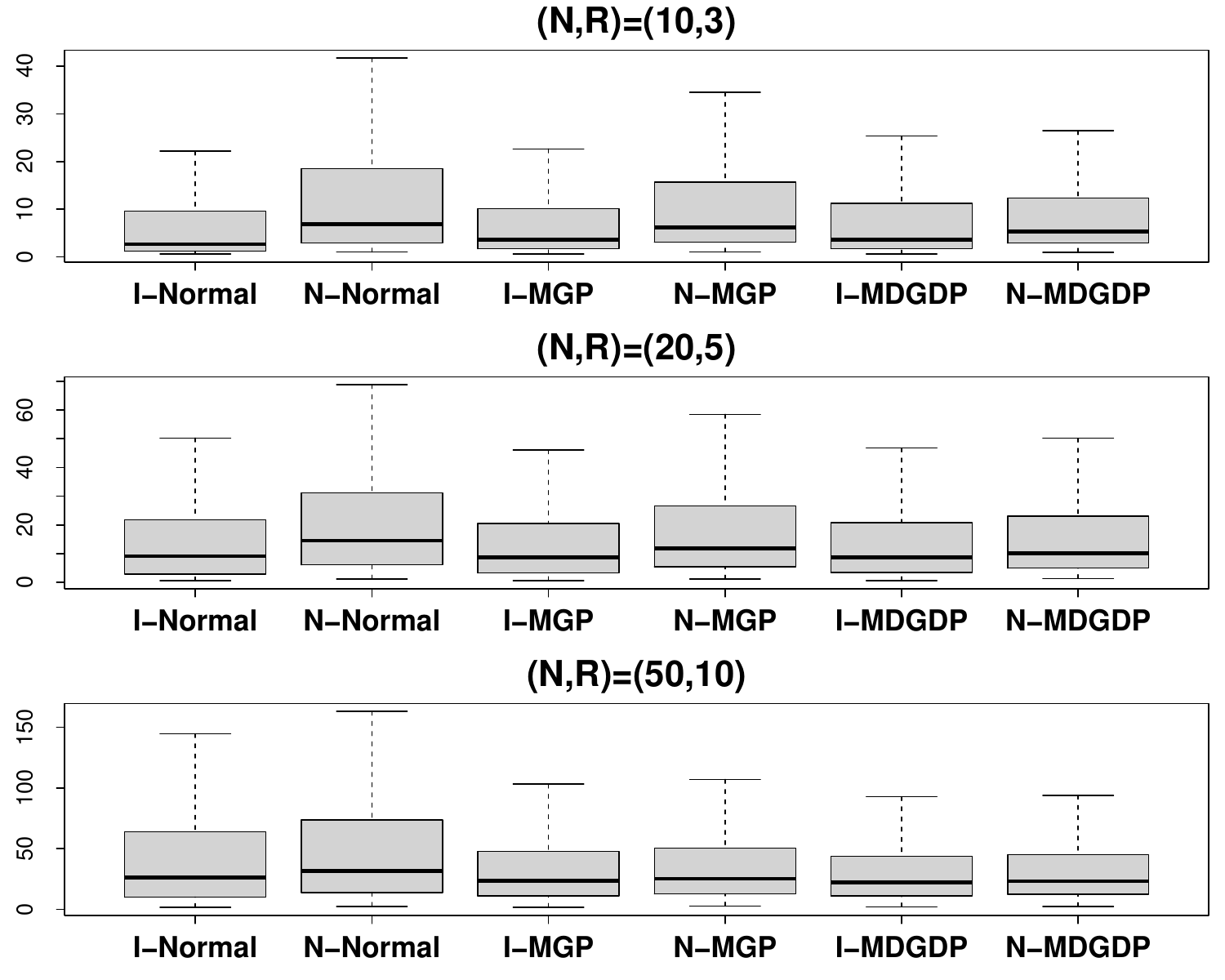}
        \caption{Boxplots of inefficiency factors of coefficient matrices from different scenarios: $(N,R)=(10,3)$ (top), $(N,R)=(20,5)$ (middle) and $(N,R)=(50,10)$ (bottom). Inferential schemes with and without interweaving are represented as "I-" and "N-", respectively, followed by a prior setting. Outliers are discarded.}
        \label{fig:mixing coefficient}
\end{figure}

The section contains the following tables and figures based on the simulation study in Section \ref{sec5}:
\begin{itemize}
    \item Table \ref{rank inference table sensitivity} provides the sensitivity test to select thresholds $\gamma_1$ and $\gamma_2$.
    \item Figure \ref{acf insight} presents the autocorrelations (acfs) of all draws of $\boldsymbol{\beta}^{(1)}_{1,1}$ after the burn-in period.
    \item Table \ref{convergence test geweke} shows the convergence diagnostic based on the Geweke diagnostic \citep{geweke1991evaluating}.
    \item Figure \ref{fig: mixing insight different pivots} demonstrates trace plots of a margin based on different pivots described in Section \ref{ppp}.
    \item Figure \ref{fig:mixing coefficient} presents the inefficiency factors of coefficients. 
\end{itemize}

To choose $\gamma_1$ and $\gamma_2$, we used $\gamma_1$ from a range of values close to 0, $\{10^{-4},10^{-3}, 5$\texttimes$ 10^{-4},10^{-3} \}$, and $\gamma_2$ from values below and close to 1, $\{0.85, 0.9, 0.95\}$, to the simulation study of the $(N, R) = (10,3)$ scenario. Table \ref{rank inference table sensitivity} provides the inferential results based on different combinations of $\gamma_1$ and $\gamma_2$. The inference of coefficients is not sensitive to the combination, but the rank inferred is. We choose $\gamma_1 = 0.001$ and $\gamma_2=0.9$ because this combination leads to the lowest rank value and narrowest 90\% credible interval. 
\begin{table}[!htbp]
\centering
\scriptsize
\begin{tabular}{lllll}
\hline
($\gamma_1$, $\gamma_2$) & MSE                 & R               & ESS                          & Running Time (hr)   \\
\hline
\addlinespace[0.2cm]
(0.0001,0.85)          & 0.011 (0.003,0.037) & 5.4 (3.6,9.4)   & 4199.721 (2641.187,7027.556) & 0.387 (0.344,0.445) \\
\addlinespace[0.1cm]
(0.0001,0.9)           & 0.011 (0.003,0.037) & 5.28 (3,9.4)    & 4203.355 (2668.541,6850.881) & 0.397 (0.336,0.471) \\
\addlinespace[0.1cm]
(0.0001,0.95)          & 0.011 (0.003,0.037) & 6.16 (3.6,11.4) & 4186.997 (2629.508,7058.076) & 0.42 (0.379,0.505)  \\
\addlinespace[0.1cm]
(0.0005,0.85)          & 0.011 (0.003,0.037) & 4.24 (3,6.4)    & 4168.846 (2739.378,6953.491) & 0.369 (0.328,0.4)   \\
\addlinespace[0.1cm]
(0.0005,0.9)           & 0.011 (0.003,0.037) & 4.4 (3,6.4)     & 4228.833 (2624.555,7160.893) & 0.368 (0.286,0.4)   \\
\addlinespace[0.1cm]
(0.0005,0.95)          & 0.011 (0.003,0.037) & 3.96 (3,5.8)    & 4222.431 (2645.81,7013.176)  & 0.372 (0.326,0.403) \\
\addlinespace[0.1cm]
(0.001,0.85)          & 0.011 (0.003,0.037) & 3.96 (3,6)      & 4181.059 (2722.343,6912.24)  & 0.412 (0.366,0.445) \\
\addlinespace[0.1cm]
(0.001,0.9)           & 0.011 (0.003,0.038) & 3.84 (3,5)      & 4128.954 (2603.196,6871.378) & 0.429 (0.361,0.504) \\
\addlinespace[0.1cm]
(0.001,0.95)          & 0.011 (0.003,0.037) & 3.84 (3,5.8)    & 4143.039 (2707.385,6898.21)  & 0.404 (0.347,0.509)\\
\hline
\end{tabular}
\vskip 0.5cm
\caption{Sensitivity table of the MGP with different threshold ($\gamma_1$) and proportion ($\gamma_2$). Each cell correspond to averaged value over 25 simulations with a 90\% credible interval in the parenthesis.}
\label{rank inference table sensitivity}
\end{table}

Figure \ref{acf insight} displays autocorrelations (acfs) of all draws of $\boldsymbol{\beta}^{(1)}_{1,1}$ after the burn-in period. The acfs from the interweaving strategy decay quickly, with only the one from the standard normal prior showing non-negligible values by 100 lags. All of these three acfs without interweaving remain large for many lags.

Figure \ref{fig: mixing insight different pivots} suggests using $\boldsymbol{B}_3$ as candidates for the pivot matrix, rather than $\boldsymbol{B}_1$, $\boldsymbol{B}_2$ and $\boldsymbol{B}$, because the result corresponding to $\boldsymbol{B}_3$ has the best mixing. Note that numbers of rows in $\boldsymbol{B}_1$, $\boldsymbol{B}_2$, $\boldsymbol{B}_3$ and $\boldsymbol{B}$ are $N$, $N$, $P$, $2N\texttt{+}P$, respectively. One possible reason for this best performance of using $\boldsymbol{B}_3$ is that $N$ and $2N\texttt{+}P$ are greater than $P$ in our simulation and real data experiments, so 
it is easier to correctly match columns in $\boldsymbol{B}_3$ to those in $\boldsymbol{B}^\text{pivot}_3$, compared to similar procedures using $\boldsymbol{B}_1$, $\boldsymbol{B}_2$ and $\boldsymbol{B}$. 

\begin{table}[H]
\scriptsize
\centering
\begin{tabular}{lllllllll}
\hline
\textbf{N=10\_R=3} &  Interweaving & Non-interwoven & \textbf{N=20\_R=5} &    Interweaving & Non-interwoven   & \textbf{N=50\_R=10} &    Interweaving & Non-interwoven\\
\hline
Normal  & 0.950 &	0.940	&  Normal &  	0.910 & 	0.921 & 	Normal	& 0.913	&  0.911\\
MGP & 	0.665	& 0.752	& MGP &	0.627	& 0.615	& MGP&	0.636	&0.665\\
MDGDP &	0.912	& 0.680	& MDGDP&	0.881&	0.760&	MDGDP	& 0.813&	0.787\\
\hline
\end{tabular}
\vskip 0.5cm
\caption{Averaged proportions of margins which are convergent according to Geweke’s Diagnostics.}
\label{convergence test geweke}
\end{table}

\subsection{Additional Descriptions and Results about Forecasting} \label{Additional Results Real}
\begin{sidewaysfigure}[!htb]
    \centering
    \includegraphics[width=0.95\textwidth]{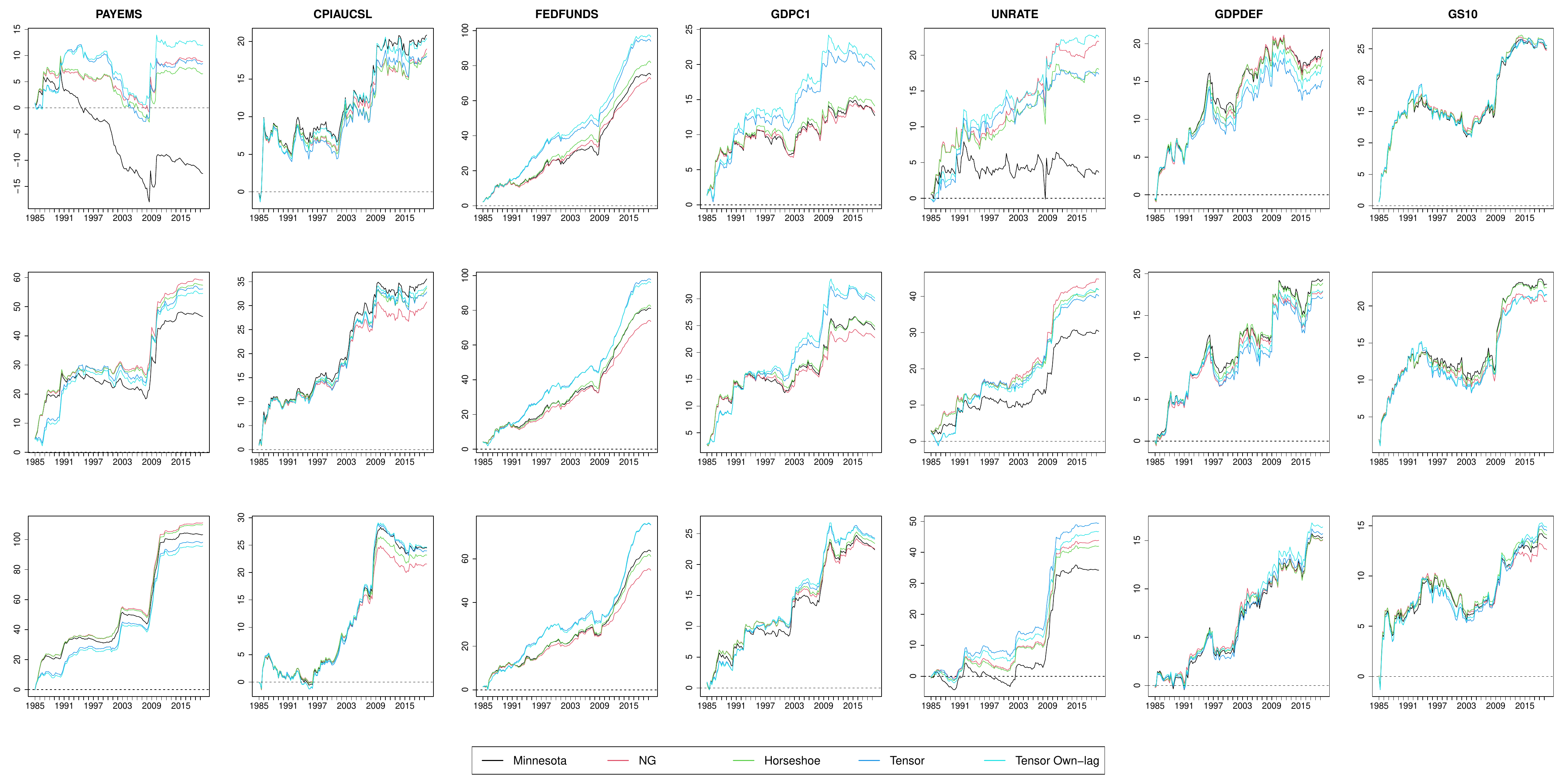}
    \caption{Cumulative ALPL relative to the flat prior benchmark.}
    \label{fig:cumulative alpl}
\end{sidewaysfigure}

\begin{figure}[!htbp]
     \centering
     \begin{subfigure}[b]{0.48\textwidth}
         \centering
    \includegraphics[width=\textwidth, height=0.5\textwidth]{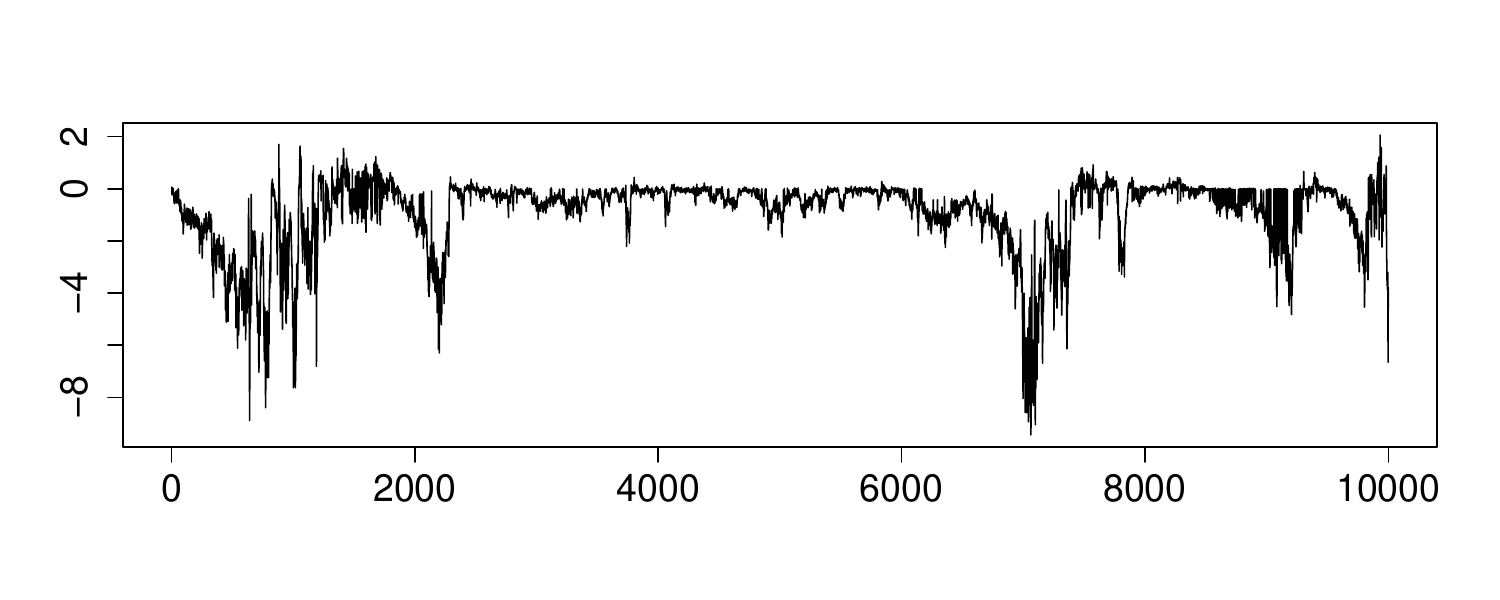}
         \caption{Non-restrictive $\boldsymbol{H}_t$}
         \label{traceplot order invariant}
     \end{subfigure}
     \hfill
     \begin{subfigure}[b]{0.48\textwidth}
         \centering
         \includegraphics[width=\textwidth, height=0.5\textwidth]{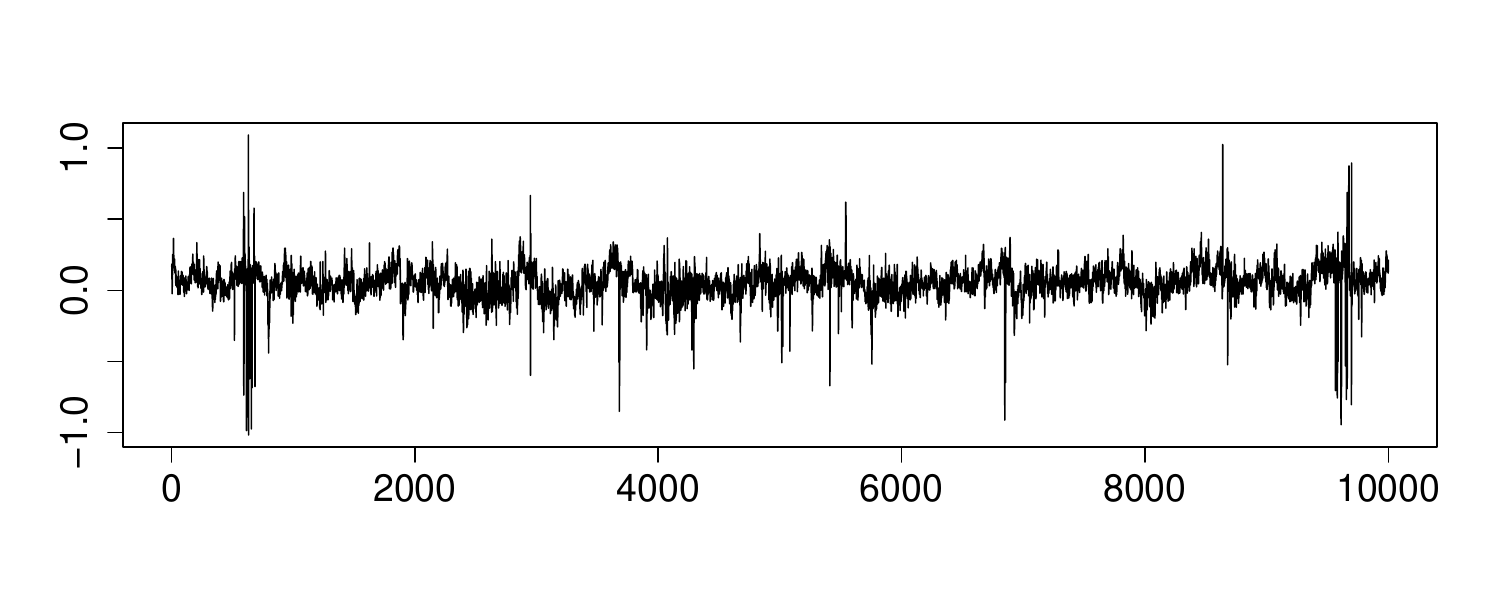}
         \caption{Lower-triangular $\boldsymbol{H}_t$}
         \label{traceplot order-variant}
     \end{subfigure}
        \caption{Trace plots of the 27-2 entry of the tensor matrix $\boldsymbol{B}$. }
        \label{ordering issue}
\end{figure}

This subsection includes supplementary materials of Section \ref{Forecasting Results}:
\begin{itemize}
    \item Table \ref{RMSFE medium} and Table \ref{RMSFE large}: MSFE of the medium and large data sets, respectively.
     \item Table \ref{MAE medium} and Table \ref{MAE large}: MAE of the medium and large data sets, respectively.
     \item Figure \ref{fig:cumulative alpl}: Cumulative marginal ALPL from the large data set.
     \item Table \ref{MSFE medium alternative}, \ref{MAE medium alternative} and \ref{ALPL medium alternative}: MSFE, MAE and ALPL from the alternative medium data set.
     \item Table \ref{RMSFE medium order-invatiant}, \ref{MAE medium order-invatiant} and \ref{ALPL medium order-invatiant}: MSFE, MAE and ALPL from the ordering-invariant model fitted from the medium data set.
     \item Table \ref{RMSFE large order-invatiant}, \ref{MAE large order-invatiant} and \ref{ALPL large order-invatiant}: MSFE, MAE and ALPL from the ordering-invariant model fitted from the large data set.
     \item Figure \ref{ordering issue}: Trace plots of margins using order-invariant model and the Tensor VAR described in Equation \labelcref{standard_VAR}
\end{itemize}
We use mean squared forecast error (MSFE), mean absolute error (MAE) and average log predictive likelihood (ALPL) to assess the point and density forecasting performance. Both joint and marginal forecasting performance are evaluated. The joint MSFE is the averaged MSFE over the 20 and 40 time series for medium- and large-scale data sets, respectively:
\begin{equation*}
    \text{MSFE}_{m,\text{joint}}=\frac{1}{N}\sum_{i=1}^N\text{MSFE}_{m,i}\text{, with }\text{MSFE}_{m,i}=\frac{1}{\overline{T}-h-T+1}\sum_{t=T}^{\overline{T}-h}\left(\boldsymbol{y}_{t+h,i}-\mathop{\mathbb{E}}\left(\boldsymbol{y}_{t+h,i}\mid \boldsymbol{y}_{1:t} ,m\right)\right)^2,
\end{equation*}
where  $\overline{T}$ is the total number of time points in the data set, and $h$ is the horizon. $\mathop{\mathbb{E}}(\boldsymbol{y}_{t+h,i}\mid \boldsymbol{y}_{1:t} ,m)$ is the Monte Carlo estimate of posterior predictive mean. 

Similarly, the joint MAE is written as:
\begin{equation*}
    \text{MAE}_{m,\text{joint}}=\frac{1}{N}\sum_{i=1}^N\text{MAE}_{m,i}\text{, with }\text{MAE}_{m,i}=\frac{1}{\overline{T}-h-T+1}\sum_{t=T}^{\overline{T}-h}\lvert\boldsymbol{y}_{t+h,i}-\mathop{\mathbb{E}}\left(\boldsymbol{y}_{t+h,i}\mid \boldsymbol{y}_{1:t} ,m\right)\rvert,
\end{equation*}

We follow \cite{billio2023bayesian} to approximate the joint ALPL, $\text{ALPL}_{joint}$, by its Monte Carlo estimate in terms of stochastic volatility and lower triangular matrix sampled over the $L$ iterations ($L=10,000$ in this case),
\begin{equation*}
     \text{ALPL}_{joint}\approx\frac{1}{\overline{T}-h-T+1}\sum_{t=T}^{\overline{T}-h} \log \, \left(\frac{1}{L}\sum_{l=1}^L p\left(\boldsymbol{y}_{t}\mid \boldsymbol{y}_{1:t},\boldsymbol{S}^{(l)}_{(t+1):(t+h)},\boldsymbol{H}^{(l)}\right)\right).
\end{equation*}

Marginal MSFE and MAE for $i$-th variable from model $m$ relative to the benchmark are defined as
\begin{equation*}
    \text{RMSFE}_{m,i}=\frac{ \text{MSFE}_{m,i}}{ \text{MSFE}_{\text{benchmark},i}},\,
    \text{RMAE}_{m,i}=\frac{ \text{MAE}_{m,i}}{ \text{MAE}_{\text{benchmark},i}}.
\end{equation*}

Similarly, the relative ALPL for the $i$-th variable from model $m$ is
\begin{equation*}
\text{RALPL}_{m,i}=\text{ALPL}_{m,i}-\text{ALPL}_{\text{benchmark},i},
\end{equation*}
where $\text{ALPL}_{m,i}$ is also approximated by its Monte Carlo estimate.

Table \ref{RMSFE medium} and Table \ref{RMSFE large} show the performance of joint and marginal point forecasts inferred from data sets with different sizes. Overall, Tensor VARs achieve better joint performance than standard VARs. For the marginal performance, Tensor VARs outperform standard VARs in 11 out of 21 cases for both data sets. Forecasts of PAYEMS, UNRATE, and GDP are more favourable when using Tensor VARs, while standard VARs have better performance in forecasting CPIAUCSL, FEDFUNDS and GS10. For the point forecasts evaluated by MAE, see Table \ref{MAE medium} and Table \ref{MAE large}, most results are consistent with the MSFE. One notable difference is that Tensor VARs are better than standard VARs in forecasting FEDFUNDS and GS10. 

Comparing the difference between point and density forecasting performance, we notice that the best model for forecasting PAYEMS in longer horizons (h = 2 or 4) changes from Tensor VARs to standard VARs, if one considers density forecasts rather than the point ones. We inspects margin density forecasts of PAYEMS by looking at the cumulative log predictive likelihood shown in Figure \ref{fig:cumulative alpl}. A potential explanation for the inferior performance of Tensor VARs compared to standard VARs in forecasting PAYEMS is attribute to the volatile economic data before the Great Moderation, because the slopes of cumulative ALPLs are less steeper than the standard VAR counterparts from 1985 to 1990, and they share a similar trend afterwards.

 \begin{table}[!htbp]
\scriptsize
\centering
\begin{tabular}{llcccccccl}
\hline
\input{forecast_RMSFE_medium_1.tex}
\end{tabular}
\caption{MSFE of joint and marginal variables using the medium-scale data set. The best forecasts are in bold. }
\label{RMSFE medium}
\end{table}
\begin{table}[!htbp]
\scriptsize
\centering
\begin{tabular}{llcccccccl}
\hline
\input{forecast_RMSFE_large.tex}
\end{tabular}
\caption{MSFE of joint and marginal variables using the large-scale data set. The best forecasts are in bold. }
\label{RMSFE large}
\end{table}
\begin{table}[!htb]
\scriptsize
\centering
\begin{tabular}{llcccccccl}
\hline
\input{forecast_MAE_medium_new.tex}
\end{tabular}
\caption{MAE of joint and marginal variables using the medium-scale data set. The best forecasts are in bold. }
\label{MAE medium}
\end{table}

\begin{table}[!htb]
\scriptsize
\centering
\begin{tabular}{llcccccccl}
\hline
\input{forecast_MAE_large_new.tex}
\end{tabular}
\caption{MAE of joint and marginal variables using the large-scale data set. The best forecasts are in bold. }
\label{MAE large}
\end{table}

\begin{table}[!htb]
\scriptsize
\centering
\begin{tabular}{llcccccccl}
\hline
\input{forecast_RMSFE_medium_alternative.tex}
\end{tabular}
\caption{MSFE of joint and marginal variables using an alternative medium-scale data set. The best forecasts are in bold. }
\label{MSFE medium alternative}
\end{table}

\FloatBarrier

To check the robustness of forecasting performance across different variable choices in the medium data set, we construct an alternative medium  data set with variables selected in Appendix \ref{data}. Experimental results from this data set are available in Table \ref{MSFE medium alternative}, \ref{MAE medium alternative} and \ref{ALPL medium alternative}. These results lead to the similar conclusion as the ones from the medium data set: Tensor VARs are better than standard VARs in joint point (especially for MAE) and density forecasting. Marginal performance in Tensor VARs is also competitive. A consistent pattern in these results mirrors the findings from the original medium data set: 1) Tensor VARs yield better density forecasts for CPIAUSL compared to point forecasts; 2) standard VARs outperforms Tensor VARs in forecasting FEDFUNDS when the evaluation metric is MSFE, yet Tensor VARs exhibit superior forecasts when the metric changes to ALPL and MAE. 

\begin{table}[!htb]
\scriptsize
\centering
\begin{tabular}{llcccccccl}
\hline
\input{forecast_MAE_medium_alternative.tex}
\end{tabular}
\caption{MAE of joint and marginal variables using an alternative medium-scale data set. The best forecasts are in bold. }
\label{MAE medium alternative}
\end{table}

\begin{table}[!htb]
\scriptsize
\centering
\begin{tabular}{llcccccccl}
\hline
\input{forecast_ALPL_medium_alternative.tex}
\end{tabular}
\caption{ALPL of joint and marginal variables using an alternative medium-scale data set. The best forecasts are in bold. }
\label{ALPL medium alternative}
\end{table}

The last consideration about forecasting is the ordering issue due to the decomposition of $\boldsymbol{\Omega}_t$, the variance-covariance matrix. The Cholesky Decomposition of $\boldsymbol{\Omega}_t$ might affect the inference of parameters in (Tensor) VARs, as discussed in many papers - \cite{carriero2019large, chan2024large, arias2023macroeconomic}, among others. The basic idea is that the prior of each element in the variance-covariance matrix $\boldsymbol{\Omega}_t$ depends on the ordering of variables. For example, denote $\boldsymbol{\Omega}_t$ as the variance-covariance matrix corresponding to a particular variable order and $\boldsymbol{\tilde{\Omega}}_t$ as the one corresponding to switching the first and second variables in the original order. Then the prior of the 1-1 entry of $\boldsymbol{\Omega}_t$ is not equivalent to the prior of the 2-2 entry of $\boldsymbol{\tilde{\Omega}}_t$. 

Motivated from the ordering issue, we applied the non-restrictive $\boldsymbol{H}_t$ proposed in \cite{chan2024large} to model $\boldsymbol{\Omega}_t$. The prior of each element in $\boldsymbol{H}_t$ is a standard normal distribution. The inference of margins in Tensor VARs do not require any amendments, and the inference of $\boldsymbol{H}_t$ can be found in the original paper. Table \ref{RMSFE medium order-invatiant} - \ref{ALPL large order-invatiant} give the point and density forecasting performance using medium- and large-data sets. The same conclusion can be found from these tables: Tensor VARs outperforms standard VARs in both point and density forecasts. 

We do not replace the results using Cholesky Decomposition to $\boldsymbol{\Omega}_t$ by the ones using the ordering-invariant decomposition because some Markov chains of margins do not exhibit good mixing if we infer the model with the latter decomposition. For example, Figure \ref{ordering issue} presents the trace plots of the 27-2 entry of the tensor matrix $\boldsymbol{B}$ (corresponding to the effect of the past economy to the M2 money supply). The margin inferred from the ordering-invariant model has bad mixing issue that affects the interpretation of $\boldsymbol{B}$. 

\begin{table}[!htbp]
\scriptsize
\centering
\begin{tabular}{llcccccccl}
\hline
\input{forecast_RMSFE_medium_order_invariant.tex}
\end{tabular}
\caption{MSFE of joint and marginal variables using the medium-scale data set. The models applied are order-invariant. The best forecasts are in bold. }
\label{RMSFE medium order-invatiant}
\end{table}

\begin{table}[!htbp]
\scriptsize
\centering
\begin{tabular}{lllcccccll}
\hline
\input{forecast_MAE_medium_order_invariant.tex}
\end{tabular}
\caption{MAE of joint and marginal variables using the medium-scale data set. The models applied are order-invariant. The best forecasts are in bold. }
\label{MAE medium order-invatiant}
\end{table}

\begin{table}[!htbp]
\scriptsize
\centering
\begin{tabular}{llcccccccl}
\hline
\input{forecast_ALPL_medium_order_invariant.tex}
\end{tabular}
\caption{ALPL of joint and marginal variables using the medium-scale data set. The models applied are order-invariant. The best forecasts are in bold. }
\label{ALPL medium order-invatiant}
\end{table}

\begin{table}[!htbp]
\scriptsize
\centering
\begin{tabular}{llcccccccl}
\hline
\input{forecast_MSFE_large_order_invariant.tex}
\end{tabular}
\caption{MSFE of joint and marginal variables using the large-scale data set. The models applied are order-invariant. The best forecasts are in bold. }
\label{RMSFE large order-invatiant}
\end{table}

\begin{table}[!htbp]
\scriptsize
\centering
\begin{tabular}{llcccccccl}
\hline
\input{forecast_MAE_large_order_invariant.tex}
\end{tabular}
\caption{MAE of joint and marginal variables using the large-scale data set. The models applied are order-invariant. The best forecasts are in bold. }
\label{MAE large order-invatiant}
\end{table}

\begin{table}[!htbp]
\scriptsize
\centering
\begin{tabular}{llcccccccl}
\hline
\input{forecast_ALPL_large_order_invariant.tex}
\end{tabular}
\caption{ALPL of joint and marginal variables using the large-scale data set. The models applied are order-invariant. The best forecasts are in bold. }
\label{ALPL large order-invatiant}
\end{table}

\FloatBarrier

\newpage
\subsection{Additional Results about Interpretation} \label{Additional Results Ordering}
\begin{figure}
    \centering
    \includegraphics[width=0.5\textwidth,height=0.7\textwidth]{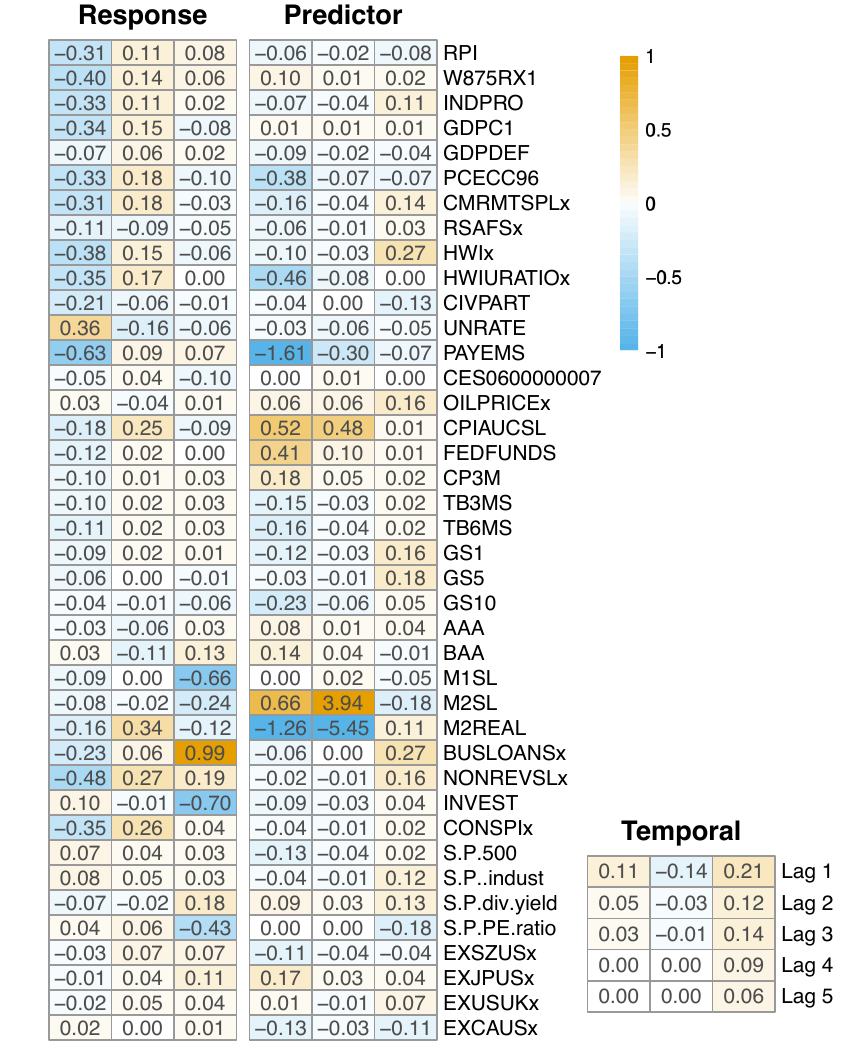}
    \caption{Posterior mean of response, predictor and temporal loadings inferred from Tensor MGP.}
    \label{fig:large tensor MGP}
\end{figure}

\begin{figure}[!htb]
     \centering
         \centering
\includegraphics[width=0.4\textwidth]{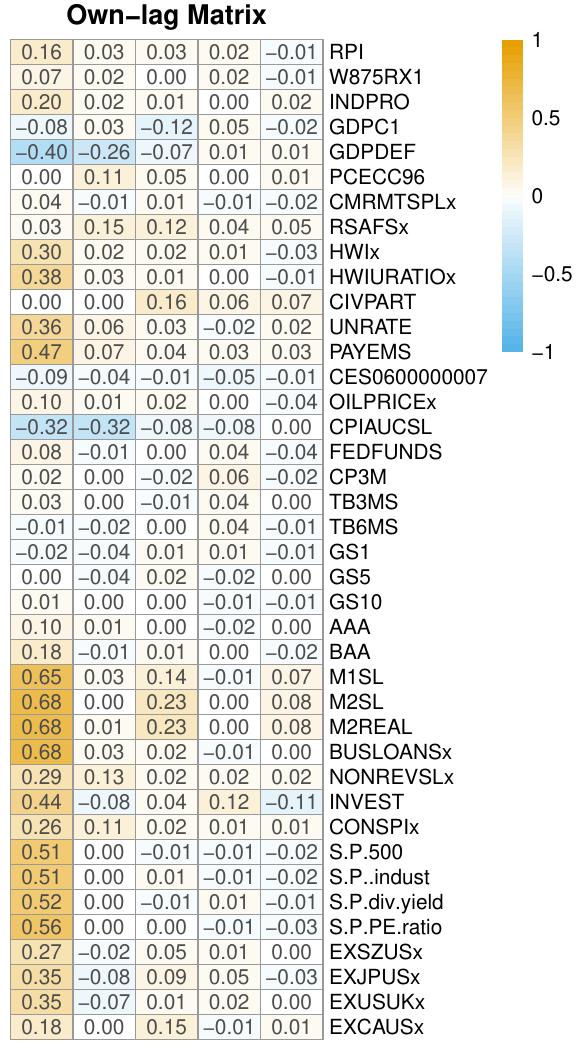}
        \caption{Posterior mean of non-zero entries in the own-lag matrix inferred using the large-scale data set. Each row correspond to a variable and each column is for a lag order.}
        \label{fig:posterior mean of own-lag}
\end{figure}

\begin{figure}[!htb]
    \centering
    \begin{subfigure}[t]{0.3\textwidth}
        \centering
\includegraphics[width=\textwidth]{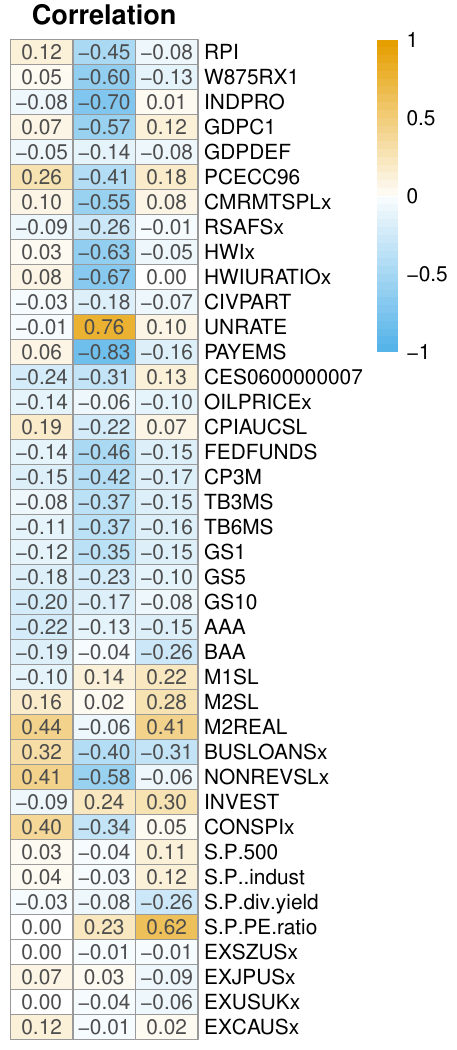}
        \caption{Tensor MGP Own-lag. }
    \end{subfigure}%
    \begin{subfigure}[t]{0.3\textwidth}
        \centering
        \includegraphics[width=\textwidth]{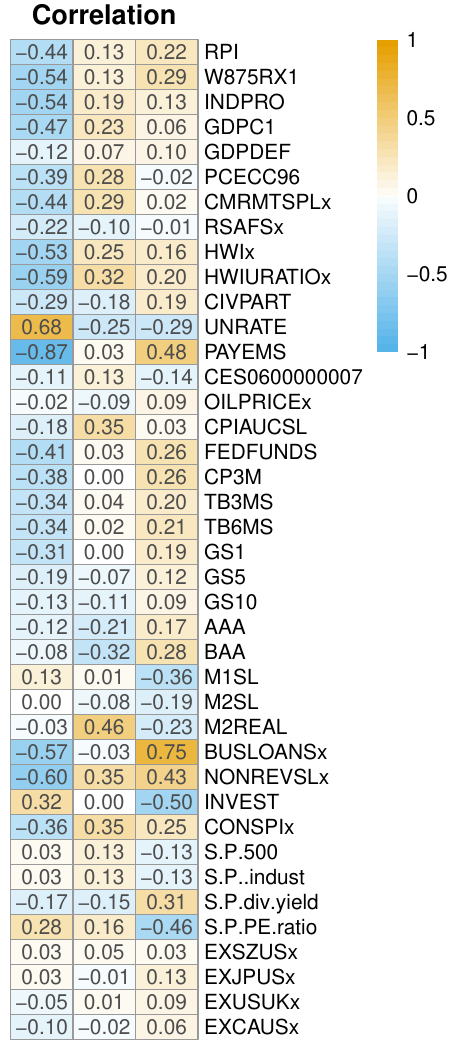}
        \caption{Tensor MGP}
    \end{subfigure}
    \caption{Correlation between variables (rows) and 3 factors (columns)}
    \label{correlation factor and variables}
\end{figure}

\begin{figure}[!htb]
     \centering
         \centering \includegraphics[width=0.6\textwidth, height=0.45\textwidth]{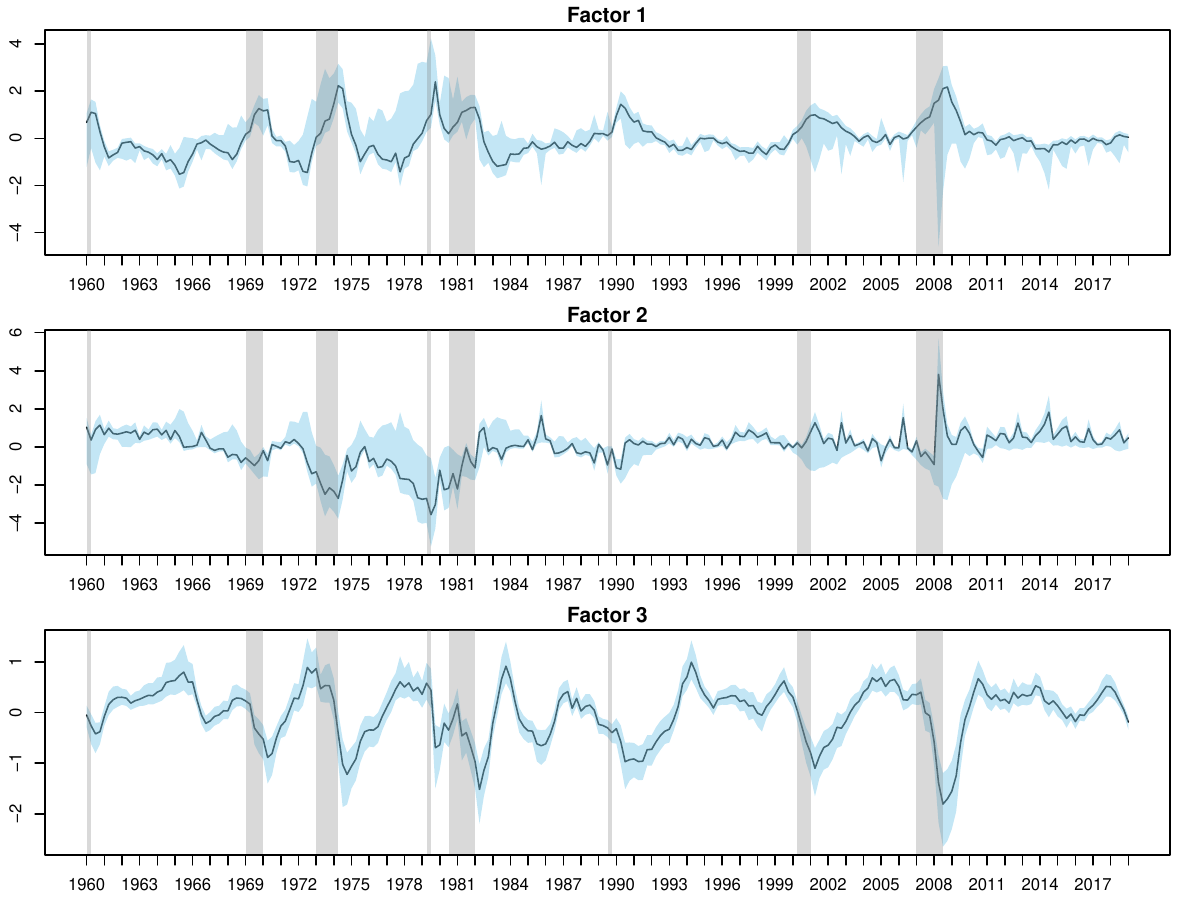}
        \caption{Time series plots of factors with median (solid line) and 80\% credible interval (dashed line). The model does not include an own-lag matrix.}
        \label{fig:factor traceplot no own-lag}
\end{figure}

The following results are the supplementary materials of Section \ref{interpretation} and \ref{effect of D}:
\begin{itemize}
    \item Figure \ref{fig:large tensor MGP} presents the posterior mean of margins inferred from Tensor MGP (without the own-lag matrix).
     \item Figure \ref{fig:posterior mean of own-lag} presents the posterior mean of non-zero elements in the own-lag matrix $\boldsymbol{D}$.
    \item Figure \ref{correlation factor and variables} shows the correlation between factors and variables.
        \item Figure \ref{fig:factor traceplot no own-lag} depicts time series plots of factors inferred from the Tensor VAR without the own-lag matrix. 
    \item Table \ref{distort} shows how the own-lag matrix can avoid over-weighting a particular group of variables.
\end{itemize}
The own-lag matrix is beneficial to model economic time series. Firstly, Figure \ref{fig:posterior mean of own-lag} displays the posterior mean of non-zero elements in the own-lag matrix. Each row and column correspond to one variable and a lag order, respectively. Own-lag effect is found in all categories except interest rate, with the first lag being the most significant. Secondly, Figure \ref{fig:large tensor MGP} shows that the same variable (PAYEMS) is associated with the largest margin magnitudes in the first columns of response and predictor loadings. This pattern holds true for the second and third columns as well, with corresponding variables being M2REAL and BUSLOANS. These large margins have the potential to distort the coefficients in such a manner that the rows and columns corresponding to these three variables in the coefficient matrix exhibit a higher proportion of large magnitudes compared to their counterparts associated with other variables. The evidence can be found in Table \ref{distort}, wherein the former proportions mentioned are 2 to 6 times larger than the latter proportions corresponding to other variables, when the results of the Tensor MGP are considered. However, when we apply Tensor MGP Own-lag, the proportions across variables appear to be similar.
\FloatBarrier

\begin{table}[!htb]
\centering
\footnotesize
\begin{tabular}{llllll}
\hline
 &                     & \multicolumn{2}{c}{Response}               & \multicolumn{2}{c}{Predictor}              \\
 \hline
                         &                     & 3 Variables & Other Variables & 3 Variables & Other Variables \\
                         \hline
Tensor MGP               & \textgreater{}0.001 & 0.687                    & 0.381           & 0.720                    & 0.378           \\
                         & \textgreater{}0.01  & 0.183                    & 0.049           & 0.265                    & 0.043           \\
                         & \textgreater{}0.1   & 0.01                     & 0.003           & 0.027                    & 0.002           \\
                         \hline
Tensor MGP Own-lag       & \textgreater{}0.001 & 0.595                    & 0.438           & 0.472                    & 0.478           \\
                         & \textgreater{}0.1   & 0.04                     & 0.027           & 0.052                    & 0.025           \\
                         & \textgreater{}0.1   & 0                        & 0               & 0                        & 0 \\
                         \hline
\end{tabular}
\caption{Proportion of coefficient magnitudes greater than certain values (0.001, 0.01 and 0.1) corresponding to a group of variables. "Response" and "Predictor" means whether this group of variables are considered as responses or predictors. If they are responses, we split coefficients by rows; otherwise, we split them by columns. "3 Variables" means the group of PAYEMS, M2REAL and BUSLOANS, and "Other Variables" corresponds to the rest variables.}
\label{distort}
\end{table}
\FloatBarrier

\FloatBarrier

\section{Data} \label{data}
\begin{center}
\input{data_appendix.tex}
\end{center}
\begin{center}
\input{data_appendix_1.tex}
\end{center}
\putbib
\end{bibunit}

\end{document}

%% file: computational_cost.tex
 &\multicolumn{2}{c}{Number of Parameters}   & \multicolumn{2}{c}{Runnning Time (hr)}\\
 \hline
\textbf{}                         & Medium             & Large             & Medium             & Large                 \\
\hline 
\rule[1pt]{0pt}{10pt}
Tensor MGP                  &       187.18             & 257.361     & 0.95               & 3.14                      \\
Tensor MGP Own-lag          &       272.19             & 456.18       & 1.07               & 3.28                     \\
Standard VAR              &        2000               & 8000   & 1.30                & 10.39                 \\
\hline

%% file: forecast_ALPL_medium.tex
Model &
  Horizon &
  ALPL &
   &
   &
   &
   &
   &
   &
   \\
   \hline
 &
   &
  Joint &
  PAYEMS &
  CPIAUCSL &
  FEDFUNDS &
  GDP &
  UNRATE &
  GDPDEFL &
  GS10 \vspace*{0.5mm}\\
\multirow{3}{*}{Tensor   MGP} &
  1 & -16.378	&0.170	&0.151	&0.637	&0.177	&0.150	&0.124	&0.160 \\
 &
  2 &\textbf{-17.820}&	0.416	&0.227	&0.634	&0.240	&0.284	&0.141	&0.128  \\
 &
  4 &\textbf{-19.460}&	0.671&	\textbf{0.179}&	0.498&	0.196&	0.306&	0.110&	0.077 \vspace*{-2mm}\\
 &
   &
  \textbf{} &
   &
   &
   &
   &
   &
   &
   \\
\multirow{3}{*}{Tensor   MPG Own-lag} &
  1 &-16.184&	\textbf{0.190}&	0.147	&\textbf{0.682}	&\textbf{0.191}	&\textbf{0.172}	&0.133&	0.163\\
 &
  2 &-17.852	&0.424	&\textbf{0.229}	&\textbf{0.656}	&\textbf{0.249}	&0.289	&0.144	&0.127\\
 &
  4 &-19.567	&0.702	&0.171	&\textbf{0.526}	&\textbf{0.207}	&\textbf{0.310}	&\textbf{0.113}	&0.081 \vspace*{-2mm}\\
 &
   &
   &
   &
   &
   &
   &
   &
   &
   \\
\multirow{3}{*}{Minnesota} &
  1 &\textbf{-15.921}&	0.129&	\textbf{0.183}&	0.519	&0.141	&0.164	&\textbf{0.181}	&\textbf{0.187}  \\
 &
  2 &-18.126	&\textbf{0.443}	&0.210	&0.507	&0.202	&\textbf{0.301}	&0.134	&\textbf{0.141}  \\
 &
  4 &-19.897&	\textbf{0.754}&	0.142&	0.379&	0.152&	0.291&	0.086&	0.082   \vspace*{-2mm}\\
 &
   &
   &
   &
   &
   &
   &
   &
   &
   \\
\multirow{3}{*}{NG} &
  1 &-16.463&	0.126&	0.126&	0.640&	0.131&	0.153&	0.149&	0.162  \\
 &
  2 &-18.277&	0.402&	0.193&	0.588&	0.183&	0.272&	0.130&	0.126  \\
 &
  4 &-19.995&	0.724&	0.140&	0.448&	0.170	&0.281	&0.096	&0.081 \vspace*{-2mm}\\
 &
   &
   &
   &
   &
   &
   &
   &
   &
   \\
\multirow{3}{*}{Horseshoe} &
  1 &-17.333	&-0.164	&0.090	&0.633	&0.112	&0.048	&0.168	&0.152\\
 &
  2 &-18.394	&0.214	&0.199	&0.626	&0.162&	0.223	&\textbf{0.146}	&0.130 \\
 &
  4 &-19.464	&0.632	&0.141	&0.495	&0.156	&0.257	&0.104	&\textbf{0.108}\\
  \hline

%% file: forecast_ALPL_large.tex
Model &
  Horizon &
  ALPL &
   &
   &
   &
   &
   &
   &
   \\
   \hline
 &
   &
  Joint &
  PAYEMS &
  CPIAUCSL &
  FEDFUNDS &
  GDP &
  UNRATE &
  GDPDEFL &
  GS10 \vspace*{0.5mm}\\
\multirow{3}{*}{Tensor   MGP} &
  1 &-24.520	&0.078	&0.126	&0.670	&0.151	&0.135	&0.103	&\textbf{0.178} \\
 &
  2 &-29.790&	0.401&	0.231	&\textbf{0.686}&	0.213&	0.286&	0.133&	0.151\\
 &
  4 &-33.847&	0.703&	0.171&	0.532&	\textbf{0.172}&	\textbf{0.353}&	0.108&	0.099 \vspace*{-2mm}\\
\multicolumn{1}{l}{} &
   &
  \textbf{} &
   &
   &
   &
   &
   &
   &
   \\
\multirow{3}{*}{Tensor MPG Own-lag} &
  1 &\textbf{-23.809}&	\textbf{0.101}	&0.143	&\textbf{0.688}	&\textbf{0.159}	&\textbf{0.172}	&0.116	&0.175     \\
 &
  2 &-30.338	&0.389	&0.240&	0.673	&\textbf{0.217}	&0.298&	0.138	&0.151  \\
 &
  4 &-35.631&	0.686&	\textbf{0.176}	&\textbf{0.533}	&0.171&	0.334	&\textbf{0.113}	&\textbf{0.101}\vspace*{-2mm} \\
\multicolumn{1}{l}{} &
   &
   &
   &
   &
   &
   &
   &
   &
   \\
\multirow{3}{*}{Minnesota} &
  1 &-26.576	&-0.073	&\textbf{0.147}	&0.534	&0.103	&0.035	&\textbf{0.133}	&0.174\\
 &
  2 &\textbf{-29.600}&	0.330&	\textbf{0.252}&	0.570	&0.173	&0.212&	\textbf{0.148}	&\textbf{0.162} \\
 &
  4 &\textbf{-32.545}&	0.736&	0.175&	0.445	&0.157&	0.243	&0.105&0.095  \vspace*{-2mm} \\
\multicolumn{1}{l}{} &
   &
   &
   &
   &
   &
   &
  \textbf{} &
   &
   \\
\multirow{3}{*}{NG} &
  1 &-28.455	&0.081	&0.133	&0.518	&0.107	&0.167	&0.130	&0.172 \\
 &
  2 &-32.823	&\textbf{0.421}	&0.218	&0.518	&0.163	&\textbf{0.316}	&0.136	&0.145 \\
 &
  4 &-36.715&	\textbf{0.793}	&0.154	&0.386&	0.159	&0.312	&0.104	&0.085 \vspace*{-2mm} \\
\multicolumn{1}{l}{} &
   &
   &
   &
   &
   &
   &
   &
   &
   \\
\multirow{3}{*}{Horseshoe} &
  1 &-27.915	&0.064	&0.129	&0.584	&0.114	&0.138	&0.124	&\textbf{0.178} \\
 &
  2 &-31.462	&0.408	&0.238	&0.580	&0.178&	0.295&	0.144	&0.158\\
 &
  4 &-34.874&	0.784	&0.165&	0.431&	0.165&	0.299&	0.104	&0.097\\
  \hline

%% file: forecast_RMSFE_medium_1.tex

Model                 & Horizon & MSFE          &                &                &                &                &                &                &                \\
\hline
                      &         & Joint          & PAYEMS            & CPIAUCSL       & FEDFUNDS           & GDP       & UNRATE         & GDPDEFL        & GS10          \vspace*{0.5mm} \\
\multirow{3}{*}{Tensor   MGP}         & 1 & 0.714	&0.624	&1.020	&0.336	&0.639	&0.735	&0.869	&0.759      \\
                      & 2       & 0.776	&0.663	&0.974	&0.374	&0.596	&0.690	&0.741	&0.823         \\
                      & 4       & 0.853	&0.678	&0.988	&0.477	&0.639	&0.680	&0.827	&0.898         \\
                      &         &                &                &                &                &                &                &                &             \vspace*{-2mm}     \\
\multirow{3}{*}{Tensor   MPG Own-lag} & 1 & 0.703&	\textbf{0.588}	&1.178	&0.323	&\textbf{0.608}&\textbf{0.683}	&0.846	&0.750\\
                      & 2       & \textbf{0.773}	&\textbf{0.653}	&0.991	&0.361	&\textbf{0.594}	&\textbf{0.677}	&\textbf{0.737}	&0.818 \\
                      & 4       & \textbf{0.852}	&\textbf{0.656}	&1.008	&0.442	&\textbf{0.623}	&\textbf{0.663}	&\textbf{0.821}	&0.892 \\
                      &         &                &                &                &                &                &                &                &         \vspace*{-2mm}       \\
\multirow{3}{*}{Minnesota} & 1 & \textbf{0.689}&	0.696&	\textbf{0.903}&	0.292&	0.697&	0.725&	\textbf{0.717}&	\textbf{0.735}   \\
                      & 2       & 0.774&	0.682	&0.968	&0.362	&0.653	&0.721	&0.743	&\textbf{0.810}\\
                      & 4       & 0.904&	0.675&	\textbf{0.983}	&0.457&	0.727&	0.725&	0.853	&0.902  \\
                      &         &                &                &                &                &                &                & \textbf{}      &    \vspace*{-2mm}         \\
\multirow{3}{*}{NG} & 1       & 0.710&	0.719&	0.965	&\textbf{0.285}	&0.740	&0.770	&0.795	&0.757    \\
                      & 2       & 0.784	&0.758	&\textbf{0.955}	&\textbf{0.349}	&0.700	&0.780	&0.744	&0.814  \\
                      & 4       & 0.859&	0.718	&0.988	&\textbf{0.437}	&0.656	&0.757	&0.836	&0.888   \\
                      &         &                &                &                &                &                &                &                &          \vspace*{-2mm}      \\
\multirow{3}{*}{Horseshoe}   & 1       & 0.790	& 1.621	& 1.178	& 0.323	& 0.780	&1.004	&0.817	&0.780   \\
                      & 2       & 0.825	&1.009	&0.991&	0.361	&0.716	&0.867	&0.742	&0.822    \\
                      & 4       & 0.873	&0.800	&1.008	&0.442	&0.667	&0.762	&0.839	&\textbf{0.850}\\
                      \hline

%% file: forecast_RMSFE_large.tex
Model &
  Horizon &
  MSFE &
   &
   &
   &
   &
   &
   &
   \\
   \hline
 &
   &
  Joint &
  PAYEMS &
  CPIAUCSL &
  FEDFUNDS &
  GDP &
  UNRATE &
  GDPDEFL &
  GS10 \vspace*{0.5mm}\\
\multirow{3}{*}{Tensor   MGP}         & 1 & 0.720	&0.856	&0.984&	0.333&	0.711&	0.835	&0.887	&\textbf{0.767} \\
 &
  2 &0.757	&0.658	&0.946	&0.352	&\textbf{0.636}	&0.708	&0.740	&0.835\\
 &
  4 &0.810	&\textbf{0.623}	&1.006	&0.442	&0.682	&\textbf{0.648}	&0.820	&0.903\vspace*{-2mm}\\
\multicolumn{1}{l}{} &
   &
   &
   &
   &
   &
   &
   &
  \textbf{} &
  \textbf{} \\
\multirow{3}{*}{Tensor   MPG Own-lag} & 1 & \textbf{0.690}&	\textbf{0.794}	&0.961&	0.301&	\textbf{0.697}&	\textbf{0.738}	&0.860	&0.770       \\
 &
  2 &\textbf{0.748}	&\textbf{0.656}	&\textbf{0.935}	&0.347	&0.644	&\textbf{0.678}&	0.736&	0.832  \\
 &
  4 &0.810	&0.630	&1.001	&0.436	&0.693	&\textbf{0.648}	&\textbf{0.818}	&\textbf{0.897} \vspace*{-2mm}\\
\multicolumn{1}{l}{} &
   &
   &
   &
   &
  \textbf{} &
   &
   &
   &
   \\
\multirow{3}{*}{Minnesota} &
  1 &0.726&	1.247&	0.994	&0.362&	0.790&	1.007&	\textbf{0.844}&	0.777\\
 &
  2 &0.763	&0.847&	0.936&	0.363	&0.669	&0.822	&\textbf{0.728}	&\textbf{0.817} \\
 &
  4 &0.811&	0.703	&\textbf{0.994}&	0.446	&\textbf{0.634}	&0.738	&0.825	&0.898 \vspace*{-2mm} \\
\multicolumn{1}{l}{} &
   &
   &
   &
   &
   &
   &
   &
   &
   \\
\multirow{3}{*}{NG} &
  1 &0.691	&0.923	&\textbf{0.949}	&\textbf{0.295}	&0.820	&0.782	&0.845	&0.778 \\
 &
  2 &0.754	&0.794	&0.952	&0.343	&0.736	&0.729	&0.745	&0.829 \\
 &
  4 &\textbf{0.809}&0.698	&0.999	&\textbf{0.429}	&0.673	&0.701	&0.824	&0.904\vspace*{-2mm} \\
\multicolumn{1}{l}{} &
   &
   &
   &
   &
   &
   &
  \textbf{} &
   &
   \\
\multirow{3}{*}{Horseshoe} &
  1 &0.712	&0.980	&0.973	&0.300	&0.822	&0.842	&0.869	&0.772  \\
 &
  2 &0.759	&0.806	&0.939	&\textbf{0.339}	&0.708	&0.748	&0.738	&0.819\\
 &
  4 &0.810	&0.698	&0.998	&0.433	&0.665	&0.703	&0.824	&0.898\\
  \hline

%% file: forecast_MAE_medium_new.tex
Model &
  Horizon &
  MAE &
   &
   &
   &
   &
   &
   &
   \\
   \hline
&
   &
  Joint &
  PAYEMS &
  CPIAUCSL &
 FEDFUNDS &
  GDP &
  UNRATE &
  GDPDEFL &
  GS10 \vspace*{0.5mm}\\
\multirow{3}{*}{Tensor   MGP}         & 1 & 0.599 &	0.955&	1.021&	0.583&	0.811&	0.945&	0.927&	0.880    \\
 &
  2 & 0.627 & 0.860 &	0.956 &	0.582&	\textbf{0.794}&	0.881&	\textbf{0.883}&	0.895 \\
 &
  4 &0.656	&0.832	&\textbf{0.981}	&0.630	&0.810	&0.896&	0.898&	0.945 \vspace*{-3mm} \\
\multicolumn{1}{l}{} &
   &
   &
   &
   &
   &
   &
   &
  \textbf{} &
  \textbf{} \\
\multirow{3}{*}{Tensor   MPG Own-lag} & 1 & \textbf{0.592}&	\textbf{0.911}&	1.012	&\textbf{0.559}&	\textbf{0.803}	&\textbf{0.912}	&0.923&	\textbf{0.874}   \\
 &
  2 &\textbf{0.622}&	\textbf{0.827}&	\textbf{0.951}&	\textbf{0.567}	&\textbf{0.794}	&\textbf{0.863}&	\textbf{0.883}	&\textbf{0.892}\\
 &
  4 &\textbf{0.651}	&\textbf{0.779}&	0.989	&\textbf{0.605}	&\textbf{0.797}	&\textbf{0.876}	&\textbf{0.894}	&0.937\vspace*{-2mm} \\
\multicolumn{1}{l}{} &
   &
   &
   &
   &
  \textbf{} &
   &
   &
   &
   \\
\multirow{3}{*}{Minnesota} &
  1 &0.632&	0.968	&\textbf{0.999}&	0.771&	0.943&	0.941&	0.938&	0.930\\
 &
  2 &0.673&	0.874&	0.997&	0.795&	0.902	&0.881&	0.942	&0.943\\
 &
  4 &0.717	&0.835	&1.032&	0.818	&0.910&	0.927	&0.952&	1.004  \vspace*{-2mm} \\
\multicolumn{1}{l}{} &
   &
   &
   &
   &
   &
   &
   &
   &
   \\
\multirow{3}{*}{NG} &
  1 &0.618&	0.986&	1.029&	0.636&	0.908&	0.957&	0.954&	0.921\\
 &
  2 &0.647&	0.866&	0.979&	0.650&	0.871&	0.887&	0.928&	0.9239 \\
 &
  4 &0.662&	0.783	&1.001	&0.655	&0.812	&0.892	&0.926	&0.955 \vspace*{-2mm} \\
\multicolumn{1}{l}{} &
   &
   &
   &
   &
   &
  \textbf{} &
  \textbf{} &
   &
   \\
\multirow{3}{*}{Horseshoe} &
  1 &0.635 & 1.214 & 1.069 & 0.580 & 0.879 & 1.019 & \textbf{0.917} & 0.915   \\
 &
  2 &0.652  & 0.949 & 0.978 & 0.584 & 0.850 & 0.922 & 0.912          & 0.909    \\
 &
  4 &0.662& 0.836 & 0.994 & 0.623 & 0.811 & 0.887 & 0.917          & \textbf{0.928}\\
  \hline

%% file: forecast_MAE_large_new.tex
Model &
  Horizon &
  MAE &
   &
   &
   &
   &
   &
   &
   \\
   \hline
 &
   &
  Joint &
  PAYEMS &
  CPIAUCSL &
  FEDFUNDS &
  GDP &
  UNRATE &
  GDPDEFL &
  GS10 \vspace*{0.5mm}\\
\multirow{3}{*}{Tensor   MGP}         & 1 & 0.607 & 1.105 & 0.995 & 0.571          & 0.861 & 0.925          & 0.943 & \textbf{0.889}        \\
 &
  2 &0.623                         & 0.877          & 0.939 & \textbf{0.556} & 0.823 & 0.852          & 0.889 & 0.903\\
 &
  4 &0.640                         & 0.814          & 1.006 & 0.597          & 0.842 & \textbf{0.834} & 0.894 & 0.943  \vspace*{-2mm} \\
\multicolumn{1}{l}{} &
   &
   &
   &
   &
   &
   &
   &
  \textbf{} &
  \textbf{} \\
\multirow{3}{*}{Tensor   MPG Own-lag} & 1 & \textbf{0.595} & \textbf{1.064} & 0.994 & \textbf{0.549} & 0.872 & \textbf{0.897} & 0.937          & 0.893   \\
 &
  2 &0.620 & 0.874 & 0.941 & 0.559          & 0.834 & \textbf{0.841} & 0.889          & \textbf{0.901}   \\
 &
  4 &0.640  & 0.814 & 1.001 & 0.593          & 0.859 & 0.839          & \textbf{0.893} & \textbf{0.938}   \vspace*{-2mm} \\
\multicolumn{1}{l}{} &
   &
   &
   &
   &
  \textbf{} &
   &
   &
   &
   \\
\multirow{3}{*}{Minnesota}            & 1 & 0.606 & 1.072          & \textbf{0.993} & 0.618          & \textbf{0.838} & 0.988 & \textbf{0.922} & 0.917     \\
 &
  2 &\textbf{0.618}                & \textbf{0.828} & \textbf{0.931} & 0.582          & \textbf{0.788} & 0.869 & \textbf{0.875} & 0.906 \\
 &
  4 &\textbf{0.629}                & \textbf{0.730} & \textbf{0.977} & \textbf{0.590} & \textbf{0.760} & 0.871 & 0.892          & 0.946 \vspace*{-2mm} \\
\multicolumn{1}{l}{} &
   &
   &
   &
   &
   &
   &
   &
   &
   \\
\multirow{3}{*}{NG} &
  1 &0.642 & 1.152 & 1.049 & 0.757 & 0.965 & 0.982 & 0.983 & 0.982  \\
 &
  2 &0.660 & 0.934 & 0.988 & 0.729 & 0.906 & 0.880 & 0.935 & 0.966 \\
 &
  4 &0.658 & 0.803 & 1.012 & 0.703 & 0.833 & 0.870 & 0.927 & 0.983 \vspace*{-2mm} \\
\multicolumn{1}{l}{} &
   &
   &
   &
   &
   &
   &
  \textbf{} &
   &
   \\
\multirow{3}{*}{Horseshoe} &
  1 &0.622 & 1.084 & 1.030 & 0.651 & 0.909 & 0.954 & 0.960 & 0.935  \\
 &
  2 &0.635 & 0.871 & 0.958 & 0.631 & 0.847 & 0.849 & 0.906 & 0.924   \\
 &
  4 &0.639& 0.764 & 0.994 & 0.631 & 0.797 & 0.849 & 0.909 & 0.950 \\
  \hline

%% file: forecast_RMSFE_medium_alternative.tex
Model                 & Horizon & MSFE          &                &                &                &                &                &                &                \\
\hline
                      &         & Joint          & PAYEMS            & CPIAUCSL       & FEDFUNDS           & GDP       & UNRATE         & GDPDEFL        & GS10          \vspace*{0.5mm} \\
\multirow{3}{*}{Tensor   MGP}         & 1 & 0.594 & 0.693 & 0.956 & 0.337 & 0.709 & 0.958 & 0.887 & 0.769  \\
                      & 2       & 0.641 & 0.653 & 0.950 & 0.349 & 0.627 & 0.837 & 0.745 & 0.825      \\
                      & 4       & 0.689& 0.673 & 0.994 & 0.493 & 0.641 & 0.796 & 0.828 & 0.900       \\
                      &         &                &                &                &                &                &                &                &             \vspace*{-2mm}     \\
\multirow{3}{*}{Tensor   MPG Own-lag} & 1 & 0.564 & \textbf{0.617} & 0.870          & 0.311          & 0.642          & 0.780 & 0.854          & 0.759\\
                      & 2       & 0.629                         & \textbf{0.615} & 0.928          & \textbf{0.333} & \textbf{0.600} & 0.781 & 0.735          & 0.825    \\
                      & 4       & 0.852	&0.656	&1.008	&0.442	&\textbf{0.623}	&\textbf{0.663}	&\textbf{0.821}	&\textbf{0.892} \\
                      &         &                &                &                &                &                &                &                &         \vspace*{-2mm}       \\
\multirow{3}{*}{Minnesota} & 1 & \textbf{0.546} & 0.649          & \textbf{0.860} & 0.309 & \textbf{0.633} & \textbf{0.702} & \textbf{0.732} & \textbf{0.729} \\
                      & 2       & \textbf{0.628}                         & 0.638          & 0.957          & 0.366 & 0.625          & \textbf{0.702} & 0.741          & \textbf{0.822}\\
                      & 4       & \textbf{0.670}                         & \textbf{0.647} & 0.996          & 0.442 & 0.632 & 0.719 & 0.835          & 0.914 \\
                      &         &                &                &                &                &                &                & \textbf{}      &    \vspace*{-2mm}         \\
\multirow{3}{*}{NG} & 1       & 0.567 & 0.684 & 0.893 & \textbf{0.246} & 0.745 & 0.734 & 0.791 & 0.774  \\
                      & 2       & 0.637  & 0.702 & 0.942 & 0.337          & 0.713 & 0.736 & 0.741 & 0.844 \\
                      & 4       & 0.681  & 0.696 & \textbf{0.993 }& \textbf{0.435} & 0.678 & 0.766 & 0.824 & 0.914  \\
                      &         &                &                &                &                &                &                &                &          \vspace*{-2mm}      \\
\multirow{3}{*}{Horseshoe}   & 1       & 0.584 & 1.621 & 0.926          & 0.270 & 0.734 & 0.876 & 0.798          & 0.752   \\
                      & 2       & 0.644  & 0.899 & \textbf{0.921} & 0.334 & 0.682 & 0.742 & \textbf{0.731} & 0.830    \\
                      & 4       & 0.691                         & 0.739 & 1.016          & \textbf{0.435} & 0.675 & 0.771 & 0.823          & 0.895\\
                      \hline

%% file: forecast_MAE_medium_alternative.tex
Model &
  Horizon &
  MAE &
   &
   &
   &
   &
   &
   &
   \\
   \hline
&
   &
  Joint &
  PAYEMS &
  CPIAUCSL &
  FEDFUNDS &
  GDP &
  UNRATE &
  GDPDEFL &
  GS10 \vspace*{0.5mm}\\
\multirow{3}{*}{Tensor   MGP}         & 1 & 0.535 & 0.959 & 0.998          & 0.567 & 0.841 & 1.011 & 0.929 & 0.882   \\
 &
  2 & 0.562                         & 0.871 & 0.941          & 0.564 & 0.798 & 0.949 & 0.882 & \textbf{0.897} \\
 &
  4 &0.584                         & 0.863 & \textbf{0.994} & 0.649 & 0.821 & 0.961 & 0.896 & 0.943 \vspace*{-3mm} \\
\multicolumn{1}{l}{} &
   &
   &
   &
   &
   &
   &
   &
  \textbf{} &
  \textbf{} \\
\multirow{3}{*}{Tensor   MPG Own-lag} & 1 & \textbf{0.524} & 0.909 & 0.973          & \textbf{0.538} & 0.825          & 0.941 & 0.923          & \textbf{0.871}   \\
 &
  2 &\textbf{0.554}                         & 0.820 & \textbf{0.933} & \textbf{0.546} & \textbf{0.789} & 0.922 & \textbf{0.877} & \textbf{0.897}\\
 &
  4 &0.572                                  & 0.805 & 0.995          & \textbf{0.609} & 0.811          & 0.944 & \textbf{0.886} & \textbf{0.935}\vspace*{-2mm} \\
\multicolumn{1}{l}{} &
   &
   &
   &
   &
  \textbf{} &
   &
   &
   &
   \\
\multirow{3}{*}{Minnesota} &
  1 &0.527 & \textbf{0.875} & \textbf{0.942} & 0.638 & \textbf{0.815} & \textbf{0.895} & \textbf{0.911} & 0.911\\
 &
  2 &0.563                         & \textbf{0.779} & 0.959          & 0.649 & \textbf{0.789 }         & \textbf{0.829} & 0.909          & 0.930\\
 &
  4 &\textbf{0.569}                & \textbf{0.739} & 0.999          & 0.634 & \textbf{0.775} & \textbf{0.864} & 0.914          & 0.967   \vspace*{-2mm} \\
\multicolumn{1}{l}{} &
   &
   &
   &
   &
   &
   &
   &
   &
   \\
\multirow{3}{*}{NG} &
  1 &0.541 & 0.950 & 0.978 & 0.578 & 0.885 & 0.932 & 0.940 & 0.934\\
 &
  2 &0.570 & 0.832 & 0.962 & 0.627 & 0.857 & 0.861 & 0.916 & 0.939  \\
 &
  4 &0.576 & 0.772 & 0.996 & 0.643 & 0.810 & 0.898 & 0.912 & 0.968 \vspace*{-2mm} \\
\multicolumn{1}{l}{} &
   &
   &
   &
   &
   &
  \textbf{} &
  \textbf{} &
   &
   \\
\multirow{3}{*}{Horseshoe} &
  1 &0.543 & 1.236 & 0.978 & 0.547 & 0.866 & 0.983 & 0.946 & 0.905   \\
 &
  2 &0.566  & 0.952 & 0.933 & 0.582 & 0.811 & 0.868 & 0.899 & 0.911   \\
 &
  4 &0.577  & 0.815 & 1.003 & 0.623 & 0.797 & 0.889 & 0.907 & 0.960\\
  \hline

%% file: forecast_ALPL_medium_alternative.tex
Model &
  Horizon &
  ALPL &
   &
   &
   &
   &
   &
   &
   \\
   \hline
 &
   &
  Joint &
  PAYEMS &
  CPIAUCSL &
  FEDFUNDS &
  GDP &
  UNRATE &
  GDPDEFL &
  GS10 \vspace*{0.5mm}\\
\multirow{3}{*}{Tensor   MGP} &
  1 & -12.857 & 0.162 & 0.144          & 0.667 & 0.153 & 0.049 & 0.121 & 0.171 \\
 &
  2 &-14.551  & 0.433 & 0.240          & 0.663 & 0.217 & 0.195 & 0.140 & 0.151  \\
 &
  4 &\textbf{-15.728}  & 0.682 & \textbf{0.182} & 0.472 & 0.188 & 0.203 & 0.109 & 0.099  \vspace*{-2mm}\\
 &
   &
  \textbf{} &
   &
   &
   &
   &
   &
   &
   \\
\multirow{3}{*}{Tensor MPG Own-lag} &
  1 &\textbf{-12.471} & \textbf{0.193} & 0.186          & 0.711          & \textbf{0.184} & 0.118 & 0.130          & 0.179\\
 &
  2 &\textbf{-14.384}  & 0.460          & \textbf{0.254} & \textbf{0.691} & \textbf{0.242} & 0.223 & \textbf{0.146} & 0.147\\
 &
  4 &-15.734 & 0.692          & 0.180          & \textbf{0.507} & \textbf{0.198} & 0.202 & \textbf{0.112} & 0.103 \vspace*{-2mm}\\
 &
   &
   &
   &
   &
   &
   &
   &
   &
   \\
\multirow{3}{*}{Minnesota} &
  1 &-13.033 & 0.156 & \textbf{0.203} & 0.585 & 0.176 & \textbf{0.170} & \textbf{0.173} & \textbf{0.200}  \\
 &
  2 &-15.146& \textbf{0.483} & 0.220          & 0.566 & 0.221 & \textbf{0.309} & 0.139          & \textbf{0.152}  \\
 &
  4 &-16.326  & \textbf{0.814} & 0.143          & 0.454 & 0.187 & \textbf{0.289} & 0.097          & 0.091   \vspace*{-2mm}\\
 &
   &
   &
   &
   &
   &
   &
   &
   &
   \\
\multirow{3}{*}{NG} &
  1 &-13.447 & 0.168 & 0.174 & \textbf{0.719} & 0.138 & 0.166 & 0.157 & 0.176\\
 &
  2 &-15.612  & 0.471 & 0.218 & 0.629          & 0.181 & 0.304 & 0.140 & 0.142 \\
 &
  4 &-16.826 & 0.795 & 0.149 & 0.475          & 0.165 & 0.266 & 0.109 & 0.091  \vspace*{-2mm}\\
 &
   &
   &
   &
   &
   &
   &
   &
   &
   \\
\multirow{3}{*}{Horseshoe} &
  1 &-13.918 & -0.136 & 0.143 & 0.665 & 0.106 & 0.064 & 0.127 & 0.186  \\
 &
  2 &-15.666  & 0.244  & 0.227 & 0.622 & 0.164 & 0.232 & 0.125 & \textbf{0.152} \\
 &
  4 &-16.742  & 0.635  & 0.142 & 0.496 & 0.141 & 0.192 & 0.103 & \textbf{0.104}\\
  \hline

%% file: forecast_RMSFE_medium_order_invariant.tex

Model                 & Horizon & MSFE          &                &                &                &                &                &                &                \\
\hline
                      &         & Joint          & PAYEMS            & CPIAUCSL       & FEDFUNDS           & GDP       & UNRATE         & GDPDEFL        & GS10          \vspace*{0.5mm} \\
\multirow{3}{*}{Tensor   MGP}         & 1 & 0.700 &	0.629&	\textbf{0.849}	&0.354	&0.661	&0.740	&0.860	&0.769      \\
                      & 2       & \textbf{0.765}&	0.626&	0.954&	0.358&	\textbf{0.613}	&0.675&	\textbf{0.736}&	0.839        \\
                      & 4       & \textbf{0.844}&	0.668&	0.999&	0.475&	0.636&	\textbf{0.681}&	0.827	&0.909        \\
                      &         &                &                &                &                &                &                &                &             \vspace*{-2mm}     \\
\multirow{3}{*}{Tensor   MPG Own-lag} & 1 & \textbf{0.670}&	\textbf{0.583}&	0.855&	\textbf{0.285}&	\textbf{0.638}&	\textbf{0.664}&	\textbf{0.692}&	\textbf{0.752}\\
                      & 2       & 0.772&	\textbf{0.609}&	0.945&	0.357&	0.640&	\textbf{0.666}&	0.737&	0.825 \\
                      & 4       & 0.849&	\textbf{0.666}&	\textbf{0.993}&	0.457&	\textbf{0.625}&	0.717&	0.834	&0.901 \\
                      &         &                &                &                &                &                &                &                &         \vspace*{-2mm}       \\
\multirow{3}{*}{Minnesota} & 1 & 0.799	&1.531	&1.090&	0.345&	0.845&	1.088&	0.819&	0.768  \\
                      & 2       & 0.816&	0.859	&\textbf{0.941}&	0.374&	0.674	&0.863&	\textbf{0.736}&	\textbf{0.816}\\
                      & 4       & 0.862&	0.715&	1.018&	0.436&	0.649&	0.749&	0.827&	\textbf{0.895}  \\
                      &         &                &                &                &                &                &                & \textbf{}      &    \vspace*{-2mm}         \\
\multirow{3}{*}{NG} & 1       & 0.700&	0.675&	0.918&	0.298&	0.721&	0.796&	0.773&	0.763  \\
                      & 2       & 0.782&	0.705&	0.957&	\textbf{0.342}&	0.679&	0.809&	0.750&	0.828 \\
                      & 4       & 0.859&	0.701	&0.997&	0.437&	0.650	&0.764&	0.831&	0.896   \\
                      &         &                &                &                &                &                &                &                &          \vspace*{-2mm}      \\
\multirow{3}{*}{Horseshoe}   & 1       & 0.717	&0.967&	0.946&	0.303&	0.765&	0.881&	0.803	&0.787  \\
                      & 2       & 0.800&	0.811	&0.983&	0.356&	0.686	&0.870&	0.761&	0.838   \\
                      & 4       & 0.862&	0.741	&1.028&	\textbf{0.433}&	0.649	&0.776&	\textbf{0.820}&	\textbf{0.895}\\
                      \hline

%% file: forecast_MAE_medium_order_invariant.tex
Model &
  Horizon &
  MAE &
   &
   &
   &
   &
   &
   &
   \\
   \hline
&
   &
  Joint &
  PAYEMS &
  CPIAUCSL &
 FEDFUNDS &
  GDP &
  UNRATE &
  GDPDEFL &
  GS10 \vspace*{0.5mm}\\
\multirow{3}{*}{Tensor   MGP}         & 1 & 0.594&	0.934&	0.950&	0.590&	\textbf{0.824}&	0.942&	0.941&	0.892  \\
 &
  2 & 0.623&	0.831&	0.957&	0.573&	0.802&	0.853&	0.889&	0.905 \\
 &
  4 &0.652&	0.805&	1.006&	0.623&	0.802&	0.878&	0.903&	0.947 \vspace*{-3mm} \\
\multicolumn{1}{l}{} &
   &
   &
   &
   &
   &
   &
   &
  \textbf{} &
  \textbf{} \\
\multirow{3}{*}{Tensor   MPG Own-lag} & 1 & \textbf{0.573}&	\textbf{0.864}&	\textbf{0.932}&	0.537&	0.843&	\textbf{0.889}&	\textbf{0.876}&	\textbf{0.885}   \\
 &
  2 &\textbf{0.619}&	\textbf{0.798}&	0.947&	\textbf{0.569}&	0.833	&\textbf{0.840}&	0.891&	0.897\\
 &
  4 &0.648&	0.778&	0.991&	0.598&	0.801&	0.889&	0.905&	0.941\vspace*{-2mm} \\
\multicolumn{1}{l}{} &
   &
   &
   &
   &
  \textbf{} &
   &
   &
   &
   \\
\multirow{3}{*}{Minnesota} &
  1 &0.628&	1.166&	1.042&	0.583&	0.867&	1.026&	0.932&	0.898\\
 &
  2 &0.633&	0.831&	\textbf{0.943}&	0.570&	\textbf{0.790}&	0.886&	\textbf{0.880}&	\textbf{0.895}\\
 &
  4 &\textbf{0.644}&	\textbf{0.738}&	\textbf{0.988}&	\textbf{0.573}&	\textbf{0.765}&	\textbf{0.872}&	\textbf{0.896}&	\textbf{0.939}  \vspace*{-2mm} \\
\multicolumn{1}{l}{} &
   &
   &
   &
   &
   &
   &
   &
   &
   \\
\multirow{3}{*}{NG} &
  1 &0.595&	0.926&	0.986&	0.590&	0.873	&0.949&	0.941	&0.912\\
 &
  2 &0.630&	0.809&	0.983&	0.593&	0.842&	0.883&	0.931&	0.917 \\
 &
  4 &0.651&	0.756&	1.013&	0.612&	0.796&	0.886&	0.924&	0.952 \vspace*{-2mm} \\
\multicolumn{1}{l}{} &
   &
   &
   &
   &
   &
  \textbf{} &
  \textbf{} &
   &
   \\
\multirow{3}{*}{Horseshoe} &
  1 &0.598&	0.988&	0.998&	0.561&	0.858&	0.964&	0.948&	0.920   \\
 &
  2 &0.632&	0.839&	0.981&	0.572&	0.830&	0.912&	0.919&	0.921    \\
 &
  4 &0.649&	0.773&	1.014&	0.588&	0.783&	0.892	&0.912&	0.944\\
  \hline

%% file: forecast_ALPL_medium_order_invariant.tex
Model &
  Horizon &
  ALPL &
   &
   &
   &
   &
   &
   &
   \\
   \hline
 &
   &
  Joint &
  PAYEMS &
  CPIAUCSL &
  FEDFUNDS &
  GDP &
  UNRATE &
  GDPDEFL &
  GS10 \vspace*{0.5mm}\\
\multirow{3}{*}{Tensor   MGP} &
  1 & -16.338&	0.095&	\textbf{0.161}&	0.519&	0.110&	0.098&	0.067&	0.125 \\
 &
  2 &\textbf{-18.191}&	0.371&	0.195&	0.538	&\textbf{0.163}&	0.270&	0.097&	0.091  \\
 &
  4 &\textbf{-19.604}&	0.616&	0.135&	0.375&	0.130&	\textbf{0.276}&	\textbf{0.056}&	0.051 \vspace*{-2mm}\\
 &
   &
  \textbf{} &
   &
   &
   &
   &
   &
   &
   \\
\multirow{3}{*}{Tensor MPG Own-lag} &
  1 &\textbf{-15.844}&	\textbf{0.140}&	0.155&	\textbf{0.584}	&\textbf{0.123}&	\textbf{0.147}&	\textbf{0.136}&	\textbf{0.139}\\
 &
  2 &-18.405&	\textbf{0.410}&	0.201&	\textbf{0.556}	&\textbf{0.163}&	\textbf{0.294}	&\textbf{0.100}&	\textbf{0.100}\\
 &
  4 &-20.173&	\textbf{0.640}	&0.151&	\textbf{0.404}	&\textbf{0.143}&	0.265&	\textbf{0.056}&	0.058 \vspace*{-2mm}\\
 &
   &
   &
   &
   &
   &
   &
   &
   &
   \\
\multirow{3}{*}{Minnesota} &
  1 &-17.936&	-0.266&	0.040&	0.405&	-0.023&	-0.058&	0.051&	0.125  \\
 &
  2 &-18.884&	0.169&	\textbf{0.211}&	0.434&	0.054&	0.153&	0.070&	\textbf{0.100}  \\
 &
  4 &-20.065&	0.509&	\textbf{0.163}&	0.308&	0.024&	0.190&	0.032&	0.051  \vspace*{-2mm}\\
 &
   &
   &
   &
   &
   &
   &
   &
   &
   \\
\multirow{3}{*}{NG} &
  1 &-18.505	&0.082&	0.112&	0.512&	0.073&	0.099&	0.095&	0.135 \\
 &
  2 &-21.045&	0.358&	0.169&	0.503&	0.118&	0.245&	0.095&	\textbf{0.100}  \\
 &
  4 &-22.763&	0.605&	0.119&	0.369&	0.099&	0.258&	0.054&	0.055 \vspace*{-2mm}\\
 &
   &
   &
   &
   &
   &
   &
   &
   &
   \\
\multirow{3}{*}{Horseshoe} &
  1 &-18.359&	0.001&	0.090&	0.496&	0.049&	0.058&	0.074&	0.123\\
 &
  2 &-21.059	&0.316	&0.158&	0.498&	0.091&	0.209&	0.076&	0.099 \\
 &
  4 &-22.015&	0.590&	0.132&	0.357	&0.090&	0.259	&0.053&	\textbf{0.059}\\
  \hline

%% file: forecast_MSFE_large_order_invariant.tex
Model &
  Horizon &
  MSFE &
   &
   &
   &
   &
   &
   &
   \\
   \hline
&
   &
  Joint &
  PAYEMS &
  CPIAUCSL &
 FEDFUNDS &
  GDP &
  UNRATE &
  GDPDEFL &
  GS10 \vspace*{0.5mm}\\
\multirow{3}{*}{Tensor   MGP}         & 1 & 0.724&	1.158&	0.928&	0.341&	0.723&	0.920&	0.881&	\textbf{0.765}  \\
 &
  2 & 0.764&	0.809&	0.940&	0.356&	0.619&	0.744&	0.737&	\textbf{0.815} \\
 &
  4 &0.814&	0.714&	1.007&	0.473&	0.642&	0.677&	\textbf{0.820}&	0.895 \vspace*{-3mm} \\
\multicolumn{1}{l}{} &
   &
   &
   &
   &
   &
   &
   &
  \textbf{} &
  \textbf{} \\
\multirow{3}{*}{Tensor   MPG Own-lag} & 1 & \textbf{0.650}&	\textbf{0.696}&	\textbf{0.830}&	\textbf{0.294}&	\textbf{0.717}&	\textbf{0.670}&	\textbf{0.711}&	0.770   \\
 &
  2 &\textbf{0.742}&	\textbf{0.609}&	0.937&	\textbf{0.336}&	\textbf{0.599}&	\textbf{0.649}&	0.733&	0.817\\
 &
  4 &\textbf{0.803}&	\textbf{0.625}&	\textbf{0.997}&	\textbf{0.436}&	\textbf{0.626}&	\textbf{0.665}&	0.833&	0.905\vspace*{-2mm} \\
\multicolumn{1}{l}{} &
   &
   &
   &
   &
  \textbf{} &
   &
   &
   &
   \\
\multirow{3}{*}{Minnesota} &
  1 &0.747&	1.621&	0.995&	0.374&	0.808&	1.049&	0.818&	0.791\\
 &
  2 &0.775&	0.910&	0.988&	0.378&	0.683&	0.838&	0.762&	\textbf{0.815}\\
 &
  4 &0.819&	0.747&	1.001&	0.446&	0.648&	0.741&	0.822&	\textbf{0.886}  \vspace*{-2mm} \\
\multicolumn{1}{l}{} &
   &
   &
   &
   &
   &
   &
   &
   &
   \\
\multirow{3}{*}{NG} &
  1 &0.728&	1.097&	0.952&	0.360&	0.820&	0.975&	0.881&	0.798\\
 &
  2 &0.769&	0.819&	0.958&	0.372&	0.674&	0.807&	0.741&	0.827 \\
 &
  4 &0.813&	0.732&	1.012&	0.462&	0.653&	0.731&	0.830&	0.921 \vspace*{-2mm} \\
\multicolumn{1}{l}{} &
   &
   &
   &
   &
   &
  \textbf{} &
  \textbf{} &
   &
   \\
\multirow{3}{*}{Horseshoe} &
  1 &0.712&	0.943&	0.916&	0.339&	0.789&	0.906&	0.893&	0.782   \\
 &
  2 &0.760&	0.776&	\textbf{0.936}&	0.362&	0.677&	0.786&	\textbf{0.726}&	0.846    \\
 &
  4 &0.810&	0.717&	1.007&	0.470&	0.647&	0.719&	0.836&	0.896\\
  \hline

%% file: forecast_MAE_large_order_invariant.tex
Model &
  Horizon &
  MAE &
   &
   &
   &
   &
   &
   &
   \\
   \hline
&
   &
  Joint &
  PAYEMS &
  CPIAUCSL &
 FEDFUNDS &
  GDP &
  UNRATE &
  GDPDEFL &
  GS10 \vspace*{0.5mm}\\
\multirow{3}{*}{Tensor   MGP}         & 1 & 0.605&	1.173&	0.957&	0.580	&\textbf{0.826}&	0.900&	0.942&	\textbf{0.891}  \\
 &
  2 & 0.619&	0.906&	\textbf{0.937}&	0.566	&\textbf{0.778}&	0.822&	\textbf{0.885}&	\textbf{0.891} \\
 &
  4 &0.637&	0.807&	1.007&	0.619&	0.778&	\textbf{0.823}&	\textbf{0.892}&	0.938 \vspace*{-3mm} \\
\multicolumn{1}{l}{} &
   &
   &
   &
   &
   &
   &
   &
  \textbf{} &
  \textbf{} \\
\multirow{3}{*}{Tensor   MPG Own-lag} & 1 & \textbf{0.578}&	\textbf{0.902}&	\textbf{0.934}&	\textbf{0.548}&	0.891&	\textbf{0.863}&	\textbf{0.894}&	0.892   \\
 &
  2 &\textbf{0.614}&	\textbf{0.794}&	0.954&	\textbf{0.550}&	0.803&	\textbf{0.806}&	0.895&	0.892\\
 &
  4 &0.640&	0.779&	1.003&	0.585&	0.819&	0.856&	0.905&	0.942\vspace*{-2mm} \\
\multicolumn{1}{l}{} &
   &
   &
   &
   &
  \textbf{} &
   &
   &
   &
   \\
\multirow{3}{*}{Minnesota} &
  1 &0.620&	1.248&	1.017&	0.590&	0.875&	1.025&	0.953&	0.904\\
 &
  2 &0.625&	0.891&	0.963&	0.560&	0.809&	0.880&	0.914&	0.889\\
 &
  4 &0.633&	0.776&	\textbf{0.974}&	\textbf{0.569}&	0.778&	0.872&	0.918&	\textbf{0.930}  \vspace*{-2mm} \\
\multicolumn{1}{l}{} &
   &
   &
   &
   &
   &
   &
   &
   &
   \\
\multirow{3}{*}{NG} &
  1 &0.612&	1.070&	0.996&	0.610&	0.886&	0.983&	0.968&	0.912\\
 &
  2 &0.625&	0.844&	0.970&	0.592	&0.812&	0.859	&0.908&	0.893 \\
 &
  4 &0.634&	\textbf{0.756}&	1.001&	0.623	&0.784	&0.863&	0.909&	0.943 \vspace*{-2mm} \\
\multicolumn{1}{l}{} &
   &
   &
   &
   &
   &
  \textbf{} &
  \textbf{} &
   &
   \\
\multirow{3}{*}{Horseshoe} &
  1 &0.605&	1.025&	0.970&	0.606&	0.874&	0.954&	0.980&	0.911   \\
 &
  2 &0.620&	0.826&	0.946&	0.586&	0.811&	0.844&	0.901&	0.904   \\
 &
  4 &\textbf{0.632}&	0.757&	0.995&	0.618	&\textbf{0.776}&	0.845&	0.921&	0.939\\
  \hline

%% file: forecast_ALPL_large_order_invariant.tex
Model &
  Horizon &
  ALPL &
   &
   &
   &
   &
   &
   &
   \\
   \hline
 &
   &
  Joint &
  PAYEMS &
  CPIAUCSL &
  FEDFUNDS &
  GDP &
  UNRATE &
  GDPDEFL &
  GS10 \vspace*{0.5mm}\\
\multirow{3}{*}{Tensor   MGP} &
  1 & -29.578&	-0.141&	0.096&	0.416&	0.054&	0.093&	0.004&	0.116 \\
 &
  2 &\textbf{-38.282}&	0.235&	0.203&	0.449&	0.116&	0.269&	0.044&	0.077  \\
 &
  4 &\textbf{-44.550}&	0.528&	0.156&	0.267&	0.076&	0.275&	0.004&	0.013 \vspace*{-2mm}\\
 &
   &
  \textbf{} &
   &
   &
   &
   &
   &
   &
   \\
\multirow{3}{*}{Tensor MPG Own-lag} &
  1 &\textbf{-27.850}&	\textbf{0.059}&	\textbf{0.162}&	\textbf{0.429}&	\textbf{0.062}&	\textbf{0.144}&	\textbf{0.086}&	\textbf{0.121}\\
 &
  2 &-41.932&	\textbf{0.384}&	\textbf{0.224}&	\textbf{0.452}&	\textbf{0.134}&	\textbf{0.306}	&\textbf{0.066}&	\textbf{0.089}\\
 &
  4 &-54.636&	\textbf{0.632}&	\textbf{0.167}&	\textbf{0.317}&	\textbf{0.093}&	\textbf{0.294}&	\textbf{0.019}&	\textbf{0.027} \vspace*{-2mm}\\
 &
   &
   &
   &
   &
   &
   &
   &
   &
   \\
\multirow{3}{*}{Minnesota} &
  1 &-39.395&	-0.276&	0.089&	0.300&	0.008&	-0.040&	0.027&	0.106  \\
 &
  2 &-45.602&	0.157&	0.199&	0.350&	0.070&	0.168&	0.042&	0.079  \\
 &
  4 &-52.200&	0.495&	0.157&	0.220&	0.040&	0.189&	0.009&	0.017  \vspace*{-2mm}\\
 &
   &
   &
   &
   &
   &
   &
   &
   &
   \\
\multirow{3}{*}{NG} &
  1 &-71.183&	-0.134&	0.082&	0.244&	-0.028&	-0.009&	-0.029&	0.074 \\
 &
  2 &-84.193&	0.240&	0.190&	0.295&	0.050&	0.194&	0.016&	0.040 \\
 &
  4 &-92.085&	0.549&	0.152&	0.163&	0.022&	0.231&	-0.021&	-0.024 \vspace*{-2mm}\\
 &
   &
   &
   &
   &
   &
   &
   &
   &
   \\
\multirow{3}{*}{Horseshoe} &
  1 &-74.687&	-0.081&	0.096&	0.282&	-0.017&	0.020&	-0.028&	0.071\\
 &
  2 &-86.357&	0.280&	0.202&	0.317&	0.054&	0.213&	0.025&	0.036 \\
 &
  4 &-100.722&	0.570&	0.148&	0.182&	0.028&	0.239&	-0.023&	-0.020\\
  \hline

%% file: data_appendix.tex
\begingroup
\footnotesize
\begin{longtable}{llllllccccc}
\multicolumn{2}{l}{\textbf{Slow Variables}}       &        &  &  &                                 &               &              &          &                \\
\hline \hline
            & \textbf{Name}                       &        &  &  & \textbf{Description}                     & \textbf{Medium} & \makecell{\textbf{Medium} \\ \textbf{(Alternative)}} & \textbf{Large} & \textbf{Category} & \textbf{Code} \\
            \hline
1           & RPI                        &        &  &  & Real Personal Income            & x   & x          & x           & 1        & 5              \\
2           & W875RX1                    &        &  &  & RPI ex. Transfers               & x  &            & x            & 1        & 5              \\
3           & INDPRO                     &        &  &  & IP Index                        &           &x    & x            & 1        & 5              \\
4           & GDP                        &        &  &  & Real Gross Domestic Product     & x    & x         & x            & 1        & 5              \\
5           & GDPDEFL                    &        &  &  & GDP deflator                    & x   & x          & x            & 1        & 6              \\
6           & \multicolumn{2}{l}{PCECC96} &  &  & Real PCE                        & x  &x            & x            & 2        & 5              \\
7           & \multicolumn{2}{l}{CMRMTSPLx}       &  &  & Real M\& T Sales                & x  & x           & x            & 2        & 5              \\
8           & RSAFSx                    &        &  &  & Retail and Food Services Sales  & x  & x           & x            & 2        & 5              \\
9           & HWI                        &        &  &  & Help-Wanted Index for US        &      &         & x           & 3        & 2              \\
10          & HWIURATIO                  &        &  &  & Help Wanted to Unemployed ratio &     &          & x            & 3        & 2              \\
11          &CIVPART                    &        &  &  & Civilian Labor Force            &      &         & x            & 3        & 5              \\
12          & UNRATE                     &        &  &  & Civilian Unemployment Rate      & x  & x           & x            & 3        & 2              \\
13          & PAYEMS                    &        &  &  & All Employees: Total nonfarm    & x   & x          & x           & 3        & 5              \\
14          & \multicolumn{2}{l}{CES0600000007}   &  &  & Hours: Goods-Producing          &   &            & x            & 3        & 5              \\
15          & OILPRICEx                  &        &  &  & Crude Oil Prices: WTI           &    & x           & x            & 4        & 5              \\
16          & CPIAUCSL                   &        &  &  & CPI: All Items                  & x    & x         & x            & 4        & 6              \\
            &                            &        &  &  &                                 &               &              &          &                \\
 \hline
 \multicolumn{9}{l}{Transformation code: 2 - first differences; 5 - first differences of logarithms; 6 - second differences of logarithms.} &  \\
 \hline
\caption{Description of slow variables.}
\label{data appendix}
\end{longtable}
\endgroup

%% file: data_appendix_1.tex
\begingroup
\footnotesize
\begin{longtable}{llllllccccc}
\multicolumn{2}{l}{\textbf{Fast   Variables}}
&        &  &  &                                 &               &              &          &       &         \\
\hline \hline
            & \textbf{Name }                      &        &  &  & \textbf{Description}                     & \textbf{Medium} & \makecell{\textbf{Medium} \\ \textbf{(Alternative)}}  & \textbf{Large}  & \textbf{Category} & \textbf{Code} \\
            \hline
1          & FEDFUNDS                   &        &  &  & Effective Federal Funds Rate     & x  & x           & x            & 5        & 2              \\
2          & CP3Mx                      &        &  &  & 3-Month AA Comm. Paper Rate     & x    &         & x            & 5        & 2              \\
3          & TB3MS                      &        &  &  & 3-Month T-bill                  &      & x         & x            & 5        & 2              \\
4          & TB6MS                      &        &  &  & 6-Month T-bill                  &   & x            & x            & 5        & 2              \\
5          & GS1                        &        &  &  & 1-Year T-bond                   &    & x           & x            & 5        & 2              \\
6         & GS5                        &        &  &  & 5-Year T-bond                   &    &           & x            & 5        & 2              \\
7         & GS10                       &        &  &  & 10-Year T-bond                  & x      & x     & x            & 5        & 2              \\
8          & AAA                        &    &    &  &  Aaa Corporate Bond Yield        &      &         & x            & 5        & 2              \\
9         & BAA                        &     &   &  & Baa Corporate Bond Yield        &     &          & x            & 5        & 2              \\
10          & M1SL                       &        &  &  & M1 Money Stock                  &     &          & x            & 6        & 5              \\
11         & M2SL                           &   &  &  & M2 Money Stock                  &      &         & x            & 6        & 5              \\
12         & M2REAL                     &    &    &  &  Real M2 Money Stock             &    & x           & x            & 6        & 5              \\
13         & BUSLOANS                   &        &  &  & Commercial and Industrial Loans & x   &          & x            & 6        & 5              \\
14          & NONREVSL                   &        &  &  & Total Nonrevolving Credit       & x     &         & x            & 6        & 5              \\
15         & INVEST                     &        &  &  & Securities in Bank Credit       &    &            & x            & 6        & 5              \\
16          & CONSPI                     &        &  &  & Credit to PI ratio              & x  & x        & x            & 6        & 2              \\
17         & S\&P 500                   &        &  &  & S\&P 500                        &     & x          & x            & 7        & 5              \\
18         & S\&P: indust               &        &  &  & S\&P Industrial                 &      &         & x            & 7        & 5              \\
19          & \multicolumn{2}{l}{S\&P div yield}  &  &  & S\&P Divident yield             &      &         & x            & 7        & 2              \\
20         & \multicolumn{2}{l}{S\&P PE ratio}   &  &  & S\&P Price/Earnings ratio       &      &         & x            & 7        & 5              \\
21          & EXSZUSx                    &        &  &  & Switzerland / U.S. FX Rate      & x     &        & x            & 8        & 5              \\
22         & EXJPUSx                    &        &  &  & Japan / U.S. FX Rate            & x    &         & x            & 8        & 5              \\
23         & EXUSUKx                    &        &  &  & U.S. / U.K. FX Rate             & x   & x          & x            & 8        & 5              \\
24          & EXCAUSx                    &        &  &  & Canada / U.S. FX Rate           & x   &          & x            & 8        & 5              \\
            &                            &        &  &  &                                 &               &              &          &  \\
 \hline
 \multicolumn{9}{l}{Transformation code: 2 - first differences; 5 - first differences of logarithms; 6 - second differences of logarithms.} &  \\
 \hline
\caption{Description of fast variables.}
\label{data appendix}
\end{longtable}
\endgroup